%% file: MWM31like.tex
\documentclass[useAMS, usenatbib, 10pt]{mnras}

\usepackage{graphicx}
\usepackage{comment}
\usepackage{newtxtext,newtxmath}

\def \MSUN{\rm M_{\odot}}
\def \RVIR{R_{\rm vir}}
\def \RTWOC{R_{\rm 200c}}

\def \MTWOC{M_{\rm 200c}}

\def \MS{M_{*}}

\title[TNG50 MW/M31 analogs]{Milky Way and Andromeda analogs from the TNG50 simulation}

\author[Pillepich et al.] {Annalisa Pillepich$^1$$^,$\thanks{E-mail: pillepich@mpia-hd.mpg.de}, Diego Sotillo-Ramos$^1$, Rahul Ramesh$^2$, Dylan Nelson$^2$, Christoph Engler$^1$
\newauthor
Vicente Rodriguez-Gomez$^3$, Martin Fournier$^1$, Martina Donnari$^1$, Volker Springel$^4$, and Lars Hernquist$^5$
\\
$^{1}$ Max-Planck-Institut f{\"u}r Astronomie, K{\"o}nigstuhl 17, 69117 Heidelberg, Germany\\
$^{2}$ Universität Heidelberg, Zentrum für Astronomie, Institut für theoretische Astrophysik, Albert-Ueberle-Str. 2, 69120 Heidelberg, Germany\\
$^{3}$ Instituto de Radioastronom\'ia y Astrof\'isica, Universidad Nacional Aut\'onoma de M\'exico, Apdo. Postal 72-3, 58089 Morelia, Mexico \\
$^{4}$ Max Planck Institut f\"ur Astrophysik, Karl-Schwarzschild-Stra\ss e 1, D-85748 Garching bei M\"unchen, Germany\\
$^{5}$ Center for Astrophysics $|$ Harvard \& Smithsonian, 6P Garden St., Cambridge, MA 02138, USA
}

\begin{document}

\maketitle

\begin{abstract}
We present the properties of Milky Way- and Andromeda-like (MW/M31-like) galaxies simulated within TNG50, the highest-resolution run of the IllustrisTNG suite of $\Lambda$CDM magneto-hydrodynamical simulations. We introduce our fiducial selection for MW/M31 analogs, which we propose for direct usage as well as for reference in future analyses.
TNG50 contains 198 MW/M31 analogs, i.e. galaxies with stellar disky morphology, with a stellar mass in the range of $M_* = 10^{10.5 - 11.2}~\MSUN$, and within a MW-like Mpc-scale environment at $z=0$. These are resolved with baryonic (dark matter) mass resolution of $8.5\times10^4\MSUN$ ($4.5\times10^5\MSUN$) and $\sim150$ pc of average spatial resolution in the star-forming regions: we therefore expand by many factors (2 orders of magnitude) the sample size of cosmologically-simulated analogs with similar ($\times 10$ better) numerical resolution. 
The majority of TNG50 MW/M31 analogs at $z=0$ exhibit a bar, 60 per cent are star-forming, the sample includes 3 Local Group (LG)-like systems, and a number of galaxies host one or more satellites as massive as e.g. the Magellanic Clouds. 
Even within such a relatively narrow selection, TNG50 reveals a great diversity in galaxy and halo properties, as well as in past histories. Within the TNG50 sample, it is possible to identify several simulated galaxies whose integral and structural properties are consistent, one or more at a time, with those measured for the Galaxy and Andromeda. 
With this paper, we document and release a series of broadly applicable data products that build upon the IllustrisTNG public release and aim to facilitate easy access and analysis by public users. These include datacubes across snapshots ($0 \le z \le 7$) for each TNG50 MW/M31-like galaxy, and a series of value-added catalogs that will be continually expanded to provide a convenient and up to date community resource.
\end{abstract}



\section{Introduction}
\label{sec:intro}

\subsection{The landscape of $\Lambda$CDM simulations of MW/M31 analogs}

About a decade ago, the first numerical realizations of thin stellar disks in galaxies with mass similar to our Milky Way (MW) have marked the beginning of a new era for hydrodynamical simulations within the full $\Lambda$CDM cosmological context. 

Cosmological {\it zoom-in} galaxy simulations such as Eris \citep[][]{Guedes.2011} and others (see Table~\ref{tab:sims}) have shown that it is possible to solve the ``angular momentum problem'' or ``overcooling catastrophe'' \citep[e.g.][]{Ceverino.2009} by introducing appropriate (stellar) feedback and
star formation recipes \citep{Agertz.2011, Stinson.2013}. A number of successes followed, with simulations capable of returning large, disk-dominated galaxies that resemble our Galaxy or Andromeda in many respects \citep{Martig.2012,Few.2012, Marinacci.2014a}.

Since then, roughly a dozen cosmological zoom-in projects have aimed to simulate the entire Local Group  \citep[e.g. CLUES, APOSTLE, ELVIS on FIRE, HESTIA, with or without constrained initial conditions:][]{Nuza.2014, Sawala.2016, Fattahi.2016, Garrison-Kimmel.2019, Garrison-Kimmel.2019b,Libeskind.2020} or to enhance the sophistication of the underlying physical models by reaching stellar particle mass resolution of a few thousand solar masses and below \citep[e.g. Latte and the FIRE-2 suite, Auriga Level 3, VINTERGATAN, DC Justice League:][]{Wetzel.2016, Hopkins.2018, Grand.2017, Agertz.2021, Applebaum.2021}. However, only recently, other projects have taken steps towards simulated samples larger than one or, at most, a handful of galaxies \citep[e.g. ][]{Buck.2020} -- the Auriga \citep{Grand.2017} and the ARTEMIS \citep{Font.2020} suites currently comprise 30 and 45, respectively, MW-like galaxies and their satellite systems with stellar particle mass resolution of a few tens of thousand solar masses.

In the meantime, cosmological {\it uniform-resolution} large-volume simulations like Illustris \citep{Vogelsberger.2014a, Vogelsberger.2014b, Genel.2014, Sijacki.2015} and EAGLE \citep{Schaye.2015, Crain.2015} have demonstrated for the first time (see also other suites in Table~\ref{tab:sims}) that it is possible, starting from a well-posed cosmological model and from basic physical laws, to approximately reproduce the fundamental properties, diversity, and scaling relations of entire galaxy populations. This has been done by simultaneously following the co-evolution of thousands of galaxies within the hierarchical growth of structure scenario as well as the processes therein (such as gas cooling, star formation, stellar and supermassive black hole (SMBH) feedback), down to subgrid spatial scales of approximately one kiloparsec and stellar particle masses of $\sim$ a million solar masses. 

Importantly, large-volume cosmological simulations \citep{Vogelsberger.2020} have shown that it is possible to realize, within the same simulation and physical model, galaxies that are star-forming, blue and disky as well as galaxies that are quiescent, red and with elliptical or spheroidal stellar morphology. This has been recently achieved, not only in a qualitative, but also in a quantitative sense by the IllustrisTNG simulations\footnote{\url{https://www.tng-project.org}} \citep[hereafter TNG:][]{Nelson.2018, Nelson.2019, Nelson.2019b, Pillepich.2018b,Pillepich.2019, Springel.2018, Marinacci.2018, Naiman.2018}, which have been demonstrated to be broadly consistent with observations of the separation of galaxies in star-formation states, optical colors, and stellar morphologies \citep[e.g.][]{Nelson.2018, Rodriguez-Gomez.2019, Huertas-Company.2019, Donnari.2021b, Zanisi.2021, Varma.2021, Guzman.2023}.

\begin{figure}
	\includegraphics[width=\columnwidth]{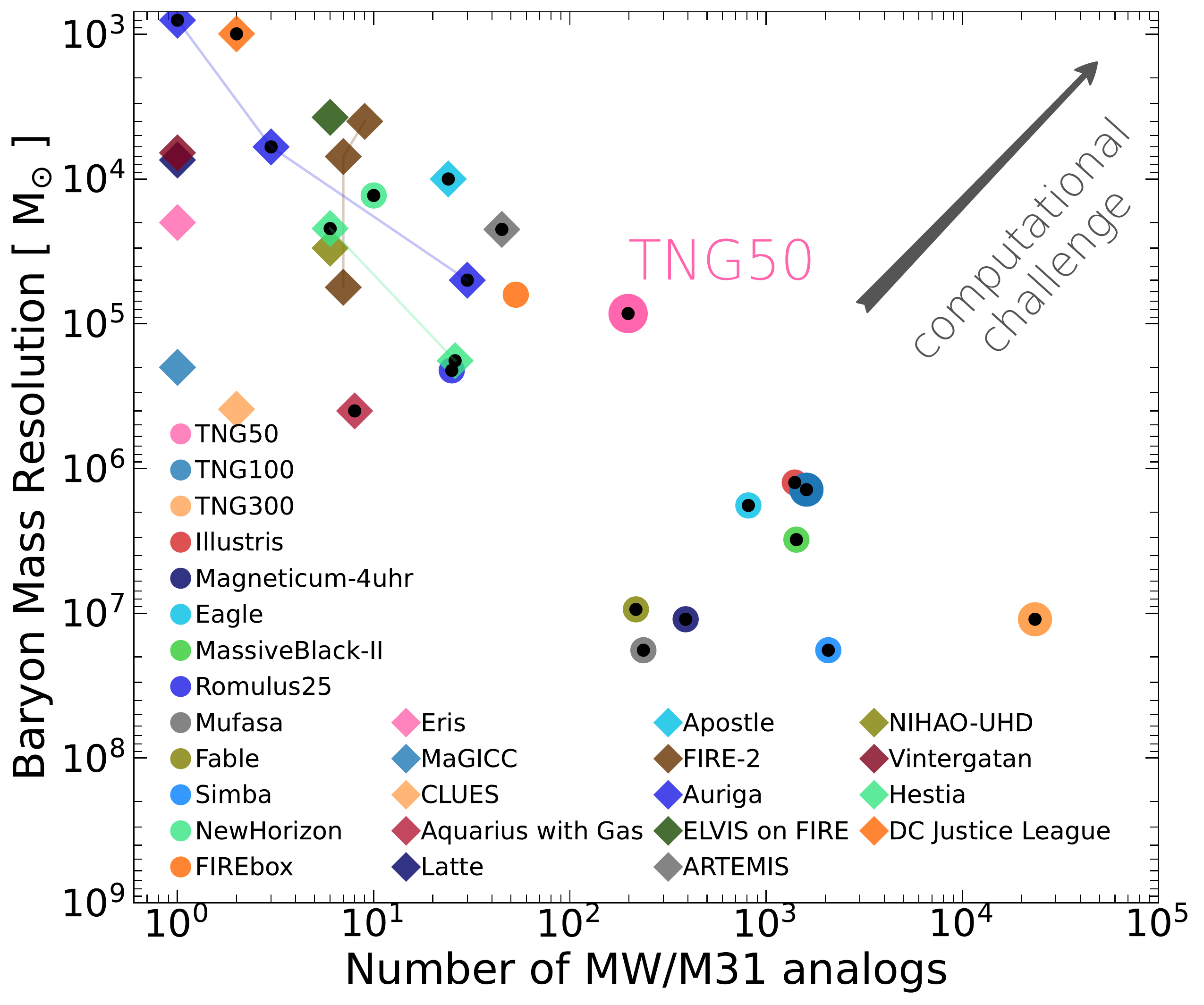}
    \caption{{\bf The landscape of $\Lambda$CDM cosmological simulations of MW/M31-like galaxies as of 2023.} We contrast numerical projects in terms of baryonic mass resolution vs. number of simulated MW/M31 analogs. The TNG50 simulation this paper focuses on (magenta circle) is compared to state-of-the-art cosmological zoom-in simulations (diamonds) and so-called box or uniform-resolution large-volume simulations (circles): this comparison is deliberately similar to that in Fig. 1 of \citet{Nelson.2019b} but focuses on MW/M31-like systems: see Table~\ref{tab:sims} for more details on these numerical models and the definition of MW/M31 analogs. Black dots in the symbols denote inclusion of SMBH feedback. TNG50 bridges the gap between zoom-in and box simulations, with about 200 MW/M31-like galaxies at $z=0$.}
    \label{fig:comparison}
\end{figure}

Large-volume cosmological hydrodynamical galaxy simulations naturally return galaxies that at $z\sim0$ are similar to our Galaxy and Andromeda e.g. in total stellar mass and global stellar morphology (see Fig.~\ref{fig:comparison}). Importantly, whereas their underlying galaxy-formation models are designed with the goal of producing realistic galaxies and galaxy populations across types, mass scales, and cosmic epochs, they are not tuned in any explicit way to replicate in detail the global or inner properties of any particular observed galaxy. Moreover, the effects of the astrophysical processes included therein have been tested against large numbers of observables and of observed galaxies across the mass spectrum, i.e. not only for MW-type galaxies.

Therefore, such simulations allow us to study the formation of disk galaxies with mass similar to our Galaxy as well as Andromeda. They also enable us to contextualize, from a theoretical perspective, what our closest most massive and best known galaxies can tell us about galaxy formation and evolution in general, and how representative our own Galaxy and Andromeda are of the general galaxy population. Importantly for the scope of this paper, they permit to do so without the biases of the a-priori selections that have often been imposed in the case of zoom-in simulations. In fact, zoom-in projects have typically been designed by intentionally selecting what halo or system to (re)simulate. The selection criteria to simulate MW-like galaxies -- and rarely, if not ever, M31-like galaxies -- have differed from simulation to simulation and have included constraints on the past assembly and merger history, environment, and dark matter (DM) host halo mass: see a partial summary of the various ``galaxy''-selection criteria adopted in past zoom-in projects in Table 1 of \citet{Engler.2021}. This necessary and a-priori choice can affect the resulting scientific messages, but this limitation can now be overcome thanks to full-volume simulations such as TNG50 \citep{Pillepich.2019, Nelson.2019}, which this paper is based upon.

The clearest drawback of large-volume cosmological hydrodynamical galaxy simulations in comparison to zoom-in ones is their limited numerical resolution (see Fig.~\ref{fig:comparison}). However, recent numerical efforts such as the aforementioned TNG50, \textsc{NewHorizon} \citep{Dubois.2021}, and FIREbox \citep{Feldmann.2022} are bridging the computational gap between zoom-in and uniform-box simulations. Whereas the latter two include more sophisticated and realistic treatments of the interstellar medium (ISM) and of feedback from stars, the TNG50 simulation remains an unmatched combination of volume and resolution, with about two hundred MW/M31-like galaxies at $z=0$. It has an average spatial resolution in the star-forming regions of galaxies of about 150 pc, and its data is already fully publicly released \citep{Nelson.2019}. Crucially, and differently from FIREbox and from most high-resolution simulations of MW/M31-like galaxies \citep{Wetzel.2016, Agertz.2021, Applebaum.2021}, TNG50 includes SMBH feedback, in addition to magnetic fields, a shock finder, and a model for the production sites of Europium. 

Finally, the outcome of the TNG50 simulation has already been studied and contrasted to observational findings across a wide range of applications, showing an unprecedented level of realism: from the inner stellar and star-formation structural morphologies of galaxies \citep[e.g.][]{Pillepich.2019, Zanisi.2021, NelsonE.2021, Motwani.2022} to the presence of small-scale, cold gas structures in the otherwise-hot circumgalactic medium (CGM) of massive ellipticals \citep{Nelson.2020} and of X-ray eROSITA-like bubbles above and below the disk of MW/M31-like galaxies \citep{Pillepich.2021}. All this makes the TNG50 simulation a unique and rich laboratory to study the formation and evolution of MW/M31-like galaxies in the full cosmological context and to interpret observational findings of the Local Group's galaxies, within the boundary conditions of the $\Lambda$CDM scenario.

\subsection{The scope of this paper}

The scope of this paper is twofold. From a practical perspective, we aim to provide the community with an all-encompassing guide to facilitate the interpretation and usage of the data pertaining the MW/M31-like galaxies simulated within TNG50. From a scientific viewpoint, we aim to provide a definite statement as to the maturity of current state-of-the-art cosmological simulations at returning MW/M31-like galaxies. More specifically, the scientific goals are:
\begin{itemize}
\item to benchmark the state-of-the-art of simulated MW/M31-like galaxies as of 2023, i.e. about a decade after the first realistic numerical realizations of late-type, disk-like galaxies; 
\item to assess to what extent today's simulations can reproduce properties of the Galaxy and Andromeda and which properties may be the most problematic to explain;
\item to expand upon the discussion on how analogs of our closest massive galaxies can be selected and on how different selections can point to different implications (Section~\ref{sec:selectionthoughts});
\item to identify and hence provide a carefully constructed and well-studied sample of TNG50 MW/M31 analogs for direct usage or for reference in future scientific explorations (Section~\ref{sec:selection});
\item to demonstrate the diversity that 13.8 billion years of cosmic evolution imparts on galaxy populations selected, at $z=0$, within a relative narrow range of galaxy stellar mass, morphology, and environment (Section~\ref{sec:props});
\item to summarize and showcase the environmental, integral and structural properties of TNG50 MW/M31-like galaxies, across spatial scales and matter components (Section~\ref{sec:props});
\item to gauge how special the Galaxy and Andromeda are given the distribution of predictions made by simulation models.
\end{itemize}

To do so, in practice we also:

\begin{itemize}
\item document and release a series of broadly-useful data products that build upon the TNG public release and aim to maximally facilitate access and analysis by the public user (Appendix~\ref{sec:data});
\item collect in a `live' table the existing cosmological $\Lambda$CDM hydrodynamical simulations available in the literature (Table~\ref{tab:sims});
\item build a reference list for more detailed analyses and studies of TNG50 MW/M31-like galaxies (Table~\ref{tab:papers}).
\end{itemize}

All the material introduced and discussed here will be maintained and released online.\footnote{\url{https://www.tng-project.org/data/milkyway+andromeda/}}

After reviewing key aspects of the TNG50 simulation in Section~\ref{sec:tng50}, we present and discuss our identification of MW/M31-like galaxies in Sections~\ref{sec:selectionthoughts} and \ref{sec:selection}. We hence provide in Section~\ref{sec:props} a zeroth-order overview of a selection of their environmental, integral and structural properties, by focusing mostly on $z=0$ results and by contextualizing the TNG50 outcome with observational constraints available for our Galaxy and Andromeda. 

It should be noted that this is not a paper whereby the full outcome of the TNG50 model (i.e. of the TNG model) is assessed globally, first and foremost because the comparison to two individual galaxies would not alone ensure realism of the underlying galaxy formation and physical model in general. Rather, we refer the reader to the many comparison papers where the outcome of the TNG simulations, and of TNG50 in particular, have been contrasted to the results from galaxy population surveys: see Section 5 of \cite{Nelson.2019} and Section 2.1 of \cite{Pillepich.2021} for partial compendiums, and online for an up-to-date list.\footnote{\url{https://www.tng-project.org/results/}} In this paper, instead, we juxtapose properties of the TNG50 MW/M31-like galaxies to those measured or inferred from observations of the Galaxy and Andromeda, and by doing so, we also contrast and highlight the similarity and differences between the two real-Universe galaxies.
Data products and added-value catalogs are described in Appendix~\ref{sec:data} and we conclude and summarize in Section~\ref{sec:summary}.

\begin{table*}
\input{table_sims}
\end{table*}

\section{TNG50 and its MW/M31-like galaxies}

\subsection{The TNG50 simulation}
\label{sec:tng50}
TNG50 \citep[aka TNG50-1;][]{Pillepich.2019, Nelson.2019b} is a cosmological magneto-hydrodynamical (MHD) simulation of the formation and evolution of galaxies in a $\Lambda$CDM universe \citep{Planck2015_Cosmology}. It is run with the moving-mesh code {\sc AREPO} \citep{Springel.2010, Weinberger.2020} and uses the fiducial TNG galaxy formation model \citep{Weinberger.2017, Pillepich.2018}. 

TNG50 evolves cold dark matter (CDM), gas, stars, super massive black holes (SMBHs), and magnetic fields in a cubic volume of 51.7 comoving Mpc a side from $z=127$ to the current epoch: it includes thousands of galaxies across different environments, from galaxies at the centers (or orbiting within) galaxy clusters as massive as $10^{14.3}\MSUN$, to dwarf galaxies in isolation or in small groups. For example, within the TNG50 volume at $z=0$, there are about 37'000 galaxies with $\ge10^6\MSUN$ in stars and about 900'000 haloes and subhaloes with total gravitationally-bound mass of $\ge10^8\MSUN$. 

Those are resolved with a target gas mass cell and stellar particle mass of $8.5\times10^4\MSUN$ (DM particle mass of $4.6\times10^5\MSUN$), placing TNG50 at the confluence between large-volume and zoom-in hydrodynamical galaxy simulations: see Fig.~\ref{fig:comparison} and more details on numerical resolution in Section~\ref{sec:res} and in \citet{Nelson.2019, Nelson.2019b, Nelson.2020} and \citet{Pillepich.2019, Pillepich.2021}. In fact, TNG50 is accompanied by three lower-resolution runs with the same initial conditions and unchanged physical model: these are called TNG50-2, TNG50-3, and TNG50-4 and enable quantitative convergence studies. TNG50-1 and TNG50 are equivalent names for the same simulation, where the former is used only when specifically referring to its resolution level. Each run of the series also has a DM-only i.e. gravity-only counterpart, which allows to study the effects of baryonic processes on the properties of CDM. 

The astrophysical processes accounted for in TNG50 include primordial and metal-line cooling down to 10$^4$ K; heating from a spatially homogeneous UV/X-ray background, and from localized X-ray radiation from SMBHs; density-threshold based star formation; galactic-scale stellar-driven outflows; seeding, growth and feedback of SMBHs; evolution and amplification of cosmic magnetic fields via ideal MHD; stellar evolution, stellar mass loss, and enrichment from AGB, SNII, and SNIa tracked via nine elements (H, He, C, N, O, Ne,  Mg, Si, Fe) in addition to a subgrid model for mergers of neutrons stars as injection sites of r-process material. All details can be found in the TNG method papers \cite{Weinberger.2017, Pillepich.2019} and in the TNG50 introduction papers \citep{Pillepich.2019, Nelson.2019b}. 

Importantly, in TNG50, the cold and dense phase of the ISM is not directly resolved. Instead, star-forming gas is treated with an effective model \citep[the two-phase model of][with ensuing effective equation of state for gas above 0.13 atoms cm$^{-3}$]{Springel.2003}, and all gas in the simulation has temperatures of $\gtrsim10^4$ K. Stellar particles represent mono-age stellar populations with a Chabrier initial mass function. Although the TNG50 model for stellar feedback neglects small-scale interactions due to the hydrodynamically-decoupled wind particle scheme, even for SN-driven winds, the calculations do capture disordered motions indirectly induced by stellar feedback outflows, i.e. complex galactic-scale fountain flows. On the other hand, the energy injections from the central SMBHs are directly coupled and affect the coldest and densest gas in galaxies, producing multiphase gas ejecta with complex properties \citep{Nelson.2019b, Pillepich.2021}.

Overall, the solution to the coupled equations of gravity, MHD and galactic astrophysics in an expanding universe is such that phenomena like cosmological gas accretion, gas outflows, shocks, galaxy mergers, galaxy interactions, gravitational  instabilities, resonances, gravitational tides, ram pressure, gravitational heating, etc. are all emergent and self-consistent phenomena within the simulation.


\begin{figure*}
	\includegraphics[width=1\textwidth]{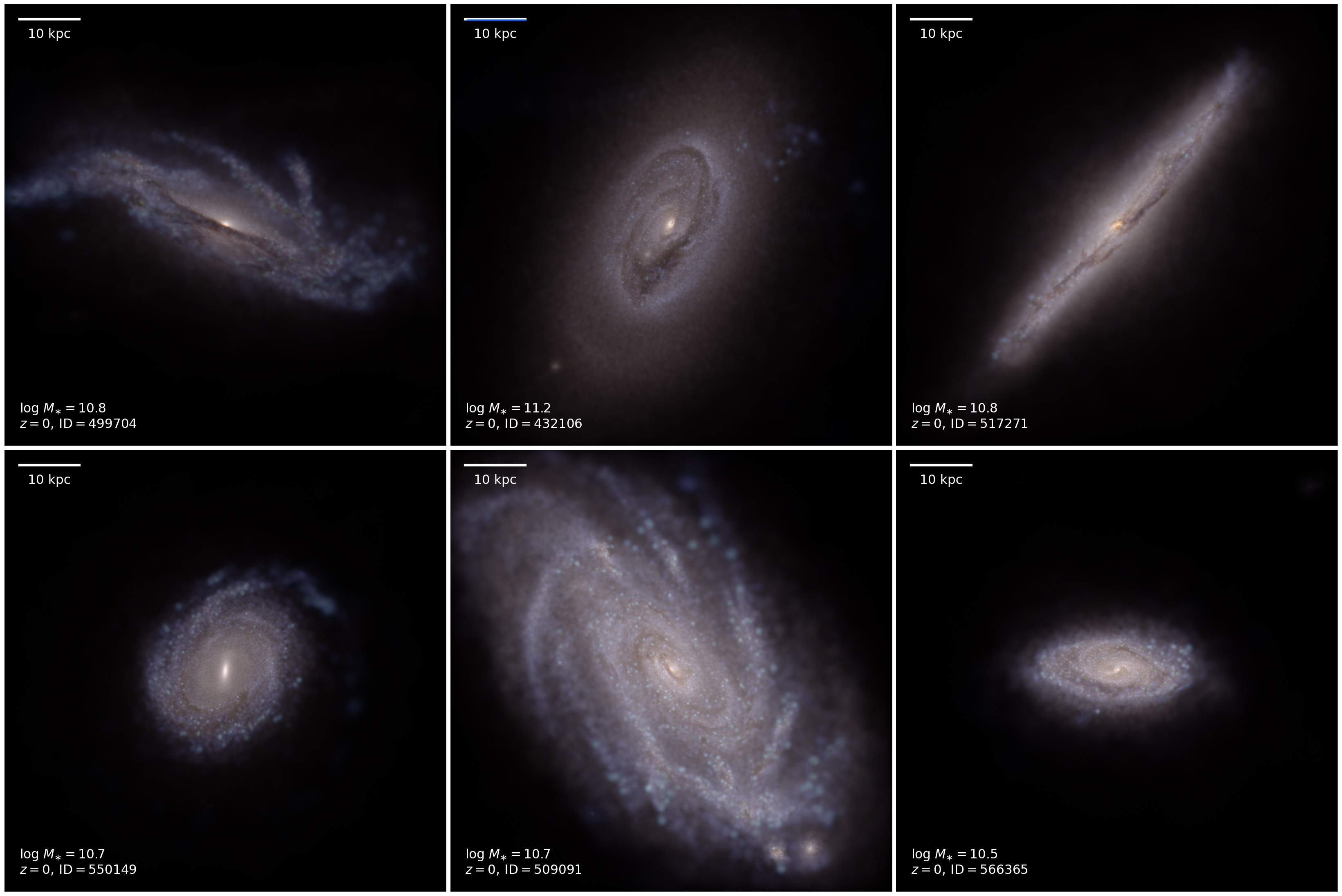}
    \caption{\textbf{Example galaxies from the TNG50 simulation at $z=0$ with stellar mass and/or disk structure similar to the Milky Way or to Andromeda, among the 198 TNG50 MW/M31-like galaxies.} These are shown in random projections, across 70 kpc a side, in a composite stellar light image of three HST/ACS bands (F435W, F606W, F775W; 0.049 arcsec/pixel) including radiative transfer in post-process with \textsc{SKIRT} \citep{Camps.2015} and placing the galaxies at $z\sim0.08$. A large gallery of all the TNG50 MW/M31-like galaxies can be found at \url{https://www.tng-project.org/explore/gallery/pillepich23a}.}
    \label{fig:examples}
\end{figure*}


\subsection{Preliminary considerations: what is an analog of the Milky Way or Andromeda?}
\label{sec:selectionthoughts}
Given its large cosmological volume, TNG50 simulates thousands of galaxies across 6 orders of magnitude in stellar mass and encompassing diverse environments. Therefore, the very first goal of this paper is to select, among the many in TNG50  at $z=0$, galaxies that resemble our Galaxy and Andromeda. 

In past works, a number of relevant properties have been adopted to define ``analogs'' of e.g. the Milky Way: total or stellar mass in a given range, stellar disky morphology, quiet merger history either recently or even since $z\sim1.5$, but also specific requirements on bulge-to-total (B/T) ratios, thin vs. thick morphological disks, the presence of companions like the Large and/or Small Magellanic Clouds or of a Sagittarius-like stream, etc. 

The definition of what constitutes an analog of our Galaxy and/or Andromeda strictly depends on the intended methodology and on the scientific question at hand. For example, in the context of large galaxy surveys, it is common to select as MW/M31-like galaxies those with stellar mass in a given range, in addition to constraints on star formation rate (SFR), B/T ratio, exponential disk scale length, the presence of a bar and spiral features, or a combination thereof \citep[e.g.][]{Boardman.2020b}. In the context of large-scale DM-only simulations as well as DM-only and hydrodynamical zoom-in projects aimed to simulate MW-like haloes or galaxies, the common approach is to focus on haloes with total mass of about $10^{12}\MSUN$ \citep[e.g.][and zoom-in simulations of Table~\ref{tab:sims}]{Stewart.2008, Springel.2008}. There, further constraints have also often been placed on the more or less recent merger history of the halo to zoom on \citep[e.g.][and  Section~\ref{sec:intro}]{Scannapieco.2009}. On the other hand, if the scientific goal is to understand e.g. the gas mass flows in the central molecular zone of {\it our own Galaxy} \citep{Tress.2020}, then it may be of the essence to work with model galaxies (and, in this case, necessarily idealized and not cosmological) whose stellar disk structure is as similar as possible to that of the Milky Way, including the height of the stellar disk, the length and features of the bar and bulge, the dynamics of the gas, and the geometry of the spiral arms. 

In fact, any galaxy (real or simulated) looked at in ever-specific and in an ever-growing number of details may be, to some extent, unique. This has been shown with observational data, where overly-strict definitions of ``analog'' have return very small, if not vanishing, samples of MW or M31-like candidates even among parent samples of thousands external galaxies \citep[such as with SDSS:][]{Fraser.2019, Boardman.2020, Boardman.2020b}. 

The Galaxy and Andromeda themselves present different observed features: Andromeda exhibits a thicker and more extended disk, a larger and more massive bulge, a more massive central SMBH, a younger and more metal-rich stellar halo population, more numerous tidal streams and satellites, more numerous and more massive globular clusters, and overall a larger total stellar mass and lower SFR than the Milky Way -- see next sections and figures.
Yet, when considered in the broader context of the whole galaxy population, from ultra faint dwarfs to brightest cluster galaxies, the commonality between the two is also apparent.

\subsection{Fiducial selection of TNG50 MW/M31-like galaxies}
\label{sec:selection}
With this paper, we advocate for a selection of MW/M31-like galaxies from cosmological (magneto)hydrodynamical galaxy simulations that is fully based on observable global galaxy properties and that is general enough to allow for galaxies with properties similar to either the Galaxy or Andromeda or both. 

We prescribe a selection of MW/M31-like galaxies from TNG50 based on three sets of criteria evaluated at the $z=0$ snapshot: galaxy stellar mass, stellar morphology, and Mpc-scale environment. In fact, the TNG50 simulation has not been designed with the purpose of reproducing the MW nor M31; rather, its success lies on that a diverse and realistic galaxy population is realized within the simulated volume, across the mass spectrum and across cosmic epochs and environments (see Section~\ref{sec:intro}). By adopting constraints on global galaxy properties and assuming the TNG50 universe is sufficiently realistic, we get the opportunity to assess how representative the Galaxy and Andromeda are in comparison to the rest of the realm of galaxies and how diverse in the detail galaxies can be (if at all) even when selected within a relatively-narrow range of global characteristics. By selecting based on observable properties, we can, firstly, more directly connect to what is known observationally; and secondly, we can provide simulation-based implications for unobservable and hard-to-constraint characteristics, such as total halo mass.

In particular, in this paper, we require that the following conditions are simultaneously fulfilled for TNG50 $z=0$ galaxies to be selected as MW/M31 analogs:
 
\begin{enumerate}
\item Stellar mass: the galaxy stellar mass is in the following range: $M_*(<30{\rm kpc}) = 10^{10.5-11.2}\MSUN$. \\

\item Stellar morphology: a disk-like stellar morphology needs to be manifest, including the presence of spiral arms. This is implemented by requiring that one {\it or} the other following criterion is fulfilled:
\begin{itemize} 
\item the minor-to-major axis ratio of the galaxy's stellar mass distribution is smaller than 0.45, with the latter measured between 1 and 2 times the stellar half-mass radius: $c/a$ [stellar mass at 1-2 $r_{*1/2}] \le 0.45$;\\
\item the galaxy appears of disky shape and exhibits spiral arms by visual inspection of three-band images of the simulated galaxies in edge-on and face-on projections.
\end{itemize}
\item Environment: no other galaxy with stellar mass $\ge 10^{10.5}\MSUN$ is within 500 kpc distance and the total mass of the halo host is smaller than that typical of massive groups, i.e. $\MTWOC(\rm{host})< 10^{13}\MSUN$.
\end{enumerate}

We expand on these criteria below. 

Our definition returns 198 MW/M31 analogs in TNG50, of which a handful of examples are shown in Fig.~\ref{fig:examples}. The selection is made among all TNG50 gravitationally-bound objects identified by the \textsc{Subfind} algorithm \citep{Springel.2001} at $z=0$.
Inspired by \cite{Boardman.2020b}, we visualize our selection in terms of Venn diagrams in the upper panel of Fig.~\ref{fig:selection}. The IDs of the 198 TNG50 MW/M31-like galaxies selected here are released with this paper (Section~\ref{sec:data}) and we invite colleagues to make use of this selection as is or as inspiration for different or more tailored subsamples.

\subsubsection{On the stellar mass criterion (i)}
\label{sec:criterion_mstars}
The stellar mass criterion enunciated in  Section~\ref{sec:selection} encompasses the current estimates available for the Milky Way and Andromeda, including statistical errorbars and systematic effects: see Fig.~\ref{fig:selection_depth}, top panel, and considerations below. 

In fact, for Andromeda, stellar mass estimates (evaluated either within 30 kpc or by summing up the stellar mass of bulge and disk) vary between $7.8\times10^{10}$ and $1.5\times10^{11} \MSUN$, i.e. approximately in the $10^{10.9-11.2}\MSUN$ range \citep{Geehan.2006, Barmby.2006, Chemin.2009,Tamm.2012, Sick.2015}. 

The stellar mass estimates of the Galaxy are lower, and vary between $4.0\times10^{10}$ and $7.3\times10^{10} \MSUN$, including the quoted $1\sigma$ errors \citep{Licquia.2015, Licquia.2016,Flynn.2006, Bland-Hawthorn.2016}. Accounting for the fact that different authors quote constraints across different components (e.g. bulge+disk vs. total stellar mass), that in the simulation a Chabrier IMF is assumed, that we measure the mass within a spherical radius of 30 kpc as our fiducial galaxy stellar mass measure (in analogy to some available constraints for M31), and that the latter can be approximately up to 0.1 dex smaller than the total stellar mass (at least for TNG50 galaxies $\lesssim10^{11}\MSUN$), these estimates translate into the approximate range of $10^{10.5-10.9}\MSUN$ in stars. 

\begin{figure*}
		\includegraphics[width=1.1\columnwidth]{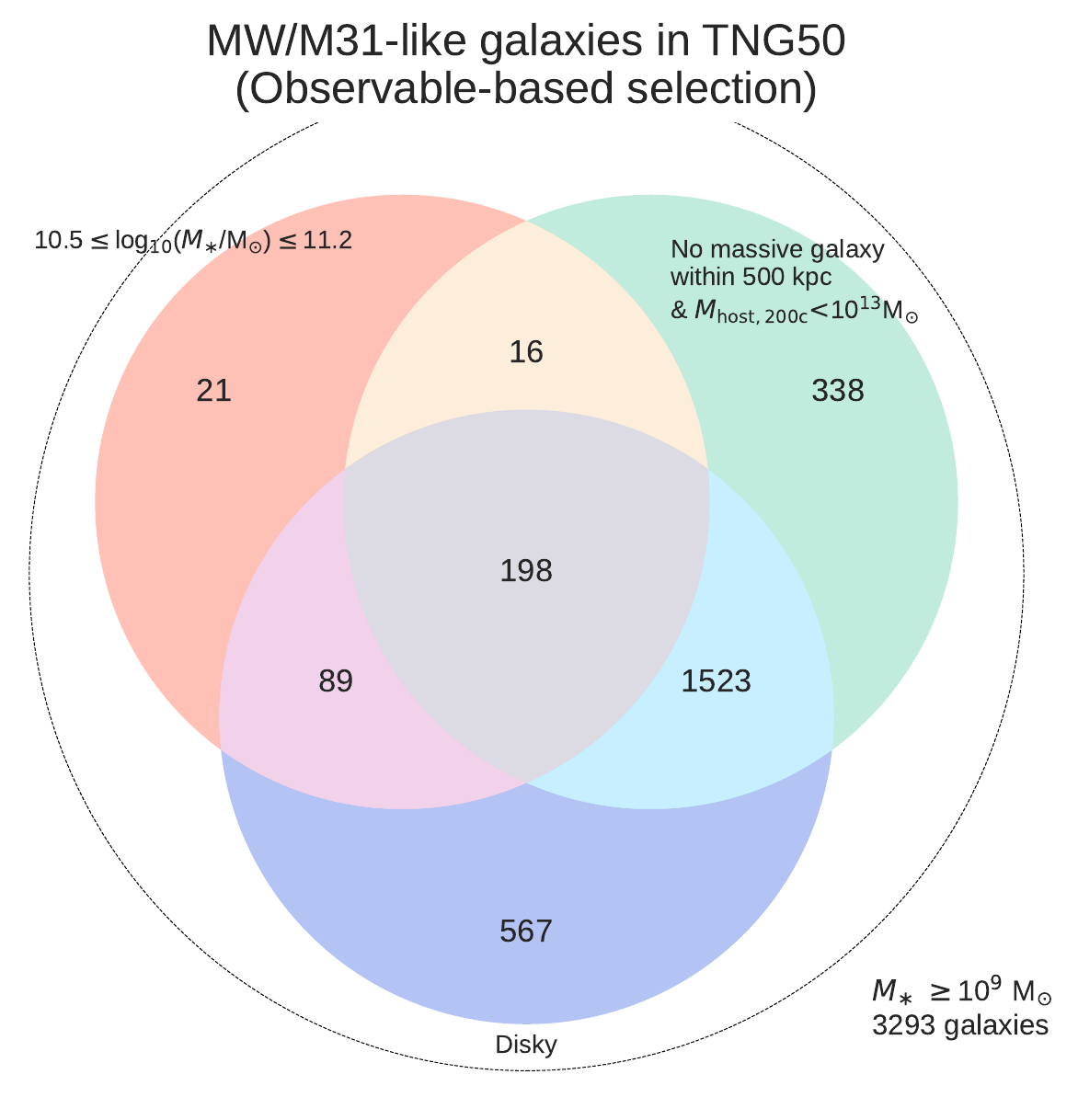}
		\includegraphics[width=\columnwidth]{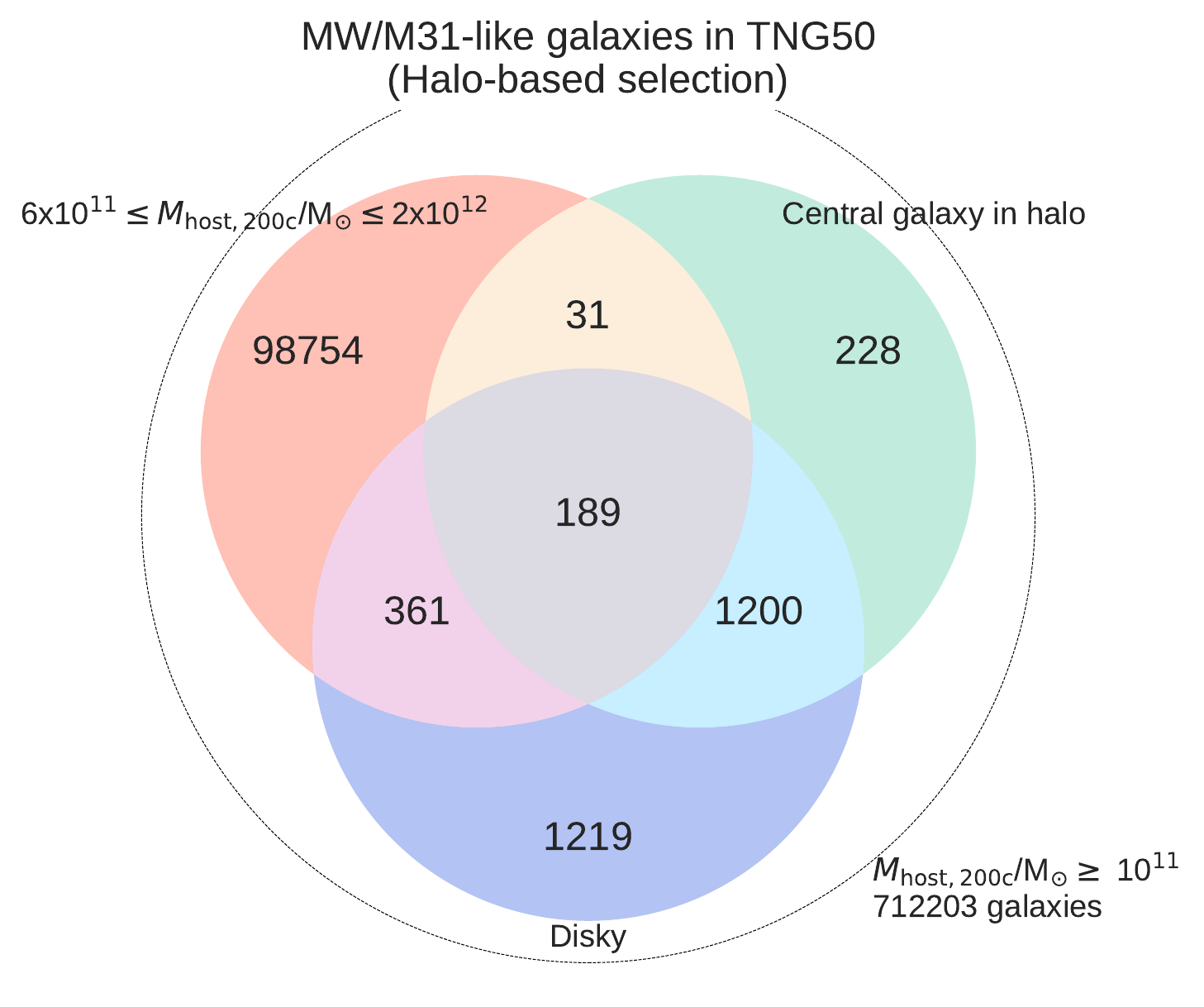}
		\includegraphics[width=\columnwidth]{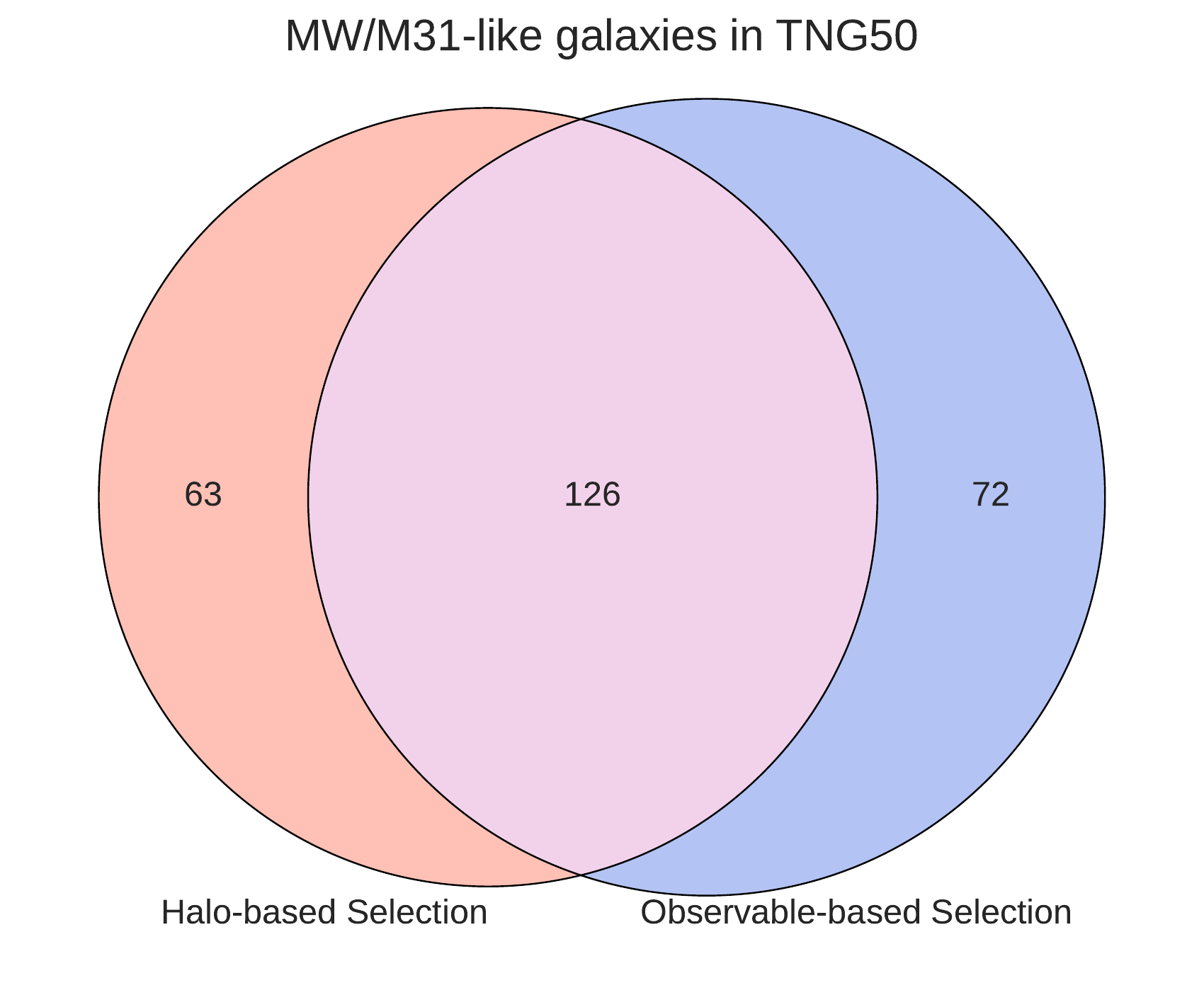}
    \caption{\textbf{Selection criteria and numbers of MW/M31-like galaxies in the TNG50 simulation at $z=0$.} Top: fiducial selection proposed in this paper, based on observable quantities: galaxy stellar mass, stellar disky morphology and Mpc-scale environment. Bottom left: alternative selection that mimics what is often done in the context of zoom-in projects, chiefly based on selecting DM haloes of a certain mass and their central galaxy. Bottom right: comparison between the two selection approaches in the context of TNG50 at $z=0$. In the top panel, by ``massive galaxy'', we mean $M_*\geq10^{10.5}\MSUN$; in the bottom left, $M_{\rm host, 200c}$ is the total virial halo mass of the host or parent halo of any galaxy, satellite or central. According to our fiducial selection (top panel), there are 198 galaxies in TNG50 that can be considered analogs of our Galaxy and Andromeda.}
    \label{fig:selection}
\end{figure*}

There are 324 galaxies in the TNG50 simulation at $z=0$ that satisfy the stellar mass constraint of $M_*(<30{\rm kpc}) = 10^{10.5-11.2}\MSUN$: see top panels of Figs.~\ref{fig:selection} and \ref{fig:selection_depth}. These include also galaxies with non-disky stellar morphology, that live in massive groups and clusters or that are merging with similarly-massive galaxies. Throughout this paper, among the 198 TNG50 MW/M31-like galaxies, the majority of them (138) have MW-like stellar mass, whereas 60 have M31-like stellar mass, with demarcation at $10^{10.9}\MSUN$: this is because TNG50 galaxies constitute a volume-limited sample, i.e. a sample that naturally includes more numerous lower-mass objects, also within the fiducial MW/M31-like selection. 

\subsubsection{On the stellar diskyness criterion (ii)} 
\label{sec:criterion_diskyness}

The selection on stellar morphology enunciated in  Section~\ref{sec:selection} is required to remove from the sample simulated galaxies that are obviously elliptical or highly distorted: these are not what it is typically understood to be the case for the Galaxy and Andromeda and yet may be realized in the TNG50 volume as, for example, MW-mass satellites orbiting in massive clusters may undergo morphological transformation and, because of environmental processes, change from disky to non-disky \citep{Joshi.2020, Galan.2022}. 

\begin{figure}
\center
	\includegraphics[width=1\columnwidth]{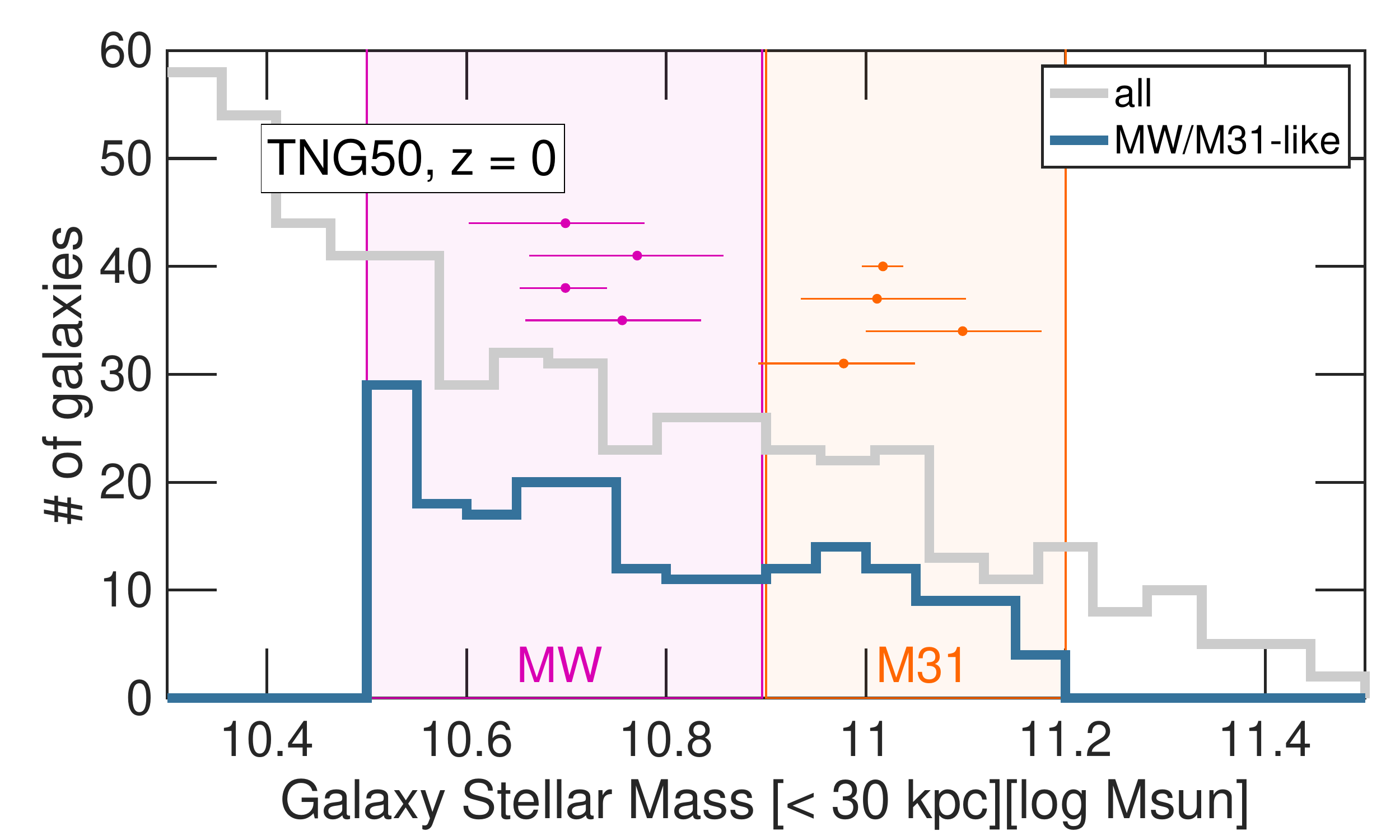}
    \vspace{0.7cm}
    \includegraphics[trim=0cm 0.7cm 0cm 0cm, clip,width=0.8\columnwidth]{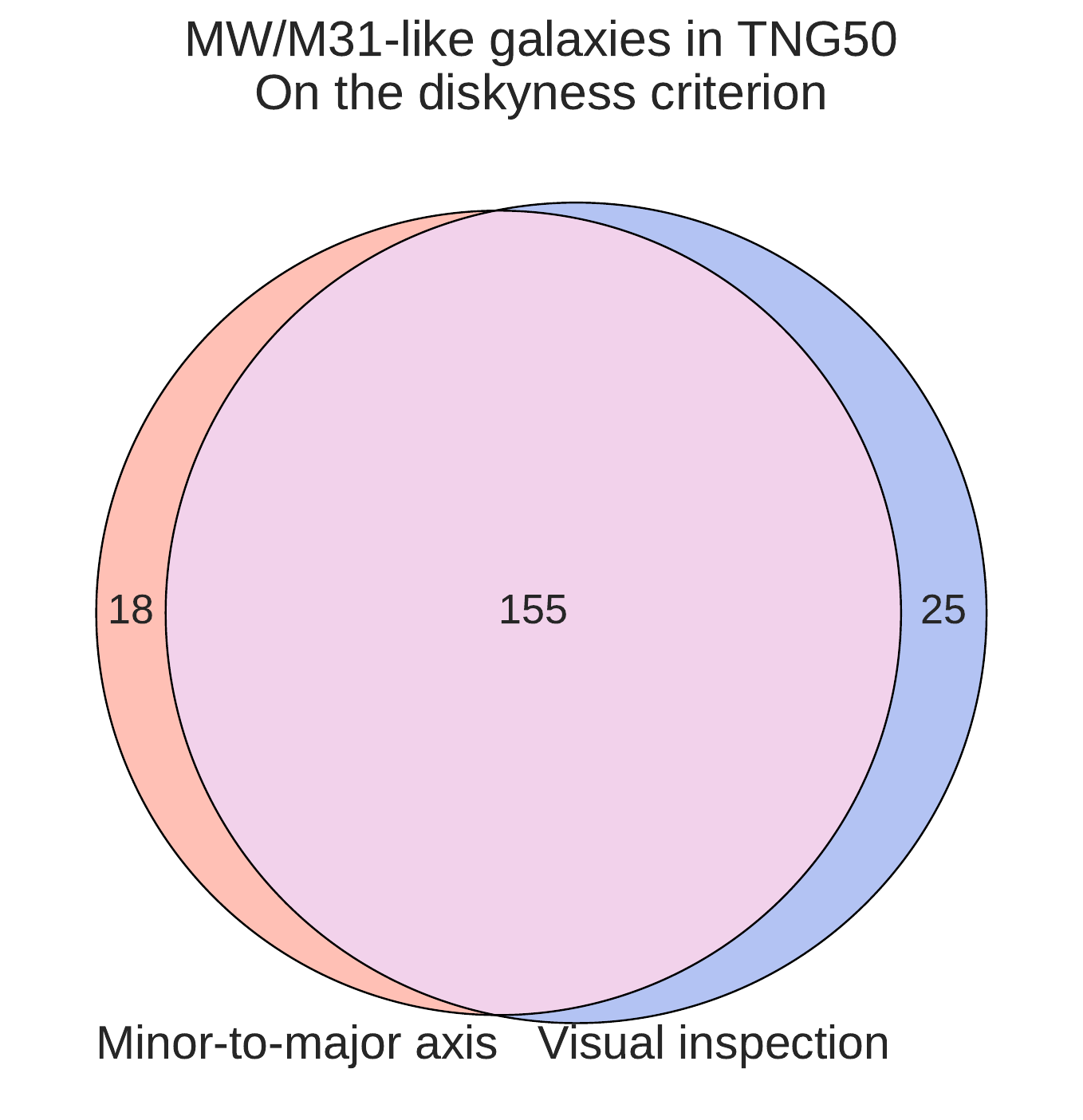}
    \caption{\textbf{Additional inputs on the selection criteria of MW/M31-like galaxies in the TNG50 simulation at $z=0$.} Top: distribution of galaxy stellar mass for all galaxies in TNG50 (gray) and for TNG50 MW/M31-like galaxies (blue histogram), overlaid to available observational constraints on the stellar mass of the Galaxy and Andromeda taken at face value (from top to bottom: \citet{Bland-Hawthorn.2016, Boardman.2020b, Flynn.2006, Licquia.2016} and \citet{Boardman.2020b, Sick.2015, Tamm.2012, Chemin.2009}, respectively). Throughout this work, we select MW/M31-like galaxies to have stellar mass in the range $10^{10.5-11.2}\,\MSUN$, to accommodate for $>1\sigma$ errors and possible systematic differences. MW-like (M31-like) galaxies are those below (above) $10^{10.9}\,\MSUN$. These masses are measured within a spherical aperture of 30 kpc throughout this paper. Bottom: visualization of how the two-pronged approach unfolds for the selection of TNG50 galaxies with disky stellar morphology. The visual inspection adds 25 galaxies to the 173 galaxies that satisfy the stellar mass, stellar $c/a$ and environment constraints. 18 galaxies that would not meet the visual selection are kept in the fiducial TNG50 MW/M31-like sample: 12 of them are S0 objects, with no evident spiral arms but with disky morphology.}
    \label{fig:selection_depth}
\end{figure}

The morphological selection is implemented in a two-pronged approach. On the one hand, we impose a constraint on the 3D ellipticity of the stellar mass distribution, i.e. on the minor-to-major stellar axis ratio, measured as in \citet{Pillepich.2019}. On the other hand, as we do not want to miss \textit{any} disky and spiral galaxy of the relevant mass, we also visually inspect all the TNG50 galaxies that satisfy the stellar mass and environment constraints enunciated above. The fiducial set of disky galaxies is given by the union of the two lists, as shown in terms of Venn diagram in the bottom panel of Fig.\ref{fig:selection_depth}. 

Now, the reasons to choose the stellar minor-to-major axis ratio as morphological indicator are multifold. Firstly, this is a popular proxy for morphology with large galaxy samples, such as in extra-galactic population studies \citep[e.g.][]{Wel.2014b, Zhang.2019}. Secondly, differently than circularity-based selections \citep{Abadi.2003, Scannapieco.2009, Emami.2021}, it does not require kinematic information of the stars: on the one hand, this would not be easily available for e.g. M31; on the other hand, disk-to-total (D/T) or B/T ratios typically provided with observations, also of the Milky Way, are based on photometry or spatial-distribution measurements rather than kinematics, and the mapping between the two is complex \citep{Du.2020}. Finally, by measuring the elliptical shapes of galaxies at galactocentric radii larger than their stellar half-mass radius \citep[and hence beyond the typical bulge radius:][]{Zhu.2022}, we get the opportunity to study the diversity of B/T ratios in the presence of a stellar disk \citep[][and Section~\ref{sec:inner}]{Gargiulo.2022}.

Typically, disky galaxies are defined as those with a minor-to-major axis ratio $\leq 0.33$ (and middle-to-major axis ratio $\ge0.66$): see e.g. Fig. 8 of \citet{Pillepich.2019} and observational references therein. In the case of the Galaxy, the ratio of the stellar disk height to the stellar disk length varies in the ranges of $0.05 - 0.11$ and $0.14 - 0.81$ for the geometrically thin and thick disks, respectively, if we allow for the full range of the various different measurements in the literature. If instead we take exclusively the best values reported by \cite{Bland-Hawthorn.2016}, namely thin and thick disk lengths of $2.6\pm0.5$ and $2\pm0.2$ kpc, respectively, and thin and thick disk heights at the Sun location of $300\pm50$ and $900\pm180$ pc, respectively, the ratios of the stellar disk height to the stellar disk length for the Galaxy are 0.11 and 0.45, for the thin and thick geometrical components, respectively. The latter value for the Milky Way is the reason for which we adopt here a less strict condition on $c/a$, in addition to the fact that the geometrically-thick scale height correlates more closely, in comparison to the geometrically-thin component, with the stellar half-mass height -- namely, for the average TNG50 galaxy, the geometrically-thick component is more closely related to where the majority of the stellar mass is distributed. 

Here we deliberately decide to ignore the value of the middle-to-major axis ratio in order to keep the selection blind with respect to the presence of a stellar bar. 

How the stellar $c/a$ ratio translates to other measures of ``diskyness'' -- such as cuts in the amount of stars in circular orbits via circularity, in triaxiality, or in the ratio between rotational velocity and dispersion ($V/\sigma$), etc. -- really depends on the definition of the latter and on where, within a galaxy’s body, the measurements are taken. Additional inputs and discussion on what this fiducial selection implies in terms of other morphological descriptors for TNG50 MW/M31-like galaxies can be found in the next Sections and in \citet{Sotillo.2022} for the relationships between stellar disk lengths vs. stellar disk heights, kinematic bulge mass fraction, and stellar $V/\sigma$, and in \citet{Gargiulo.2022} for photometrically-defined bulges and bars. 

Now, whereas the stellar shape criterion can be applied automatically and hence to large galaxy samples, it cannot distinguish between galaxies with and without spiral arms, a salient feature of late-type galaxies. Moreover, we have noticed that it may still miss galaxies that, upon visual inspection, clearly exhibit a late-type morphology.

The visual-inspection step is included precisely to ensure that, within TNG50 and given the stellar mass constraints, all spiral, late-type, disky galaxies at $z=0$ are selected. This is performed with face-on and edge-on stellar light composite maps of 100 kpc a side for the JWST NIRCam F200W, F115W, and F070W filters (rest-frame), neglecting dust effects. The focus is placed in the visual morphology of the stellar light distribution in the edge-on projection and on the presence of spiral features, possibly more prominent among the younger stellar populations, in the face-on maps.

The visual inspection adds 25 galaxies to the 173 galaxies that already satisfy the stellar mass, $c/a$ and environment constraints (see bottom panel of Fig.~\ref{fig:selection_depth}). These are objects whose stellar $c/a$ ratio is larger than 0.45 and yet appear disky and with spirals: a majority of these exhibit complex structures, either with particularly-extended spheroidal central components (bulge or halo) or with multiple misaligned planes of stars of e.g. different ages, and yet with blue relatively-thin disk-like structures and spiral arms. 
On the other hand, among the 173 galaxies that satisfy the stellar mass, stellar shape and environment constraints, 18
objects would probably not be described as late-type based on visual inspection. These are \textit{not} removed from the final sample. Two objects simply have disturbed morphologies and are borderline cases (SubhaloIDs 400973 and 41961), while all others are in fact disky galaxies with no manifest spiral arms. Hence, among the 198 TNG50 MW/M31-like galaxies, there are about 12 S0-type objects (which we explicitly flag in the catalogs released in Section~\ref{sec:data}).

\subsubsection{On the Mpc-scale environment (iii)} 
\label{sec:criterion_environment}
Finally, to mimic the facts that neither the Galaxy nor Andromeda are currently merging with a similarly-massive galaxy and that, whereas they may be forming a group, such a Local Group is not expected to be very massive \citep{Benisty.2012}, a minimal isolation criterion and constraints on the Mpc-scale environment are required, as enunciated above (Section~\ref{sec:selection}). These are designed to be agnostic to the presence of an M31-like companion at a certain further distance but to allow for the possibility of having two galaxies of mass similar to the Galaxy or Andromeda in a Local Group-like configuration.

For ``similarly-massive'' or ``massive'' merging companion, we take the lower limit of the stellar mass range criterion used to select MW/M31-like galaxies (Section~\ref{sec:criterion_mstars}). Namely, we impose that no galaxy more massive than $M_*= 10^{10.5}\,\MSUN$ is within a certain distance at the time of inspection, i.e. $z=0$. The latter distance limit is set to 500 kpc, with the following reasoning. The current observed separation between the Galaxy and Andromeda is $785\pm25$ kpc \citep{McConnachie.2005} or $752 \pm 27$ kpc \citep{Riess.2012}; and the relative radial velocity between the two galaxies is estimated to be approximately $-109.2 \pm 4.4$ km s$^{-1}$ \citep{Marel.2012}. Now, there is some arbitrariness in the choice of the $z=0$ snapshot (snapshot 099) from the simulation to search for analogs: namely, we could have also inspected a previous snapshot or the one before that, with these being separated by approximately 150 Myr, or one in the future. Whereas galaxy properties may not change to a degree that is relevant for this paper in the time span of $200-300$ Myr, the distance from e.g. a merging companion would. Given the current relative velocity between the Galaxy and Andromeda and that 1000 km s$^{-1} \eqsim $ 1 kpc Myr$^{-1}$, it is reasonable to add a supplementary $\pm(200-300)$ kpc ``uncertainty'' to the MW-M31 distance to be mimicked in our selection. Therefore, we take as a minimum distance for the presence (or rather absence) of a massive companion to be 500 kpc.

It should be emphasised that our fiducial selection of MW/M31-like galaxies is blind to their central vs. satellite status within our simulation framework. In fact, the definition of the latter would need to be qualified and converted into criteria that can be applied to the available observational data. In the context of the TNG simulations, it is customary to define the central as the most massive galaxy, or the galaxy at the potential minimum, of a halo identified by the Friends-of-Friends (FoF) algorithm \citep{Davis.1985}. However, we deliberately avoid choosing central galaxies only in the selection of MW/M31 analogs as we want to keep open the possibility of finding, a posteriori, a MW+M31 pair also within the same FoF halo (see Section~\ref{sec:environment}). However, estimates on the total mass of the Local Group have been obtained with a diversity of models and techniques and they are capped at $8\times10^{12}\MSUN$, including errorbars: see Fig. 4 of \citet{Benisty.2012} for a recent compilation. Accounting for systematic deviations across different definitions of total halo mass, here we impose TNG50 MW/M31 candidates to be hosted by (FoF) haloes with $\MTWOC<10^{13}\MSUN$ -- a similar cut has been imposed for the search of MW analogs in the local Universe with e.g. the SAGA Survey \citep{Geha.2017, Mao.2021}. 

Of the 198 MW/M31-like galaxies in TNG50, 190 are centrals of their FoF host halos; the other 8 are ``FoF'' satellites of a more massive galaxy (their SubhaloIDs are: 342447, 358609, 372755, 392277, 400974, 414918, 454172, and 500577). These are satellites in the sense that they belong to a FoF halo that includes a slightly more massive object; however, given our selection criteria, such ``satellites'' are still at least 500 kpc away from their more massive companion.



\subsubsection{Comparison between our fiducial and alternative selections of MW/M31 analogs}
\label{sec:selection_comp}

Starting from N-body or DM-only simulations, zoom-in simulations of MW-like galaxies necessarily impose a-priori selections based on total halo mass rather than observable galaxy properties: this is because the zoom-in technique is based on re-simulating, typically, DM-only low-resolution simulations, from which one halo is chosen. In the absence of baryonic properties, a halo mass of about $10^{12}\MSUN$ has been the primary criterion, in addition to focusing on the resulting galaxy that is the central of such a halo.

In the bottom panels of Fig.~\ref{fig:selection}, we show how an alternative, halo-based selection would perform within TNG50, imposing constraints on total host halo mass, on central status, and on stellar diskyness of the central galaxy (bottom left). A very similar number of TNG50 galaxies would be selected, namely 189 vs. 198, with a similarly small number of galaxies being excluded because of their non-disky morphology. However, as the bottom right panel of Fig.~\ref{fig:selection} emphasizes, only about 2/3 of the TNG50 MW/M31-like galaxies are common between our fiducial observable-based selection and the traditional halo-based one. This is chiefly due to a difference in the selection of galaxies in the stellar-to-halo mass plane. Differently than the traditional approach, our fiducial observable-based selection allows for non-central galaxies (which contribute a handful objects) and also for galaxies hosted by somewhat more massive haloes than the canonical limit of $2\times10^{12}\,\MSUN$: these account for 75 per cent of the 72 MW/M31-like galaxies missed by the halo-based selection, with median host halo mass of $10^{12.4}\,\MSUN$ -- which is in fact admitted based on current inferences (see Section~\ref{sec:haloes}). On the other hand, the halo-based selection with minimum halo mass at $6\times10^{11}\,\MSUN$ would include galaxies that, according to TNG50, have too low stellar mass to be considered analogs of the Milky Way. 

This comparison, and the mere fact that halo mass is not an observable, lend even more support to our preference of proceeding with the fiducial observable selection visualized at the top of Fig.~\ref{fig:selection}.

\begin{figure*}
\center
	\includegraphics[width=0.77\columnwidth]{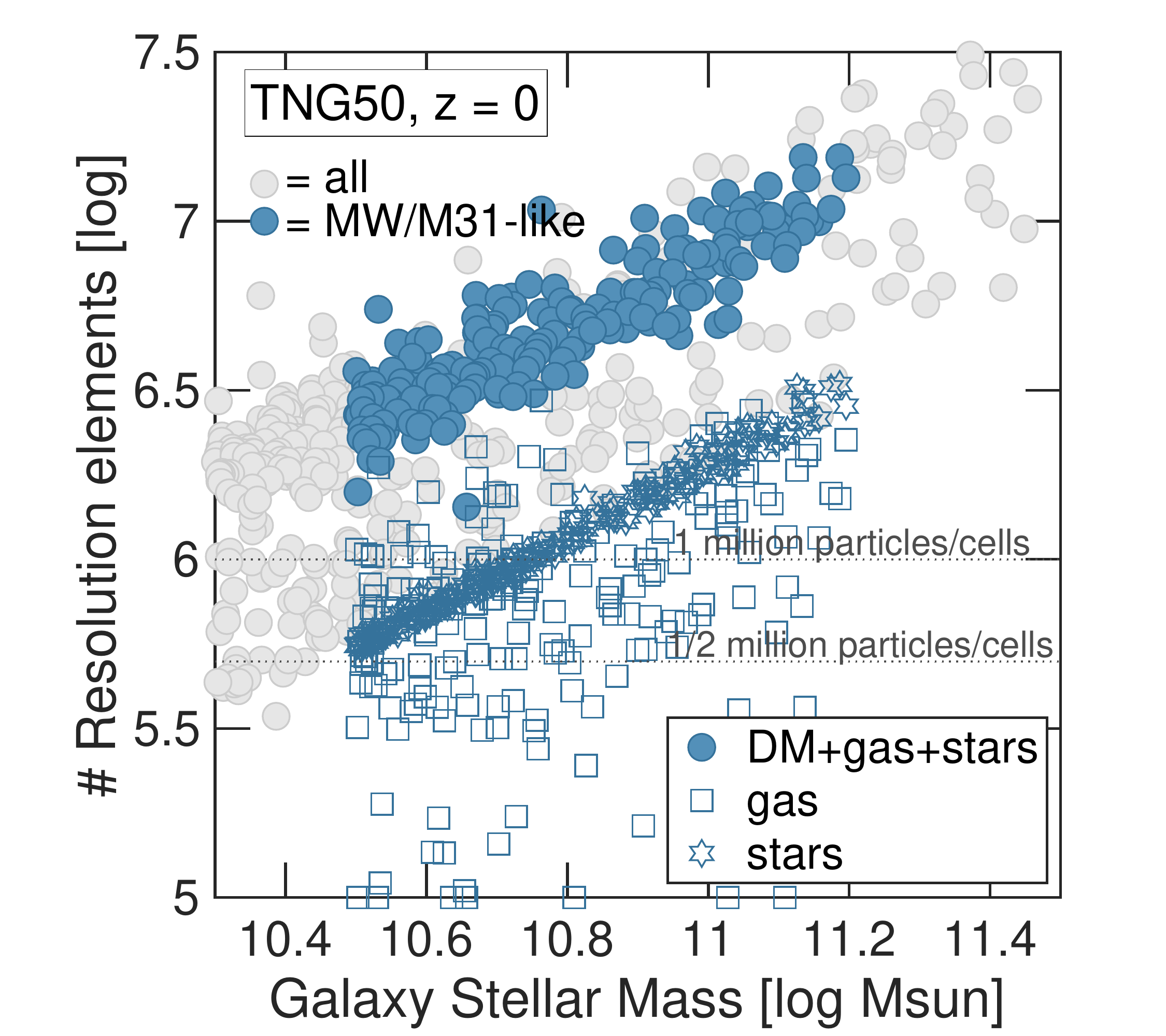}
    \includegraphics[width=0.6\columnwidth]{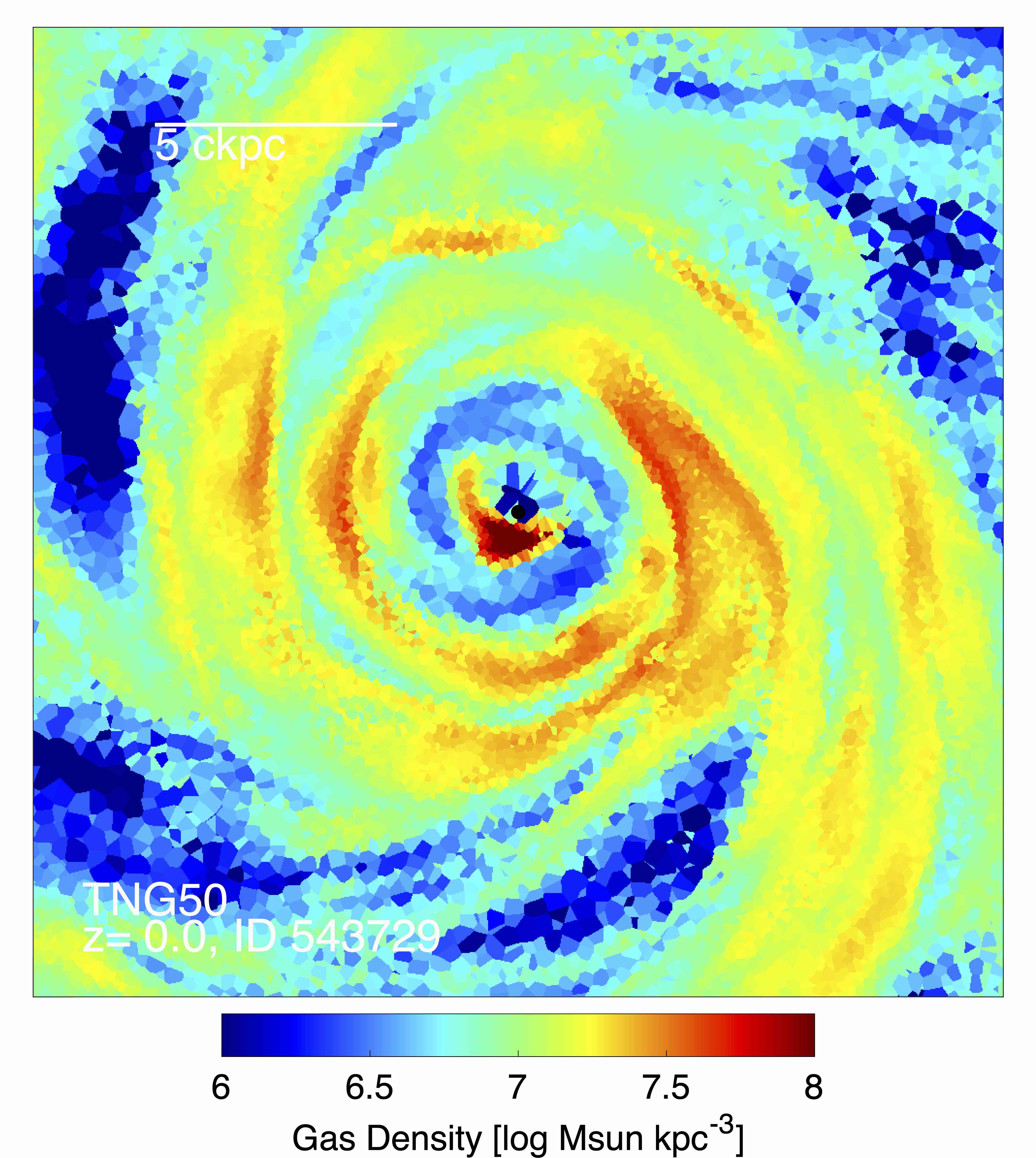}
    \includegraphics[width=0.6\columnwidth]{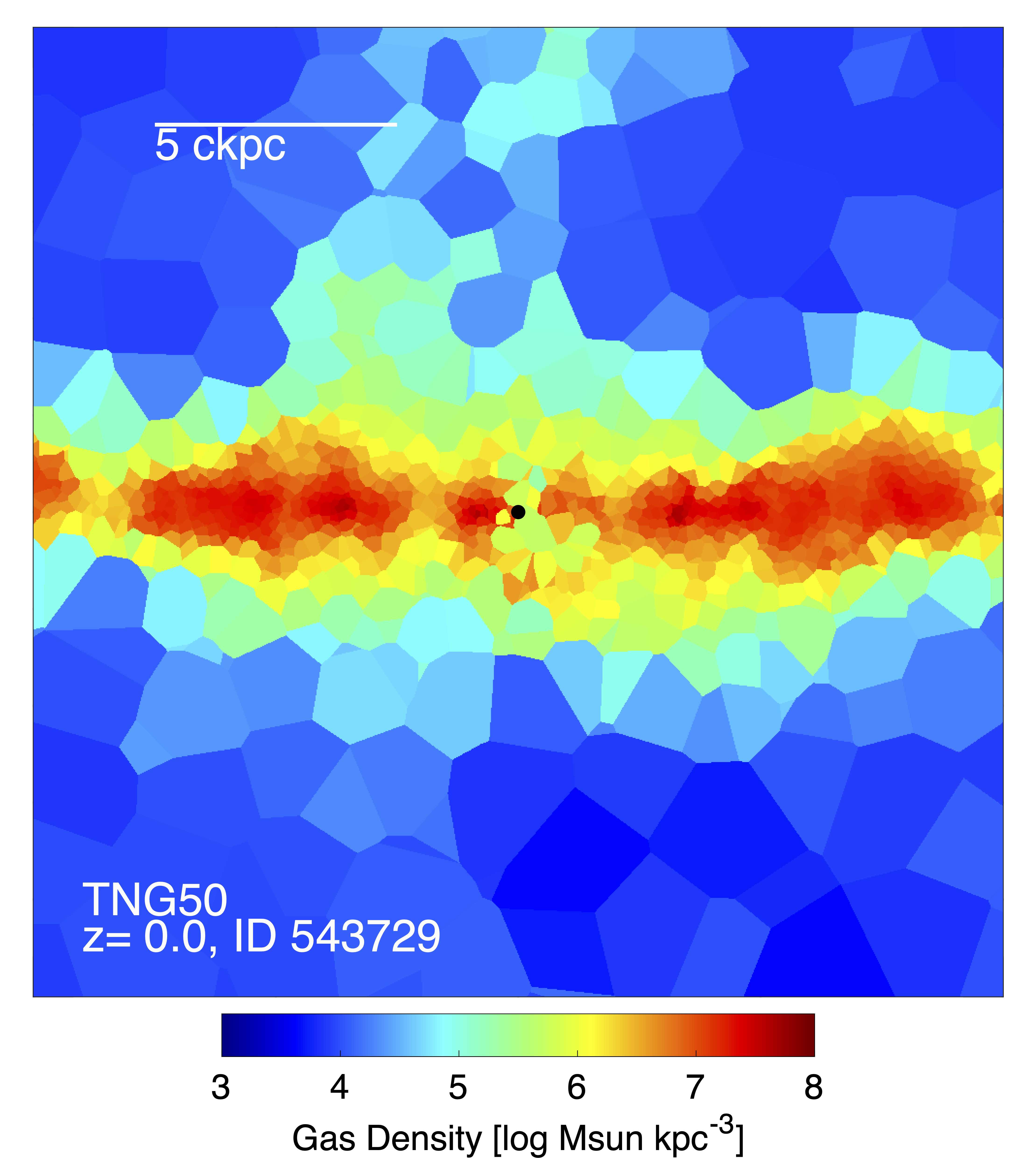}
    \includegraphics[width=1\columnwidth]{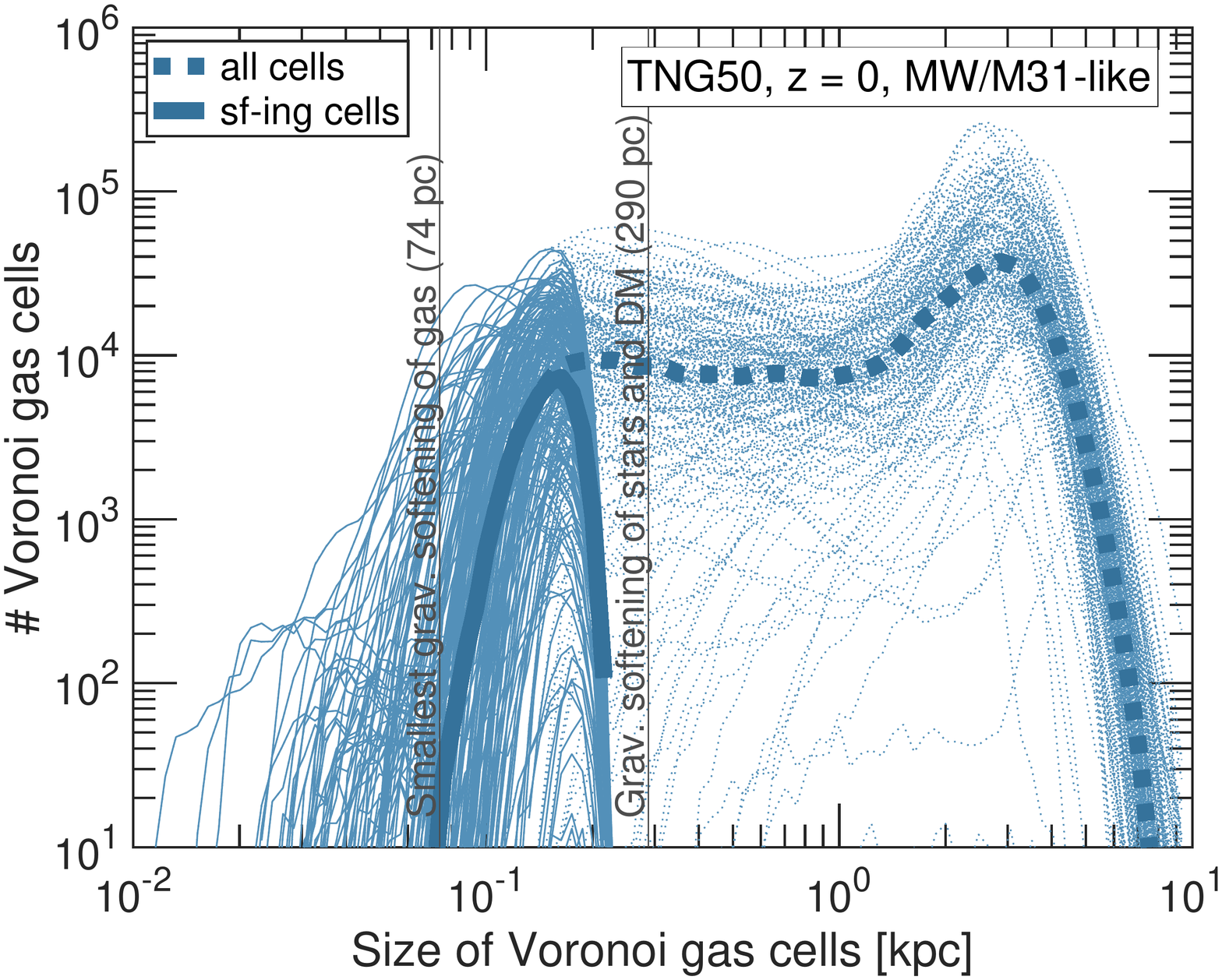}
        \includegraphics[width=1\columnwidth]{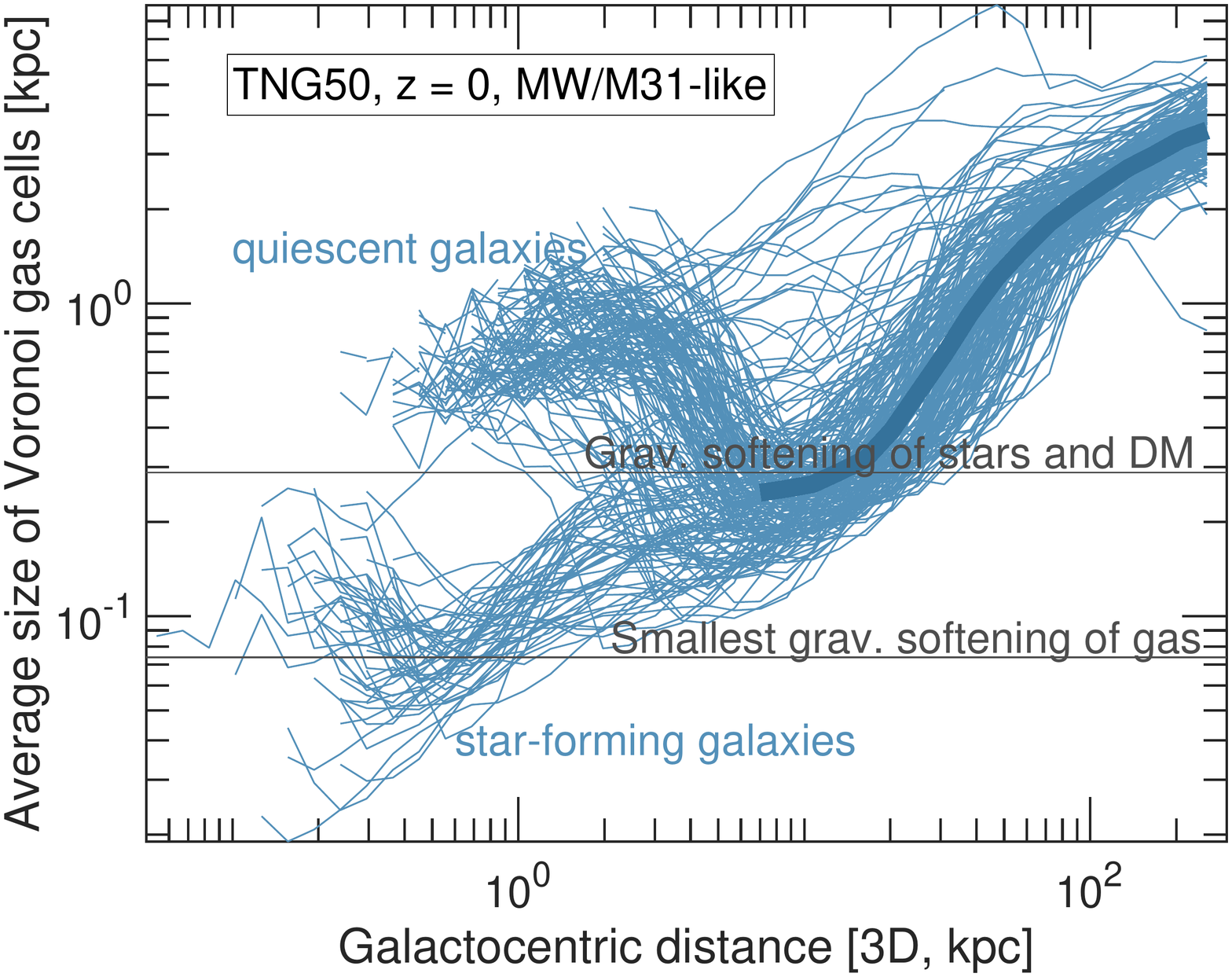}
    \caption{\textbf{Characterization of the mass and spatial resolution in the TNG50 MW/M31-like galaxies}. In the top left panel, we show the number of resolution elements (gas cells, DM and stellar particles) that are gravitationally bound to each of the 198 galaxies, identified by galaxy stellar mass on the x axis. In the top right panels, we provide maps of the Voronoi tessellation of the gas in the mid plane and in a plane perpendicular to the galactic plane and passing through the galaxy centre of an example galaxy: these show that, due to the quasi-Lagrangian nature of the AREPO code and the fixed target gas cell mass, galaxies are resolved with smaller gas cells at higher densities, making the MHD spatial resolution fully adaptive. In the bottom panels we hence quantify the whole hierarchy of Voronoi cell sizes, and hence MHD spatial resolution, in individual galaxies (thin curves) and averaging across all galaxies (thick solid blue curves). Again we consider all gravitationally bound gas cells in each galaxy and plot their size distributions (left: dashed and solid curves for all and star-forming cells, respectively) and the mean cell size as a function of galactocentric distance (right). The softening lengths with which the gravitational forces are approximated are instead given for comparison as gray annotations (see text for details). Not only all TNG50 MW/M31-like galaxies are resolved with at least half a million stellar particles and many million resolution elements among all types, but they are also characterized by gas cells that, in the densest, innermost and star-forming regions of galaxies span between $10-20$ pc and $\sim200$ pc.} 
    \label{fig:res}
\end{figure*}

\subsection{The numerical resolution of TNG50 MW/M31-like galaxies}
\label{sec:res}

Spatially resolved hydrodynamical simulations such as TNG50 are difficult to characterize in terms of numerical resolution. Given the modeled physical processes (chiefly gravity and MHD), there are three types of resolution of relevance: mass, spatial and temporal.

The spatial dynamic range spanned by the TNG50 simulation covers $\sim 6$ orders of magnitude: from the smallest Voronoi gas cell in the volume of $7-8$ pc at $z=0$ to the box size of 52 comoving Mpc. The temporal dynamic range is also large, $4-5$ orders of magnitude, from $\sim13.7$ billion years of cosmic evolution to the smallest time steps of less than 50 kyr \citep[see also][]{Nelson.2019, Pillepich.2019, Nelson.2019b, Pillepich.2021}.

In terms of mass resolution, the baryonic mass i.e. the target mass of the Voronoi gas cells and the initial mass of the stellar particles are both $8.5\times10^4\,\MSUN$ (Section~\ref{sec:tng50}). Dark matter particles are more massive, proportionally to the ratio between the dark-matter and the baryonic energy density parameters: $4.5\times10^5\,\MSUN$. SMBHs (one per galaxy) are seeded with an initial mass of about $1.2\times10^6\,\MSUN$ and then can grow via gas accretion of BH-BH mergers up to $\sim 10^{9}\,\MSUN$.

Given the fixed mass resolution of DM, gas and stars, more massive galaxies are composed of more numerous resolution elements: this is shown in the top left panel of Fig.~\ref{fig:res}, for gas cells, DM and stellar particles that are gravitationally bound to each of the 198 MW/M31 analogs in TNG50 (blue filled circles, as compared to the same measure for all TNG50 galaxies). The total number of resolution elements in these galaxies is mostly contributed by DM particles and is compared to the number of gas cells (blue squares) and stellar particles (blue pentagons). All TNG50 MW/M31-like galaxies are resolved with at least half a million stellar particles (which are mostly in the stellar disks) and many million resolution elements among all types. The number of gravitationally-bound gas cells can be more diverse depending on the galaxy: as will be clear in the next Section, this is connected to the star-formation status and the diverse manifestations and effects of stellar and SMBH feedback.

The spatial resolution is also complex, depending on the physical process under consideration. The gravitational force is softened starting at 290 (580) pc for stellar and DM particles at $z<1$ ($z\ge1$) and at $\sim3$ kpc for SMBHs. This does {\it not} imply e.g. that stellar disks cannot be thinner than about 300 pc: see next Sections and \citealt{Pillepich.2019}. The gravitational softening of the gas changes with the cell size ($\propto 2.5 \times$ the effective sphere-equivalent cell radius) and it is hence adaptive; however, a minimum is enforced at 74 pc in TNG50.

The spatial resolution of the MHD is instead set by the size of the gas cells, which are again fully adaptive: Fig.~\ref{fig:res}, bottom, provides a quantification of the full hierarchy of the effective cell radii in TNG50 MW/M31-like galaxies, whereas the top right panels of Fig.~\ref{fig:res} showcase the Voronoi-tessellation of the gas cells in an example TNG50 galaxy. See also e.g. Table 1 of the TNG data release paper \citep{Nelson.2019}.

Within any single typical MW/M31 analog, cells can span between a few tens of parsecs to $\lesssim10$ kpc. The smallest cells are those that sample the highest densities, such as in the innermost regions of galaxies or at the midplane of disks, where they can be as small as $10-10$s of pc. The most extended cells are in the outskirts of the haloes, where gas densities are low. Overall, the star-forming regions of TNG50 MW/M31-like galaxies are resolved to $70-200$ pc (average of about 150 pc: see solid curves in the bottom left panel of Fig.~\ref{fig:res}). This implies that processes like star formation and stellar feedback are subgrid below these spatial scales. Interestingly, not in all galaxies is the spatial resolution of the gas maximal (i.e. best) in the innermost regions: see the bimodality in the mean cell sizes within a few kpc distance from the galaxy centers in the bottom right panel of Fig.~\ref{fig:res}. As it will be clearer in the next Section, TNG50 MW/M31-like galaxies may undergo inside-out quenching due to the feedback of their SMBHs, which dilute the central gas reservoirs.

\section{Basic properties of TNG50 MW/M31 analogs}
\label{sec:props}

To present our main results, we echo the structure of the review by \citet{Bland-Hawthorn.2016}, but in reverse order: namely, we proceed with an overview of the TNG50 outcome from the outer and more global galaxy properties to their innermost regions. 

In what follows, we provide a sketch of the TNG50 MW/M31-like galaxies to a) address the scientific questions outlined in Section~\ref{sec:intro}, b) orient the readers and the future users of the publicly available simulation data, and c) give an idea of what can be studied, and has been already studied, with this rich dataset. We focus on quantities and galaxy properties that have already been discussed in the literature, whether from an observational or a numerical perspective. We deliberately leave more in-depth analyses and more original explorations of the simulated data to future and targeted work by this team or the community. Table~\ref{tab:papers} provides an overview of the specific aspects analyzed in current and upcoming publications based on the TNG50 MW/M31-like sample presented here.

In particular, in this paper, for the 198 TNG50 MW/M31-like galaxies that we have selected, we provide information about their local and large-scale environment (Section~\ref{sec:environment}); the few examples of Local Group analogs (Section~\ref{sec:LG}); the fundamental properties of their DM, stellar, gaseous and magnetic haloes, including a discussion of the underlying total halo mass (Section~\ref{sec:haloes}); their satellite populations (Section~\ref{sec:sats}), current and past star formation status (Section~\ref{sec:sf}) and global stellar populations (Section~\ref{sec:stellarpops}); the total, stellar, HI gas, and DM mass budget and structural distributions in the disk regions (Section~\ref{sec:disks}), including the stellar and gaseous metallicity and abundances and magnetic fields; and we conclude with a view onto the galaxies' innermost regions: namely, their SMBHs, bulges and bars (Section~\ref{sec:inner}). 

\begin{table}
\input{table_papers}
\end{table} 

Throughout and when possible, we aim at contrasting the outcome of the TNG50 simulation and the properties of the simulated MW/M31 analogs to those observed or inferred for our own Galaxy and Andromeda. For this broad analysis we simply compare the simulated and observed quantities, placing no prior on which observational result to adopt in particular: we rather attempt to consider a broad range of literature results. Our approach is motivated by the following considerations. On the one hand, not all observational results are in agreement among each other nor are the operational definitions of various galaxy properties necessarily consistent. On the other, great care would be needed to properly map simulated variables into observationally-derived quantities and to replicate the observational steps followed to measure or infer certain properties. These are tasks better left to more targeted and focused efforts.


\subsection{Local and large-scale environment}
\label{sec:environment}
\begin{figure}
		\includegraphics[width=1.04\columnwidth]{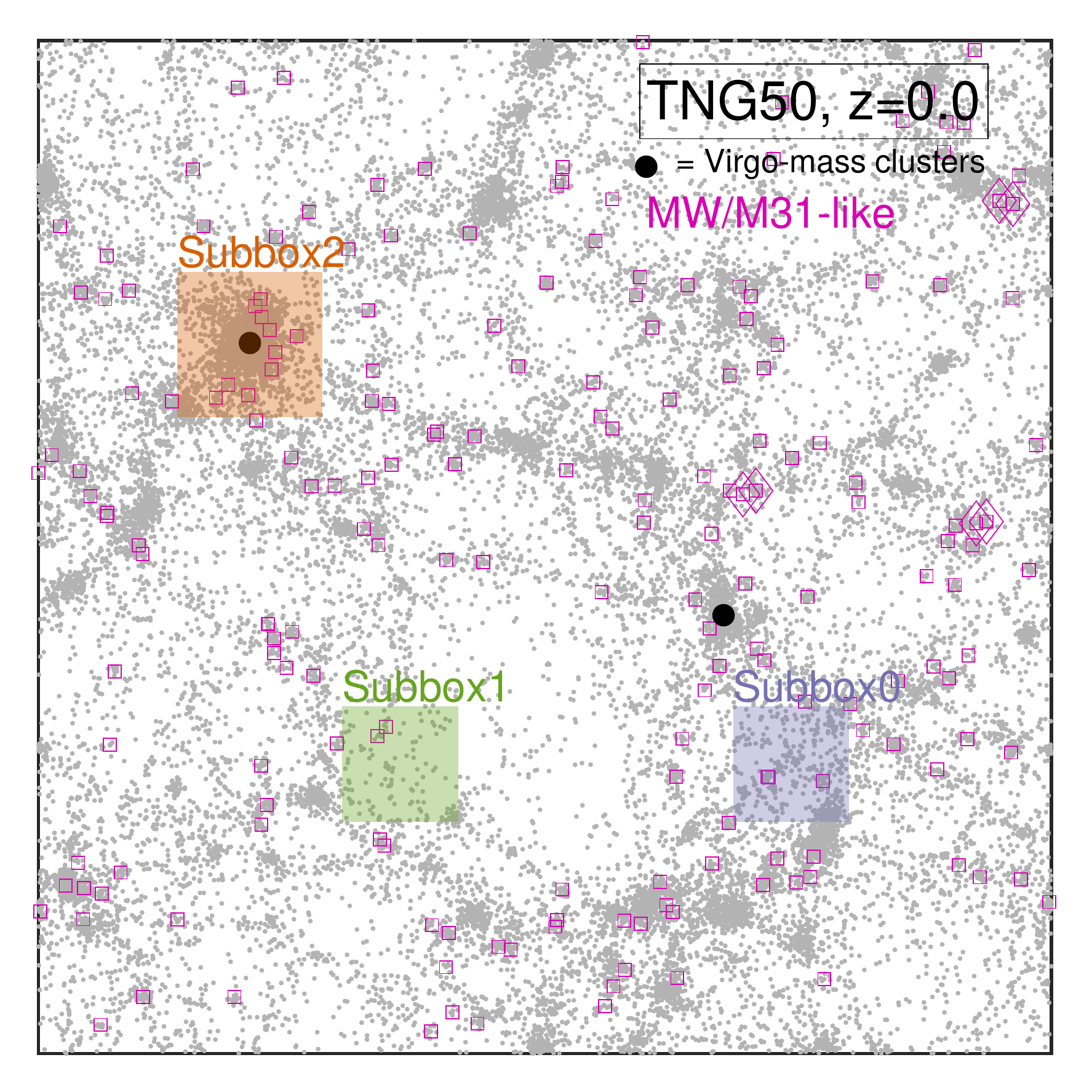}
    \caption{{\bf Location and large-scale environment of the 198 MW/M31-like galaxies (magenta) within the whole TNG50 simulated volume.} The large-scale structure throughout the comoving volume of TNG50 (which spans about 52 comoving Mpc a side) is here depicted  at $z=0$ through the position of TNG50 galaxies more massive than $10^6\,\MSUN$ in stars (gray circles). This is a random projection, along the z-axis of the box, and also shows the location of the two Virgo-mass clusters in the simulation (large black circles). A handful of the selected objects are within the so-called Subboxes (shaded comoving cubes), whereby data is available every few million years (see Section~\ref{sec:data} and Table~\ref{tab:subboxes}). Three pairs compose Local-Group analogs (magenta diamonds).}
    \label{fig:environment}
\end{figure}

The selection of the MW/M31-like galaxies provided in Section~\ref{sec:selection} is agnostic as to their large-scale location within the whole (52 cMpc)$^3$ volume, at both $z=0$ and especially at previous times. This is shown at the current epoch in Fig.~\ref{fig:environment}, with the 198 MW/M31 analogs given in magenta.

The majority of TNG50 MW/M31-like galaxies are located -- as to be expected from the hierarchical growth of structures, in higher-density regions -- along filaments or sheets. In fact, 21 among the selected galaxies lie within a 3D distance of 10 Mpc or less from a Virgo-mass cluster: there are two such haloes at $z=0$ in the TNG50 volume, with total halo mass of $\MTWOC\gtrsim10^{14}\,\MSUN$ (see large black circles in Fig.~\ref{fig:environment}). The Subhalo IDs of these MW/M31-like galaxies are released (Section~\ref{sec:data} and Table~\ref{tab:files_ids}), so that additional isolation selections may be applied to e.g. mimic the fact that our galaxy is far enough (about 16.5 Mpc) from the real Virgo cluster. It should be noted, however, that by selection none of the 198 MW/M31-like galaxies are satellites of such massive hosts. Furthermore, they are all located many Mpc away from such objects, with the closest currently at 2.6 Mpc from the center of the most massive cluster of TNG50.

In terms of smaller-scale environments, we have already pointed out (Section~\ref{sec:criterion_environment}) that 8 of the 198 MW/M31-like galaxies are in fact ``FoF satellites'' of more massive hosts (still below total mass of $10^{13}\,\MSUN$ and at $\geq500$ kpc distance). Furthermore, their Mpc-scale environment is determined by the selection criteria defined earlier (Section~\ref{sec:criterion_environment}): below we expand on the cases of Local-Group analogs (Section~\ref{sec:LG}) and on their satellite populations (Section~\ref{sec:sats}).

\begin{figure*}
    \includegraphics[width=1\textwidth]{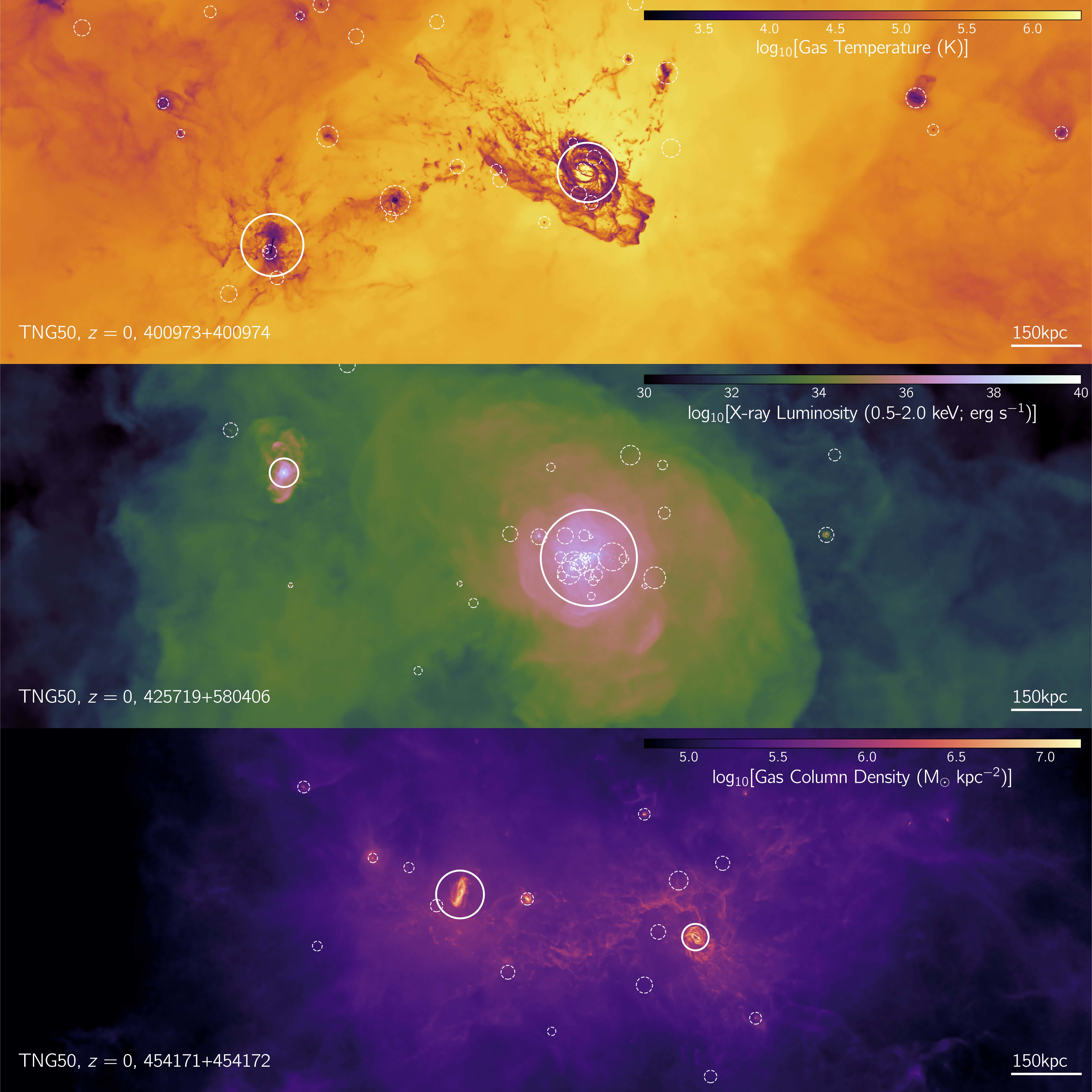}
\caption{{\bf Visualization of the three Local-Group like pairs in TNG50 at $z=0$.} We show the three systems in three different gas quantities: mass-weighted gas temperature, X-ray surface brightness and gas column density, from top to bottom. The LG analogs are shown in a random projection in a volume of  ($\pm$1.2 Mpc, $\pm$400 kpc, $\pm$1.2 Mpc) and centered at center of mass of the LG-pair. White circles (10 $\times$ the stellar half-mass radius) denote the location of galaxies with stellar mass $\geq5\times10^6\,\MSUN$: solid for the two main MW/M31-like galaxies, dashed for the others. The environments of LG-like galaxies are rich and complex.}
    \label{fig:LGlike}
\end{figure*}

\subsection{Local Group analogs}
\label{sec:LG}

Remarkably, there are three serendipitous pairs of MW/M31-like galaxies in the TNG50 volume that could be considered analogs of the Local Group (LG) at $z=0$. 

We search for LG-like pairs with the following criteria: i) there must be two (but only two) MW/M31-like galaxies within 500-1000 kpc distance from each other; and ii) the two galaxies must move with negative radial velocity towards each other. The distance condition is driven by the previous discussion about the measured distance between the Galaxy and Andromeda (see Section~\ref{sec:environment}). The requirement on the relative velocity is relaxed with respect to the observed radial relative velocity, which has been accurately measured at $-109.3 \pm 4.4$ km s$^{-1}$ \citep[][ the transverse velocity being instead more controversial]{Marel.2012}. Given the arguments above and given that the observables commonly used to characterize the LG (namely, distance and relative velocity) are in fact transient, i.e. depend on information of the primordial density field encapsulated at very small spatial scales \citep{Sawala.2021}, we believe that the approach outlined above is suitable for getting a first glimpse into how such LG-like systems could look like, behave, and impact the evolution of the galaxies therein. For more targeted studies or for analyses with possibly larger sample statistics, we refer the reader to simulations with so-called constrained initial conditions (see Table~\ref{tab:sims}). 

The Subhalo IDs of these TNG50 LG-like pairs are: 400973+400974; 425719+580406; and 454171+454172\footnote{We note that there is also one system in TNG50 that consists of three MW/M31-like galaxies within $500-1000$ kpc from each other (SubhaloID = 414917, 414918, and 528836.}. These are indicated with magenta diamonds within the TNG50 box in Fig.~\ref{fig:environment}. In the case of two of these three pairs, one of the MW/M31-like galaxy is a satellite of the other, in that the two pair galaxies are part of the same FoF halo, whereas in the case of the 425719+580406 pair, both MW/M31 analogs are centrals from the perspective of the FoF algorithm.

We visualize the TNG50 LG-like pairs in Fig.~\ref{fig:LGlike}, one panel for each LG analog depicted in gas temperature, X-ray surface brightness and gas density, from top to bottom, across $\pm$1.2 Mpc in a random projection. Gas permeates the systems out to hundreds of kpc from either MW/M31-like galaxy: some of it is cold, also outside the star-forming disks (top; see also \textcolor{blue}{Ramesh et al. 2023b}); some of it creates extended bridges between the two main galaxies (bottom); the majority is hot and volume filling, emits in X-ray, and expands through shock fronts into the intergalactic medium (middle), driven by energy injections from the central SMBHs \citep{Pillepich.2021, Ramesh.2023}.

White circles in Fig.~\ref{fig:LGlike} denote galaxies with at least $5\times10^6\,\MSUN$ in stars: not all satellite galaxies appear to contain gas. In fact, in \citet{Engler.2023}, we have tentatively shown that a LG-like environment may affect the star-formation activity of galaxies therein, with the quenched fractions of $\gtrsim 10^8\,\MSUN$ satellites being somewhat higher in the case of hosts in LG-like pairs vs. isolated ones.

\subsection{The haloes}
\label{sec:haloes}

As anticipated by Fig.~\ref{fig:LGlike} and as expected from galaxy formation in a $\Lambda$CDM scenario, the TNG50 MW/M31-like galaxies are embedded within haloes of cold DM, gas, and stars that extend for hundreds of kpc from the galaxy centers. These are composed of gravitationally-bound and non-bound material, and of matter accreted from the large-scale structure or expelled or heated up from their innermost regions, and hence exhibit complex phase-space properties. 

\subsubsection{Total halo masses and DM haloes}
TNG50 MW/M31-like galaxies live in haloes that span the range $10^{11.5-12.8}\,\MSUN$ in total mass ($\MTWOC$, the mass within a radius that encloses an average density equal to 200 times the critical density of the Universe) and $147-404$ kpc in virial radius (i.e. $\RTWOC$). 

\begin{figure}
		\includegraphics[width=\columnwidth]{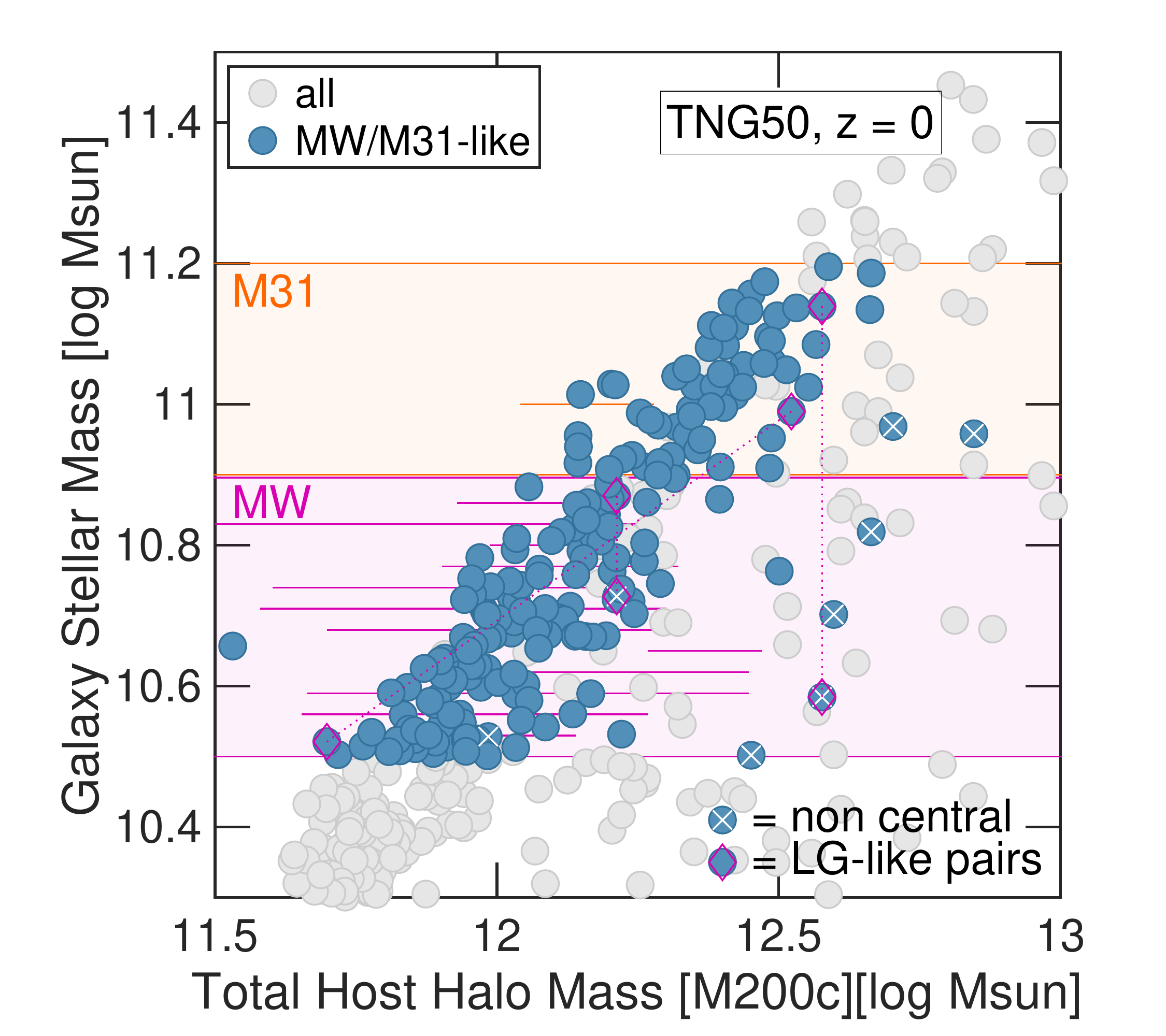}
    \caption{{\bf Stellar-to-halo mass relation of the TNG50 MW/M31-like galaxies presented in this paper.} Although we place halo mass on the x-axis, the MW/M31-like galaxies (blue circles, versus all galaxies in the simulated volume, in gray) are selected based on galaxy stellar mass and other criteria: see previous sections and the orange and magenta horizontal shaded bands. They live in haloes that span the range $10^{11.5-12.8}\,\MSUN$ in total mass, according to the relationship produced by the TNG50 simulation. However, not all depicted galaxies are centrals of their DM and gaseous \textit{FoF} haloes: the 8 MW/M31-like galaxies that are satellites of a larger object, i.e. are within the FoF halo of a more massive galaxy, are denoted with white crosses and for them we plot the total mass of their ``host halo'' (see text for details); the 3 pairs in LG analogs are shown as connected diamonds. But for a few cases, the halo masses of the TNG50 MW-like galaxies are consistent with observational inferences, as collected by \citet{Callingham.2019}: magenta horizontal lines, summarized here by methodology and shown in random order. Inferences for the total halo mass of Andromeda are harder and sparser: here we show constraints from \citet{Watkins.2010}, in terms of total mass within 300 kpc.}
    \label{fig:stellarhalomass}
\end{figure}

More massive galaxies naturally form in more massive haloes, according to a relationship between stellar mass and halo mass that is an emerging outcome of the TNG50 simulation and that is shown in Fig.~\ref{fig:stellarhalomass}. Clearly, because of their larger stellar masses, M31-like galaxies tend to form in more massive haloes than MW-like galaxies, with median halo mass of $2.5\times10^{12}\,\MSUN$ and of $1.1\times10^{12}\,\MSUN$, respectively. This is a general prediction of all galaxy formation cosmological simulations, not only TNG50, albeit the quantitative details can differ substantially across galaxy formation models.

Not all galaxies in the sample (blue circles) are the central, i.e. the most massive galaxy, of their (FoF) halo: we hence plot more generally the total mass of the host halo (i.e. the $\MTWOC$ of the most massive galaxy in their FoF) to show also the 8 galaxies among our TNG50 MW/M31-like galaxies that are satellites of more massive galaxies (white crosses, see Section~\ref{sec:environment}). But for a handful of cases, the observable-based selection adopted throughout and discussed in Section~\ref{sec:selection} also returns halo masses that are consistent with observational inferences for the Milky Way (magenta horizontal lines) -- observational constraints for the total mass of Andromeda are much harder and sparser (orange horizontal line): there we show inferences for the total mass within 300 kpc by \citet{Watkins.2010} .

In Fig.~\ref{fig:stellarhalomass} we also indicate the locus of the LG-like pairs (magenta diamonds): in two of the three pairs, the two main galaxies have stellar masses consistent with the Galaxy and Andromeda, respectively; the pairs located in the same FoF halo are those with the same host mass.

It is interesting to see that the stellar-to-halo mass relation of Fig.~\ref{fig:stellarhalomass} seems to be  described by a single power law, instead of exhibiting a change of derivative, at least in the probed mass range. This is consistent with what we showed for the case of TNG100 {\it spiral i.e. disky} galaxies \citep{Rodriguez-Gomez.2022} and what is inferred from the dynamical modeling of observed spiral galaxies in the local Universe \citep{Posti.2019}.

The total halo mass ($\MTWOC$) of MW/M31-like galaxies is dominated by DM, in particular in our model by cold DM, which accounts for $\gtrsim 85$ per cent of the mass and which arranges itself in both smooth as well as in phase-space structures such as subhaloes, streams, and shells.

It is important to highlight that the TNG galaxy formation model induces the DM haloes of MW/M31-like galaxies to be strongly contracted in comparison to the DM-only predictions \citep{Lovell.2018, Anbajagane.2021}. Whereas their DM halo profiles can still be fitted with NFW \citep{Navarro.1996} and Einasto \citep{Einasto.1984} functional forms, their DM concentration parameters are larger than those predicted along the canonical concentration-mass relation and their potential is dominated by DM also within the stellar half-mass radius \citep[][with DM fraction $\gtrsim 45-50$ per cent]{Lovell.2018}. We will study in depth the DM halo radial profiles of TNG50 MW/M31 analogs in a dedicated paper. Moreover, the TNG galaxy formation results in rounder DM haloes compared to DM-only simulations: in particular, those with MW/M31-mass galaxies are the most spherical, with an average minor-to-major axial ratio of $< s > \sim 0.75$ in the inner halo \citep[an increase of 40 per cent in comparison to their DM-only counterparts:][]{Chua.2022}.

\subsubsection{Basic properties of the stellar haloes}
\label{sec:stellarhaloes}
While the majority of the stellar mass is concentrated at the bottom of the DM halo's potential, MW/M31-like galaxies are also predicted to be surrounded by low-density cocoons of stars all the way to $0.5-0.7 \RVIR$ \citep{Deason.2020}: extended stellar haloes.

TNG50 returns a remarkable diversity of stellar halo properties across the 198 MW/M31-like galaxies: this is a reflection of their diverse underlying total halo mass and assembly histories \citep{Pillepich.2014b}. 
The mass fraction in the stellar halo to the total stellar mass ranges from 5 to 70 per cent, with a median of about 30 per cent mass in the stellar haloes \citep[][]{Sotillo.2022}: see their Fig. 13, based on the kinematic decomposition by \citet{Du.2020}\footnote{Note that different ways to define halo stars may return substantially different stellar halo mass estimates, and thus fractions: e.g. \citet{Merritt.2020}.}. The 3D slopes of the spherically-symmetric stellar mass profiles in the stellar haloes vary between $-6.8$ and $-3.5$ \citep[][their Fig. 14]{Sotillo.2022}, encompassing all empirical estimates for the stellar haloes of the Galaxy and Andromeda. We have also demonstrated that MW/M31-like galaxies with a major merger (stellar mass ratio $\ge$ 1 to 4) within the last 5 billion years exhibit more massive and somewhat shallower stellar haloes, again in qualitative compatibility with the different observed properties of the Galaxy and Andromeda, which are thought to have undergone their last major merger 8–11 and 2 Gyr ago, respectively \citep[][and refs therein]{Sotillo.2022}.

Stellar haloes have been demonstrated to be mostly composed of accreted or ex-situ stars, namely stars that formed in other (typically lower-mass) galaxies that then got accreted as satellites or mergers and got stripped via tides \citep[e.g. with TNG:][]{Rodriguez-Gomez.2016, Pillepich.2018b, Merritt.2020}. However, this does not imply that the majority of the ex-situ mass of any MW/M31-like galaxy resides in its stellar halo: this is often not the case, as shown for the Eris simulation \citep{Pillepich.2015} and for this TNG50 sample (\textcolor{blue}{Sotillo-Ramos et al. in prep}). Analogously, whereas the stellar haloes of TNG50 MW/M31 analogs always host and thus contribute the majority of the extremely metal-poor stars of the system, there are many cases whereby non-negligible amounts of very metal poor stars are within the disks, i.e. are in cold disk-like orbits \citep{Chen.2023}.

All these interesting considerations further motivate the use of this rich dataset for more detailed and additional in-depth studies, also in comparison to observational findings, and exploiting the known accretion origin of the stellar haloes. 

\begin{figure*}
\center
	\includegraphics[trim=0 4cm 0 3cm, clip, width=1\textwidth]{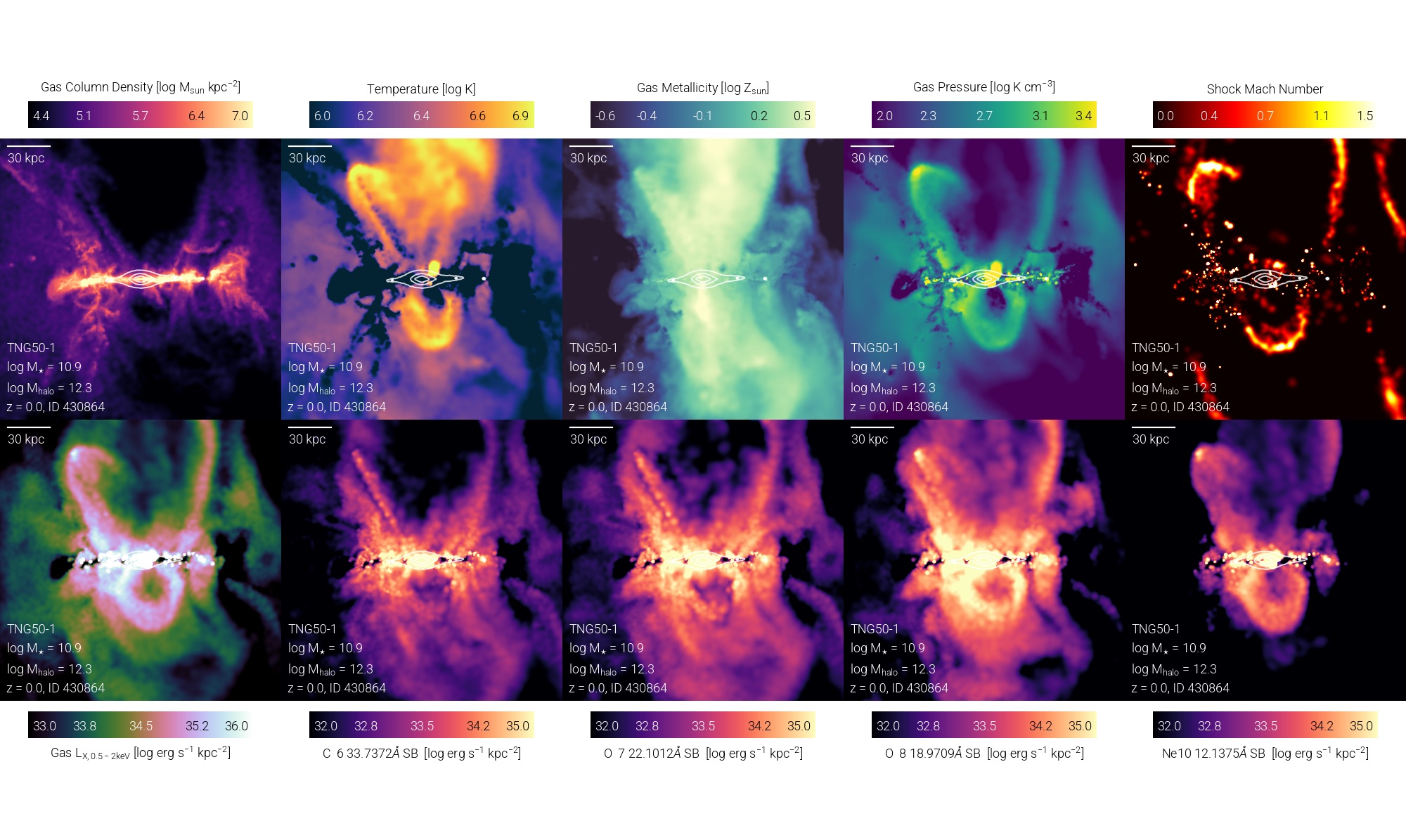}
    \caption{{\bf eROSITA-like bubbles in the CGM above and below the stellar disk of an example TNG50 MW/M31 analog.} The same galaxy (SubhaloID 430864) is shown edge-on multiple times. Physical properties of the gas are mapped in the top row (from left to right: gas column density, mass-weighted temperature and metallicity, pressure and Mach number in shocks). Intrinsic emission in X-ray from the diffuse gas is shown in the bottom row, in the [0.5-2] keV soft broad band (left) and in a series of narrow bands centered at selected CVI, OVII, OVIII and NeX metal lines. All these post-processing quantities are released with this paper (see Appendix~\ref{sec:data}). A gallery of additional images and examples can be found at \url{https://www.tng-project.org/explore/gallery/pillepich21}.}
    \label{fig:bubbles}
\end{figure*}

\subsubsection{On the cold and hot circumgalactic media}
\label{sec:cgm}
MW/M31-like galaxies are surrounded, according to TNG50, by a substantial amount of gravitationally-bound gas that extends several hundreds of kpc. The properties of such gas in this sample of TNG50 MW/M31-like galaxies have been quantified in detail by \citet{Ramesh.2023} and by \citet{Pillepich.2021} and maps of the gas in various manifestations and projections can be found online\footnote{\url{https://www.tng-project.org/explore/gallery/ramesh22/}}. Here we summarize key aspects, and refer the reader to the aforementioned papers and to Table~\ref{tab:papers} for details.

The CGM of TNG50 MW/M31-like is dominated, in terms of mass, by hot and X-ray emitting gas, with X-ray luminosities integrated within $(0.15-1)\times \RTWOC$ of $L_X(0.5-2.0~ \rm{keV}) = 10^{38.5-41.7} \rm{erg ~ s}^{-1}$ \citep[see also][and Fig.~\ref{fig:LGlike}]{Truong.2020, Truong.2021}. However, $10^6$ K gas, and even super-virial $10^{7-7.5}$ K gas, coexists with a subdominant cool component, with $<10^{4.5}$~K cold gas mass in the range of $10^{8-10.8}\,\MSUN$ throughout the CGM. This cold gas is approximately in pressure equilibrium and yet is much less volume filling than the hotter counterpart: it forms < kpc cold clouds, which are resolved due to the high resolution of TNG50 \citep{Nelson.2020} and often have high velocities, similar to the HI clouds observed in the Milky Way (\textcolor{blue}{Ramesh et al. 2023b}).

The CGM of TNG50 MW/M31-like galaxies exhibit large variations, both in terms of how properties vary across the spatial extent of each individual halo as well as how each halo-averaged property varies from galaxy to galaxy. In terms of the former, HI density, temperature, metallicity, thermal pressure, entropy, and magnetic field strength and pressure ratio vary by $3-7$ orders of magnitude across the halo of individual TNG50 galaxies \citep[see Figs. 8 and 12 of][]{Ramesh.2023}. In terms of the latter, the physical properties of the CGM of MW/M31-like galaxies are strongly modulated by the feedback from their SMBHs \citep{Truong.2020, Pillepich.2021, Ramesh.2023}: MW/M31-like galaxies having undergone kinetic SMBH feedback exhibit lower-density, higher-temperature, higher-metallicity, and X-ray dimmer CGMs than those in the thermal SMBH feedback mode, and can have outflow velocities up to $100s-2000$ km s$^{-1}$ \citep[see e.g. Figs. 8 and 10 of][respectively]{Pillepich.2021, Ramesh.2023}. 

Remarkably, TNG50 produces overpressurized, X-ray emitting bubbles, shells, and cavities in the circumgalactic gas below and above the stellar disks of its MW/M31-like galaxies \citep{Pillepich.2021} that are reminescent of the eROSITA and Fermi bubbles of our Milky Way \citep{Su.2010, Predehl.2020}. Given the novelty and complexity of the TNG50 eROSITA-like bubbles, we show a series of CGM maps of an example TNG50 galaxy (Fig.~\ref{fig:bubbles}): these include both physical properties of the gas as well as X-ray emission, particularly narrow-band X-ray imaging, which will be accessible via future large-area X-ray integral field units \citep[e.g.][]{LEM.2022}.

\begin{figure*}
\center
	\includegraphics[width=1\columnwidth]{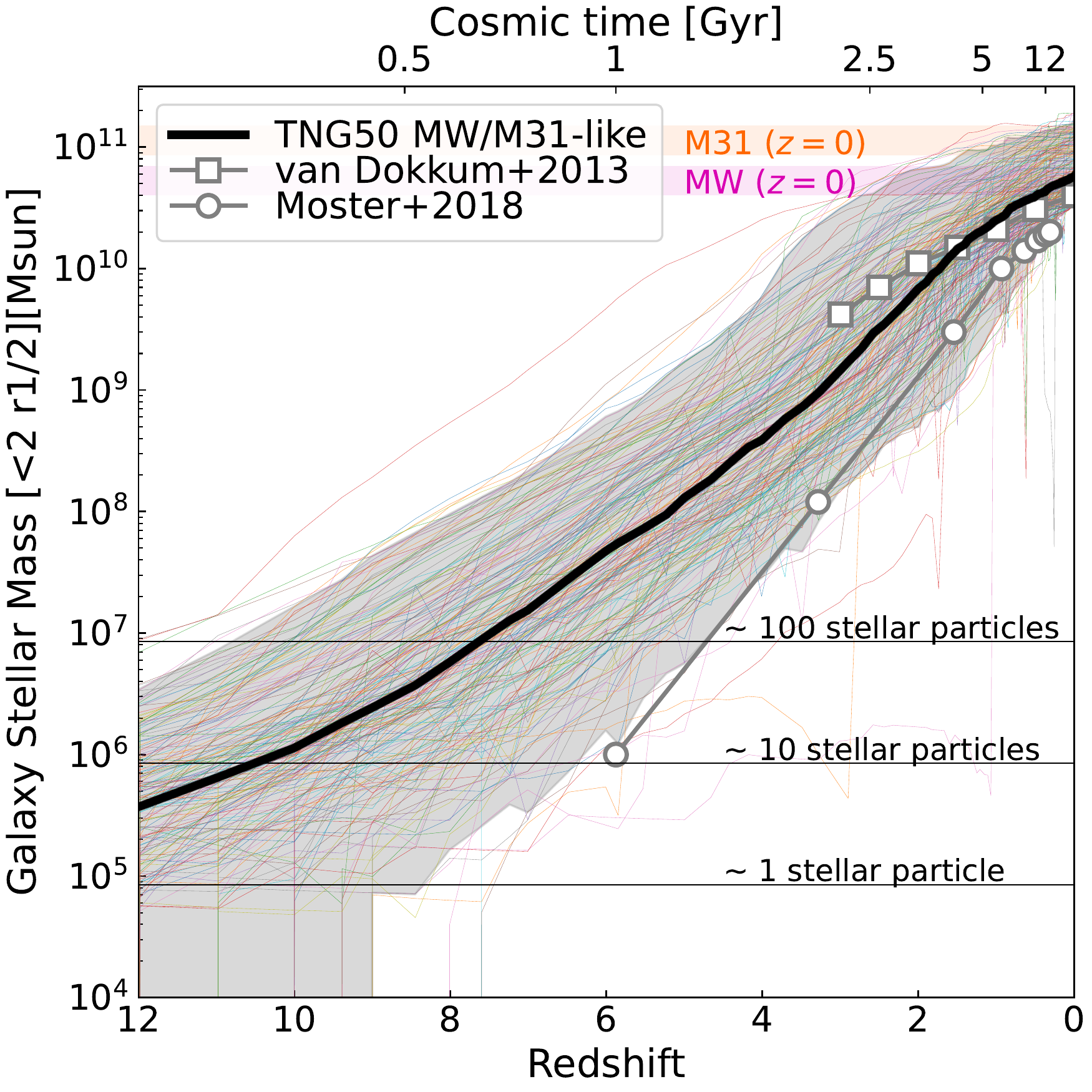}
	\includegraphics[width=1\columnwidth]{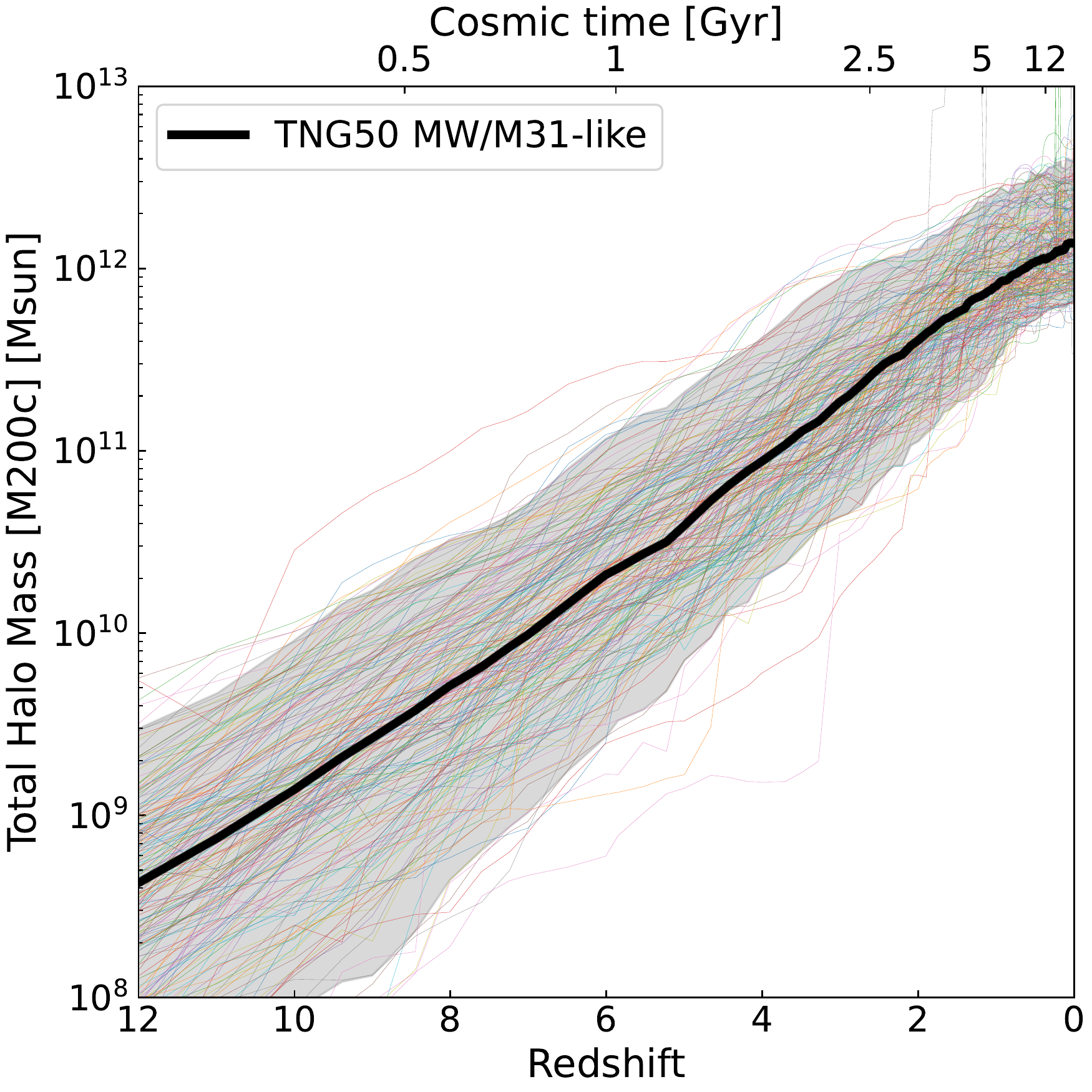}
    \caption{{\bf The high-z progenitors of TNG50 MW/M31-like galaxies.} In the top two panels, we show the mass assembly histories of the galaxies selected at $z=0$, for galaxy stellar mass (left) and total halo mass (right), according to their evolutionary tracks along the merger trees. Each galaxy is shown with a color curve, the median tracks at fixed cosmic time are marked as thick black curves, shaded gray areas denote interquantiles (10th–90th percentiles). Gray-and-white markers represent empirically-derived constraints, for mere reference. Analog plots that focus on lower redshift can be found in \citet{Sotillo.2022}.}
    \label{fig:progenitors}
\end{figure*}

\subsubsection{The magnetized haloes}
\label{sec:Bfields}
As TNG50 is an MHD simulation, it naturally predicts properties of the magnetic fields within and around the simulated galaxies. 
The average strength of the magnetic field in the typical TNG50 MW-like galaxy is as high as $1\,\mu$G in the inner CGM, but decreases rapidly with galactocentric distance \citep{Ramesh.2023}. The magnetic pressure hence dominates over thermal pressure only in the inner regions of the CGM of MW-like galaxies, and yet there can be localized regions of gas throughout the simulated haloes whereby magnetic pressure is the dominant component.

We had previously shown that, across the galaxy population, the TNG model returns weaker magnetic fields in and around red galaxies than around blue galaxies, at fixed galaxy stellar mass \citep{Nelson.2018}. In Section~\ref{sec:disks_gas} we provide further insights into the B fields in the disk regions. It will also be interesting to further characterize the properties and topology of the magnetic fields in the CGM for these TNG50 MW/M31-like galaxies, connecting to their current and past feedback activity and assembly history. 

\subsection{Diverse merger histories and progenitors}
\label{sec:mergers}
Despite being selected within 0.7 dex in galaxy stellar mass and to be disky at $z=0$, TNG50 MW/M31-like galaxies emerge after very diverse evolutionary pathways across 13.8 billion years of cosmic evolution, in qualitative agreement with previous models. 

As quantified extensively by \citet{Sotillo.2022}, 168 of the MW/M31-like galaxies in TNG50 have undergone at least one major merger since $z=5$ and 95 of them have undergone at least one major or minor merger (i.e. at least one merger with stellar mass ratio larger than 0.1) since $z=1$. In fact, interestingly, 31 of the
198 MW/M31 analogs of TNG50 (16 per cent) have experienced a major merger (stellar mass ratio $\ge$ 0.25) as recently as in the last 5 billion years and yet present a well-defined stellar disk at $z=0$. This analysis of \citet{Sotillo.2022}, demonstrates that, according to contemporary cosmological simulations, a recent quiet merger history is not a necessary condition for obtaining a relatively thin stellar disk at $z=0$.

Furthermore, the high-redshift progenitors of TNG50 MW/M31-like galaxies could not be more varied. As telescopes such as JWST are opening a new window in the very high-redshift Universe, we provide here a glance to the progenitors of TNG50 MW/M31-like galaxies at very early epochs: Fig.~\ref{fig:progenitors}. The $z=2$ progenitors of TNG50 MW/M31 analogs span almost three orders of magnitude in stellar mass across all sampled galaxies, namely vary in the range $1.8 \times 10^8 - 2.2 \times  10^{10} \, \MSUN$ (2 dex) within the 10th–90th percentiles, with a median of $4-6 \times 10^9\, \MSUN$ \citep[][and see also their Fig. 2]{Sotillo.2022}. Such a galaxy stellar mass range increases for even higher-redshift progenitors: studies with e.g. JWST data, or others, that aim to uncover the past of the Milky Way must therefore consider large and/or broad galaxy samples. Namely, according to TNG50 the $z=7-8$ progenitors of MW/M31 analogs can have produced as much (as little) as $\sim10^8\,\MSUN$ ($\sim10^5\,\MSUN$) of stars and can be hosted by haloes with total halo mass in the range $5\times10^8-5\times10^{10}\,\MSUN$ (10th–90th percentiles).

The merger statistics and assembly histories of the TNG50 MW/M31-like galaxies are further detailed by, and tabulated in the online data released with, \citet{Sotillo.2022}. See also the merger-history catalogs released with \citet{Eisert.2022}. 

\begin{figure*}
	\includegraphics[width=1.1\columnwidth]{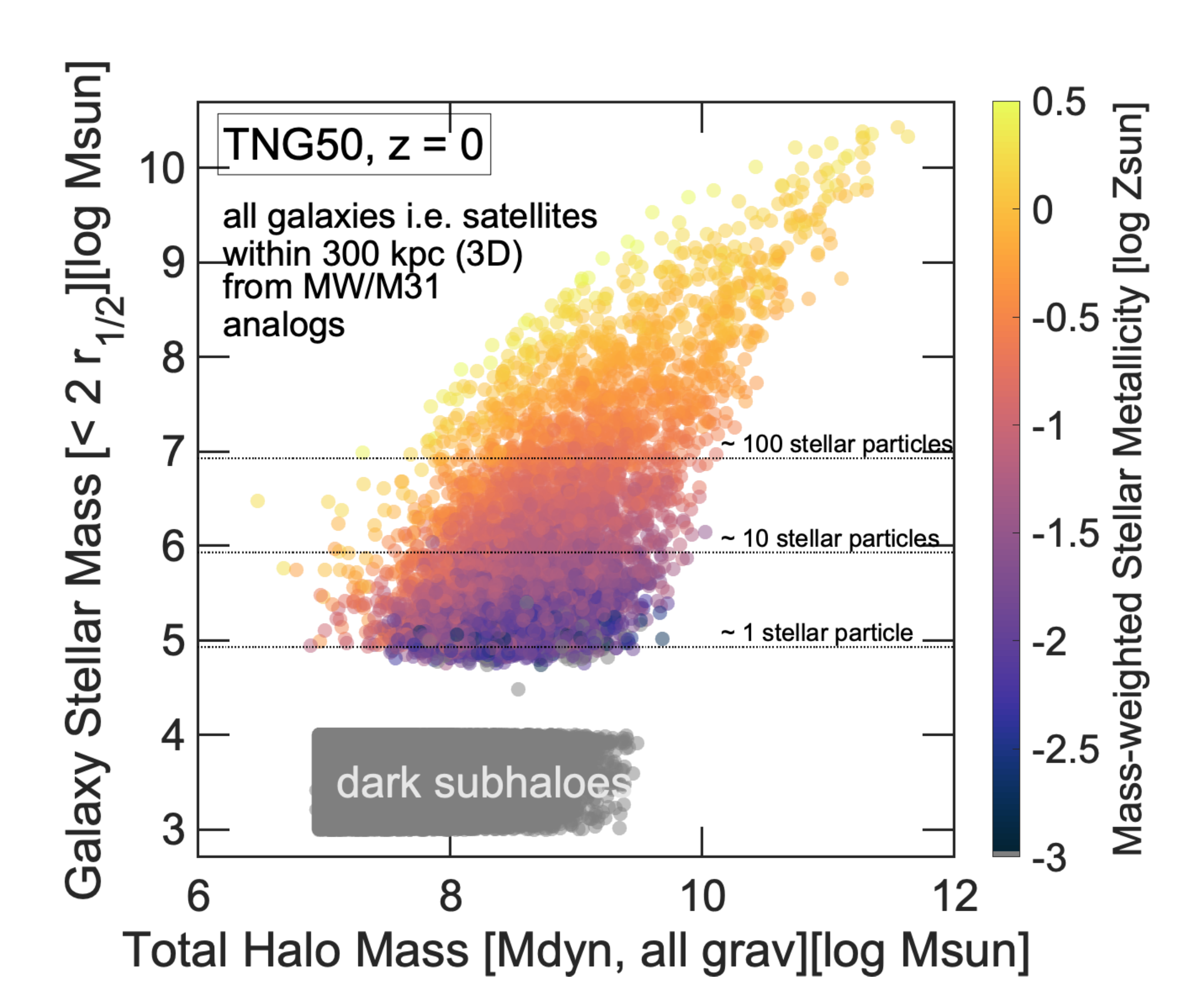}
    \includegraphics[width=1\columnwidth]{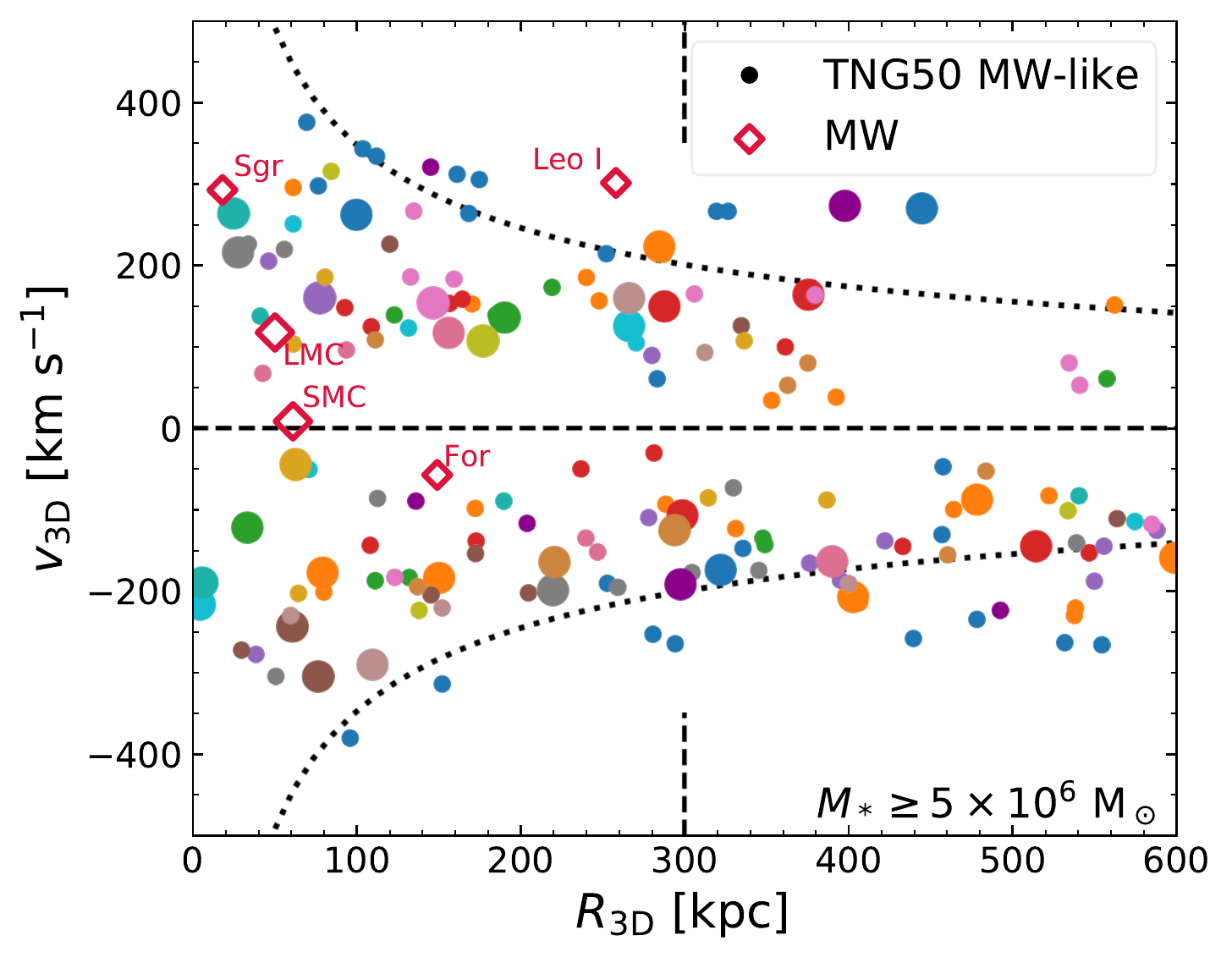}
    \includegraphics[width=1\columnwidth]{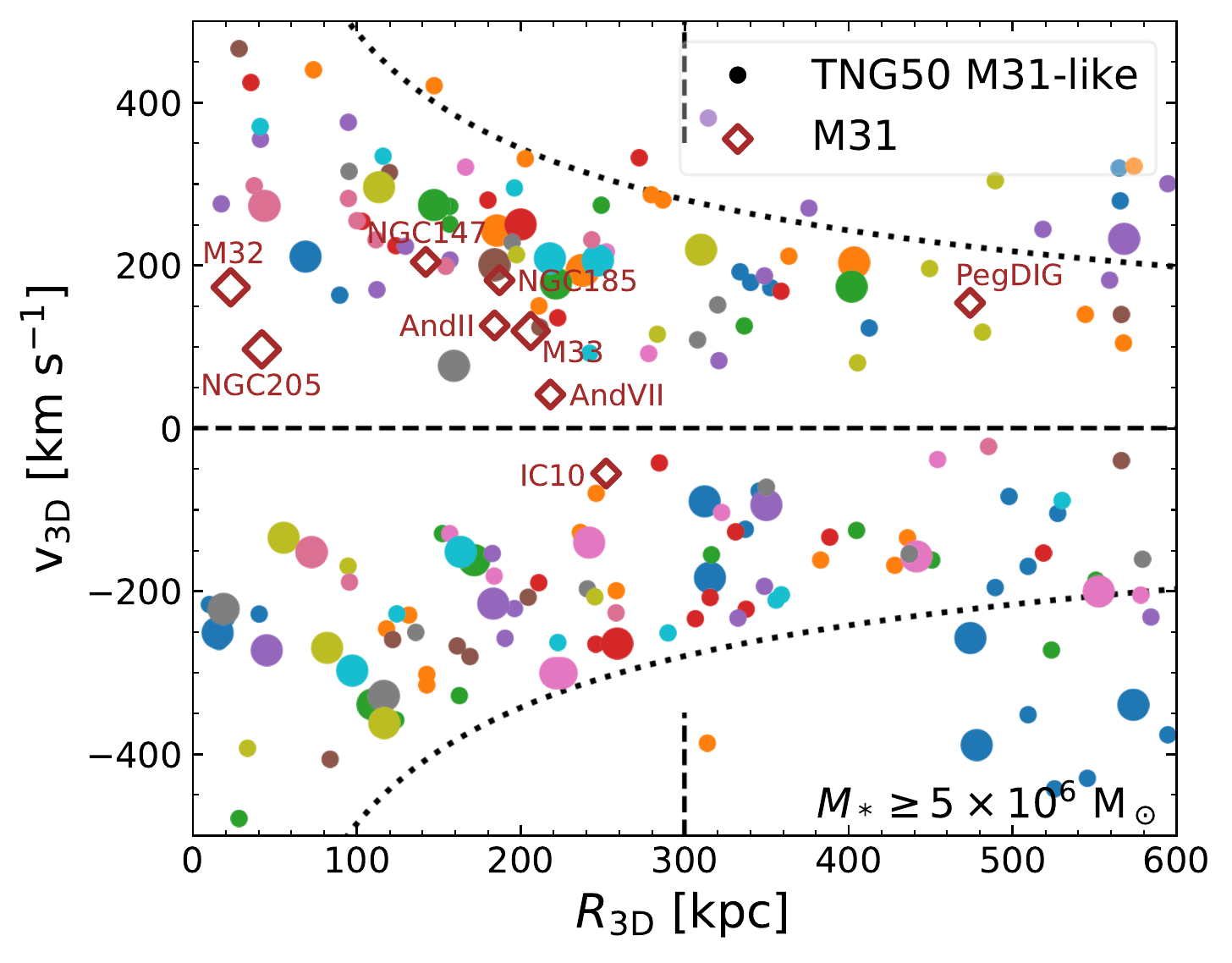}
    \caption{{\bf Satellite galaxies around TNG50 MW/M31-like galaxies at $z = 0$}. The top panel showcases the stellar-to-halo mass relation of all galaxies within a 3D distance of 300 kpc from the centers of the TNG50 MW/M31 analogs. Each satellite is color-coded by its mass-weighted average stellar metallicity. The bottom panels show the phase-space distributions of satellites i.e. galaxies in the proximity of TNG50 MW-like hosts (left) and around TNG50 M31-like hosts (right). These hosts are selected based on their galaxy stellar mass (see Section~\ref{sec:criterion_mstars}) and based on the fact that they exhibit the most similar satellite stellar mass functions to those of the Galaxy and Andromeda. Satellites are color-coded based on their closest host and their symbol size is proportional to their stellar mass. The location of the Milky Way and Andromeda satellites above the same stellar mass cut of $\ge 5\times 10^6\,\MSUN$ is also shown, from \citet{McConnachie.2012}, together with the escape velocity from a point mass of $10^{12}\,\MSUN$ for the MW and $10^{12.4}\,\MSUN$ for M31 (dotted curves).}
    \label{fig:sats}
\end{figure*} 

\subsection{Satellite populations and the case of massive satellites}
\label{sec:sats}

According to TNG50, there is no missing satellites problem, at least at the classical-dwarf scale \citep{Engler.2021}. 

In fact, TNG50 MW/M31-like galaxies host between 0 and 20 satellites with stellar mass $\ge 5\times 10^6 \MSUN$ and within a galactocentric distance of 300 kpc (3D), comfortably bracketing the observed satellite luminosity or mass functions of our Milky Way, of Andromeda and of other MW-like hosts in the nearby Universe \citep{Engler.2021}. All this corresponds to a total of 1237 TNG50 MW/M31-like satellites at $z=0$ in the aforementioned selection, i.e. with stellar mass similar or larger than that of Leo I.

The stellar-to-halo mass relation of TNG50 MW/M31-like satellites is shown in the top panel of Fig.~\ref{fig:sats}, color-coded by satellites' mass-weighted stellar metallicity.
It shows that TNG50 can in fact return galaxies as low in stellar mass as faint dwarfs, even if resolved with only a handful (or one) stellar particle. In fact, because within the TNG50 model the stellar populations can age and lose mass \citep{Pillepich.2018}, the smallest non-vanishing stellar mass in a TNG50 galaxy can be smaller than the target baryonic mass of $8.5\times10^4\,\MSUN$ (Section~\ref{sec:tng50} and \ref{sec:res}). It should be noted that, because in Fig.~\ref{fig:sats} we characterize satellite galaxies for which spherical-overdensity mass estimates are not meaningful due to stripping, we use on the x-axis the total dynamical mass of each, namely all the gravitationally-bound material with no constraint on distance -- see more discussions and alternative measures in \cite{Engler.2020}. Keeping in mind that this metric may not be the same across studies, the TNG50 satellites of MW/M31-like galaxies occupy the same stellar-to-halo-mass parameter space inferred for the classical dwarfs around the Milky Way and Andromeda (gray curves and data points in Fig.~\ref{fig:sats}, top).

Importantly, Fig.~\ref{fig:sats} (top) reiterates \citep[see also][]{Engler.2021} that MW and M31-like dwarfs can span hugely varied properties, even in their underlying total mass content given their luminous component: this is due to a combination of physical effects, including the stripping of their DM haloes after interaction with their hosts \citep{Engler.2020}. In fact, as we extensively detailed in \citet{Engler.2023}, the majority of $\ge 5\times 10^6 \MSUN$ satellites within 300 kpc from TNG50 MW/M31-like galaxies do not contain any detectable gas reservoirs at $z=0$. They are also quenched and red, unless they are massive like e.g. the Magellanic Clouds. The typical time since quenching for currently-quenched satellites is $6.9^{+2.5}_{-3.3}$ Gyr ago. Correspondingly, their stellar assembly exhibits a large degree of diversity, with more extended cumulative star formation histories for brighter, more massive satellites with a later infall, and for those in less massive hosts.  These findings are all consistent with current observational constraints for the Galaxy and Andromeda \citep{Engler.2023}. Such a diversity is mirrored in the range of average stellar metallicities, shown in Fig.~\ref{fig:sats} (top) with marker colors. The values are qualitatively in the range observed for the classical dwarfs around the Milky Way and Andromeda \citep{McConnachie.2012}. This is partly due to the large range. As expected, more massive satellites are more metal rich; however, according to TNG50, there is metallicity trend also at fixed stellar or halo mass.

Fig.~\ref{fig:sats} (bottom) visualizes instead the phase-space distribution of satellite galaxies and the host-to-host diversity mentioned above, with a few MW- or M31-like galaxies being much more satellite rich than others: satellites in the phase-space are color-coded based on their closest host. There we show the location of galaxies around selected MW/M31 analogs, specifically those with stellar mass functions most similar to the Galaxy or Andromeda. Their Subhalo IDs read: 555013, 517271, 536654, 513845, 574037, 482155, 515296, 526029, 499704, 504559; and 458470, 433289, 490814, 474008, 342447,
471248, 429471, 470345, 436932, 438148, respectively \citep{Engler.2021}.

\citet{Engler.2021} have quantified, with our fiducial TNG50 MW/M31-like sample, that hosts with larger galaxy stellar mass or brighter K-band luminosity, with more recent halo assembly, and – most importantly – with larger total halo mass typically have a larger number of surviving satellites through $z=0$: this is apparent also in Fig.~\ref{fig:sats}, with more galaxies populating the proximity of M31-like galaxies (right, $M_* = 10^{10.9-11.2}\MSUN$) than of MW-like ones (left, $M_* = 10^{10.5-10.9}\MSUN$). 

\begin{figure*}
	\includegraphics[trim=1cm 0cm 0cm 0cm, clip,width=\columnwidth]{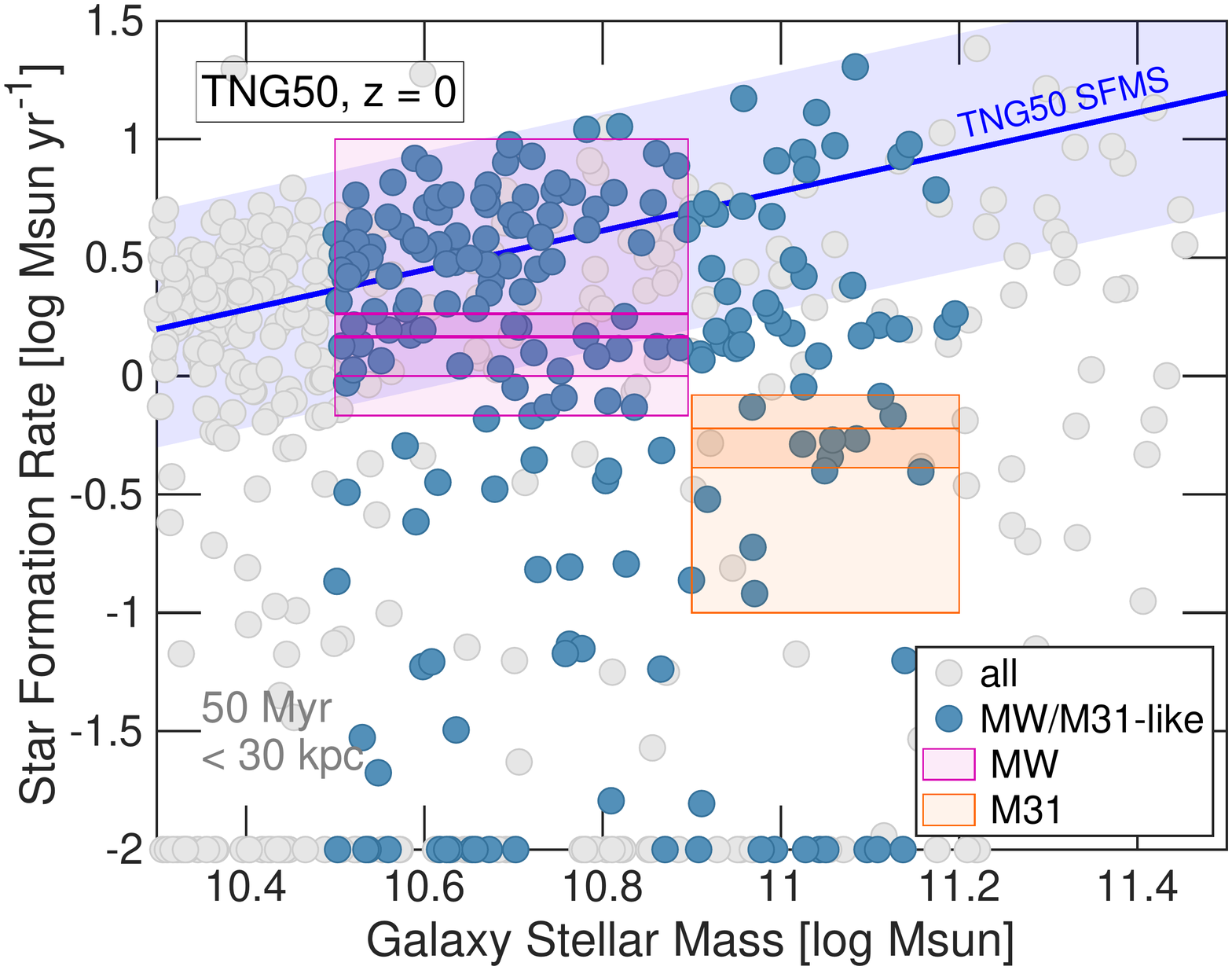}
    \includegraphics[trim=1cm 0cm 0cm 0cm, clip,width=\columnwidth]{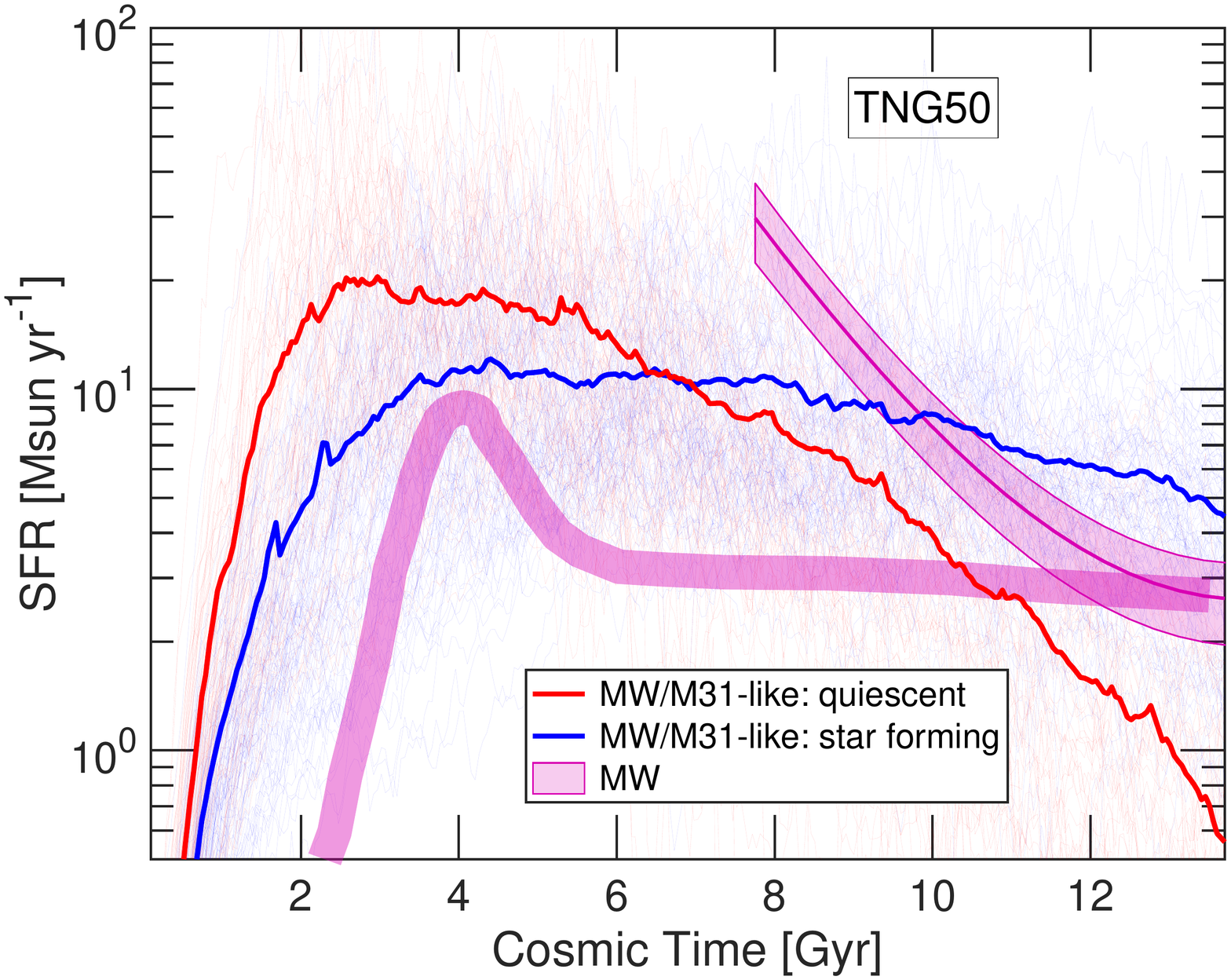}
    \caption{{\bf Star formation rates at $z=0$ (left) and star formation histories (right) of the 198 TNG50 MW/M31-like galaxies.} Our selection is agnostic to current (or past) star formation activity or stellar ages. About 40 per cent of the selected galaxies, despite being disky at the current epoch, are in fact quenched, i.e. their current SFR falls below $-1$ dex below the star forming main sequence of galaxies at similar mass. This actually allows us to have galaxies similar to Andromeda in the sample, which is known to be quenched or on the verge of being quenched: see orange constraints for M31's SFR in the left panel \citep{Williams.2003, Boardman.2020} vs. magenta ones for the Galaxy \citep{Robitaille.2010, Leitner.2011, Bland-Hawthorn.2016, Boardman.2020}. According to TNG50, the star formation histories of MW/M31-like galaxies that are star forming at $z=0$ (blue individual and blue average curves in the right panel) are different from quenched ones (red curves) already at higher redshifts. }
    \label{fig:sf}
\end{figure*}

\subsubsection{Analogs of the Magellanic Clouds and other massive satellites}
\label{sec:massivesats}

The list above and the bottom panels of Fig.~\ref{fig:sats} suggest, albeit only qualitatively, that there are MW/M31-like galaxies within TNG50 with satellites as massive as the Magellanic Clouds, M32 or NGC205. In fact, despite the environment selection criteria described in Section~\ref{sec:criterion_environment}, 15 of the 198 MW/M31-like galaxies have orbiting or merging companions within 300 kpc distance with stellar mass even as high as $10^{10-10.5}\,\MSUN$.

As tabulated and commented upon in \citet{Engler.2021}, massive satellites are indeed rare, but not vanishingly so. For example, 25 (7) TNG50 MW/M31-like hosts include {\it both} an SMC- and an LMC-like galaxy in their satellite population, i.e. within 300 kpc distance and with stellar mass within $\pm0.2  (\pm0.1)$ dex the inferred ones for the Magellanic Clouds. See corresponding IDs in Section 3.2.1 of \citet{Engler.2021}. As shown there, these hosts are not particularly biased high in terms of their underlying total halo mass and none of them is a satellite of a more massive galaxy (see Section~\ref{sec:criterion_environment}). Furthermore, this abundance of massive hosts should not be interpreted as an indication of a possibly unrealistic stellar-to-halo mass relation of satellites predicted by TNG50 (Fig.~\ref{fig:sats}, top). These frequencies are not inconsistent with those extracted from DM-only simulations and, if anything, the presence of baryons has been shown to enhance the survivability of the most massive subhaloes, irrespective of subgrid choices \citep[e.g.][]{Chua.2017}

A particularly beautiful example is the case of the galaxy with SubhaloID 511303, which exhibits a gas bridge between the two Magellanic Cloud-like satellites: see Fig. 11, top right, of \textcolor{blue}{Ramesh et al. (2023b)}. In fact, a few of these MW/M31+LMC+SMC-like systems (7 of 25) host, in addition to the SMC+LMC-like pair, also a more massive neighbor, and, more generally, a few of these hosts are surrounded by a remarkably large number of Magellanic Clouds analogs, such as SubhaloID 479938, with two SMC- and three LMC-like satellites.

Yet, in none of these MW/M31-like satellite systems, are the Magellanic Cloud-like objects in the precise phase-space locus of the observed ones, at least at $z=0$. Given the stochasticity of the hierarchical growth of structure, this is not surprising nor problematic. However, among all the 77 MW/M31-like galaxies surrounded by at least a SMC-like satellite and among the 42 ones surrounded by at least a LMC-like satellite, 4 in each grouping have the SMC- (LMC)-like satellite currently within a distance and with tangential and radial velocities comparable to the ones of the actual Magellanic Clouds. To identify these systems, we have searched for the following conditions at the $z=0$ snapshot: the satellite must be located at a distance of $56 \pm 28$ kpc \citep[putting together constraints from both the LMC and SMC:][]{Pietrzynski.2019, Graczyk.2020}, and have a radial (tangential) velocity within $57 \pm 88$ ($269 \pm 100$) km s$^{-1}$ \citep{Costa.2009}. These are the following systems, with SubhaloIDs pairs as: 342447+342458; 400973+400981; 411449+ 411470; 502995+502998; and 358609+358614; 414918+414921; 467415+467418; 471248+471251.


Less strict conditions may return more examples. In fact, by extending the search to earlier times, about 40 systems exhibit the Small/Large Magellanic Cloud analog at the ``observed'' phase-space locus at least for one snapshot in the last 5 billion years. Similar searches could be performed for the case of Triangulum, M32 and NGC205-like satellites around Andromeda (see TNG50 occurrences in \citealt{Engler.2021}). Such a richness of configurations in TNG50 lends itself to a number of interesting scientific explorations. For example, similarly to the cases of massive recent mergers of \citet{Sotillo.2022}, Magellanic Cloud analogs also undergo complex interactions with their host galaxy, leading to bursts of star formation in both the satellite and in the disk of the MW/M31-like host.

\subsection{Current and past star formation activity}
\label{sec:sf}

Disky galaxies are typically star forming, across masses and epochs. This is the case also with the TNG50 MW/M31-like sample, whose selection is however agnostic as to star formation rate, stellar color and stellar ages (Section~\ref{sec:selection}). 

In Fig.~\ref{fig:sf} we quantify the $z=0$ and past star formation rates (SFRs) of TNG50 MW/M31-like galaxies. In the left, we plot the SFRs averaged over the past 50 Myr and from within 30 kpc from the galactic center of all TNG50 galaxies (gray circles) and of those that are MW/M31-like (blue circles) vs. galaxy stellar mass. Unresolved levels of star-formation in the simulated galaxies are placed by hand at 0.01 $\MSUN$ yr$^{-1}$ \citep[see][for more details]{Donnari.2019, Donnari.2021b}. Current constraints on the Milky Way and Andromeda are shown in magenta and orange: whereas the Milky Way is a galaxy on the star forming main sequence (SFMS, blue shaded area) or just below its ridge (blue line), the observational consensus is that Andromeda is a green-valley or possibly even a quenched galaxy. Gray contours and curves denote current observational estimates of the SFMS across galaxy types, surveys and selections: they show that the TNG model, and thus TNG50, return an observationally-consistent locus of star-forming galaxies \citep[see also][]{Donnari.2019, NelsonE.2021}.

As extensively discussed in previous works also in the context of TNG and TNG50 \citep[see e.g.][]{Truong.2020, Zinger.2020, Donnari.2021a, Donnari.2021b}, the stellar mass range of the Milky Way and Andromeda is transitional, in that both in observations and in more recent successful simulations, galaxies at the $10^{10.5-11}\,\MSUN$ stellar-mass scale can be either star-forming or quenched with similar probabilities. Fig.~\ref{fig:sf}, left, shows that this is the case also for TNG50: for the case of TNG50 MW/M31-like galaxies, which were selected to be disky, 115 of 198 of them are star-forming or green valley (i.e. above SFMS $-1$ dex). Conversely and interestingly, TNG50 also returns galaxies with manifest stellar disky morphology but not recent star formation (red disks). These are objects with very little gas in the disk regions: of the 12 S0 galaxies among the TNG50 MW/M31-like ones, 9 have current SFRs below 0.1 $\MSUN$ yr$^{-1}$.

Moreover, based on the left panel of Fig.~\ref{fig:sf} and the current constraints on the SF activity of the Milky Way and Andromeda (magenta and orange boxes), TNG50 returns many galaxies as star-forming as the Milky Way and a few similar to our neighboring galaxy.

The right panel of Fig.~\ref{fig:sf} shows that the past SF histories of TNG50 MW/M31-like galaxies are also extremely diverse. These are constructed as observers would do from the stellar ages of the stars that are currently within each galaxy (all gravitationally-bound stellar particles; age bins of 50 Myr). Solid thin curves denote individual galaxies -- which we mostly show to highlight the galaxy-to-galaxy variation across cosmic epochs -- , whereas solid thick curves show the averaged SF histories across currently star-forming (blue) and currently quenched (red) MW/M31-like galaxies. 

Firstly, individual galaxies (thin curves) may undergo significant bursts of SF on time intervals of 50 million years, up to 100 $\MSUN$ yr$^{-1}$: these could be due to sudden accretion of cold gas from the large-scale structure or from flying-by and merging galaxies \citep[see][for specific examples]{Sotillo.2022}. Interestingly, such episodes of intense SF can happen as recently as in the last few billion years, depending on the assembly history of each individual galaxy. Inferred SF histories of our Milky Way cannot capture such time variability; nevertheless, the simulated galaxies fall in the ball-park of what retrieved via observations -- in Fig.~\ref{fig:sf}, right panel, we show in magenta shaded areas the SF history over the last 7 billion years of the Milky Way's {\it disk} implied by the modeling of ages, metallicities, and galactocentric radii of APOGEE red clump stars \citep{Frankel.2019} and the global one reconstructed from white dwarfs \citep{Fantin.2019}.

Importantly, it is clear from Fig.~\ref{fig:sf}, right, that TNG50 MW/M31-like galaxies that are quenched today have had, on average, rather different SF histories also prior to quenching and at $z\gtrsim1$ in comparison to those that are still star-forming (red vs. blue thick curves). In the TNG model, and hence in TNG50, massive galaxies ($\gtrsim10^{10}\,\MSUN$) quench because of the ejective and preventative effects \citep{Zinger.2020} of SMBH-driven outflows, i.e. because of AGN feedback \citep{Weinberger.2017, Weinberger.2018, Nelson.2018, Donnari.2021a}. Such SMBH feedback affects not only the SF activity of the galaxies and, eventually, their stellar structures, but also vacates the gas from the central regions of galaxies and, as already mentioned in Section~\ref{sec:cgm}, it also modulates the thermodynamical, ionization, and kinematic structure of the gaseous haloes above and below the disks \citep{Nelson.2018b, Truong.2020, Truong.2021b}. We speculate, but it should be proven explicitly, that MW/M31-like galaxies that have low levels of SF today may have grown in haloes that formed earlier, with SMBH that also grew faster and thus that reached a sufficiently-high mass to impart an effective-enough feedback to suppress SF over the past few billion years.

The TNG model hence postulates that the quenching of massive galaxies is driven by the feedback from SMBHs and, importantly, such quenching proceeds inside-out \citep{Nelson.2019b}. Such an inside-out quenching is tentatively consistent with $H\alpha$ observations of $z\sim1$ galaxies \citep{NelsonE.2021} and with HI/H2 observational constraints at low redshifts \citep{Shi.2022}. In fact, the majority of the TNG50 MW/M31-like galaxies have suppressed SF in their centre (i.e. within $1-2$ kpc) at recent epochs, even if they otherwise exhibit sustained levels of SFR globally (see next Sections and Fig.~\ref{fig:disk_HI_Halpha}. About two dozens, instead, produce stars in their centers at rates larger than 0.1 $\MSUN$ yr$^{-1}$ and potentially represent more similar candidates to our Galaxy \citep{Pillepich.2021}. We further expand on the SMBH properties and activity of TNG50 MW/M31-like galaxies in Section~\ref{sec:smbhs} and redirect the reader to \citet{Pillepich.2021} for an in-depth and quantitative description of the underlying SMBH modeling and its effects on MW/M31-like galaxies.


\begin{figure*}
\includegraphics[trim=1.4cm 0cm 1.5cm 0cm, clip, width=0.66\columnwidth]{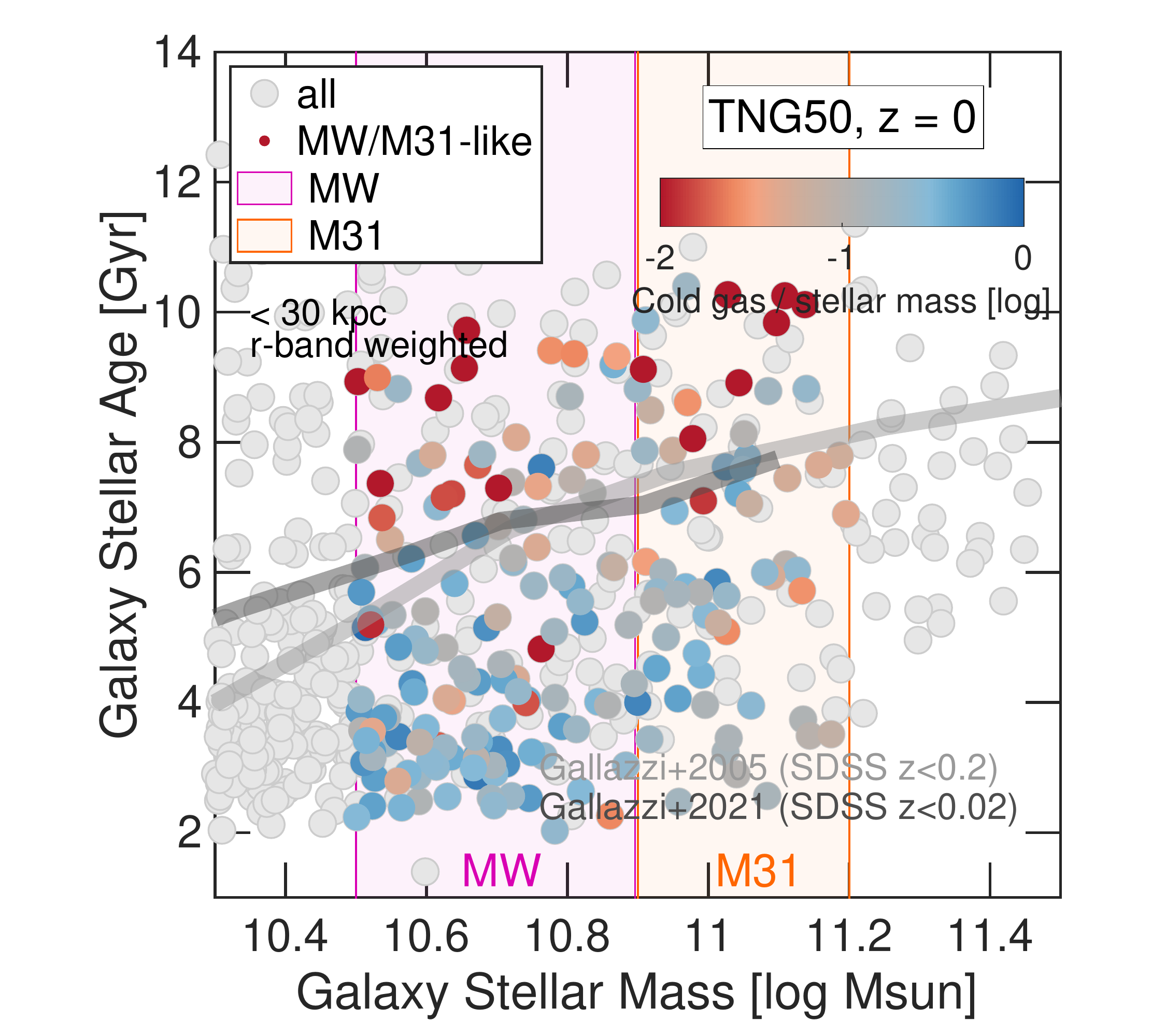}
\includegraphics[trim=1.4cm 0cm 1.5cm 0cm, clip, width=0.66\columnwidth]{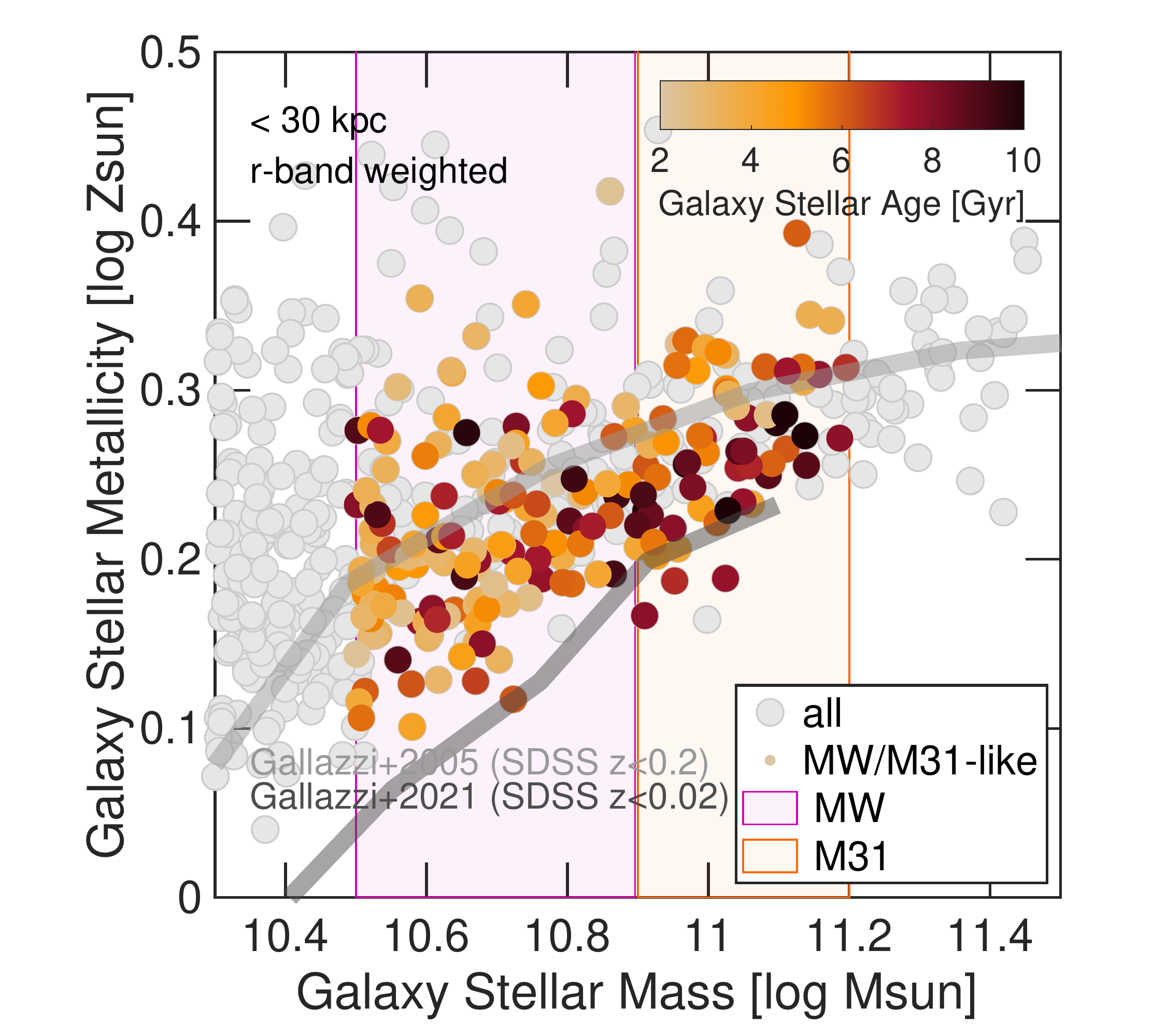}
\includegraphics[trim=1.4cm 0cm 1.5cm 0cm, clip, width=0.66\columnwidth]{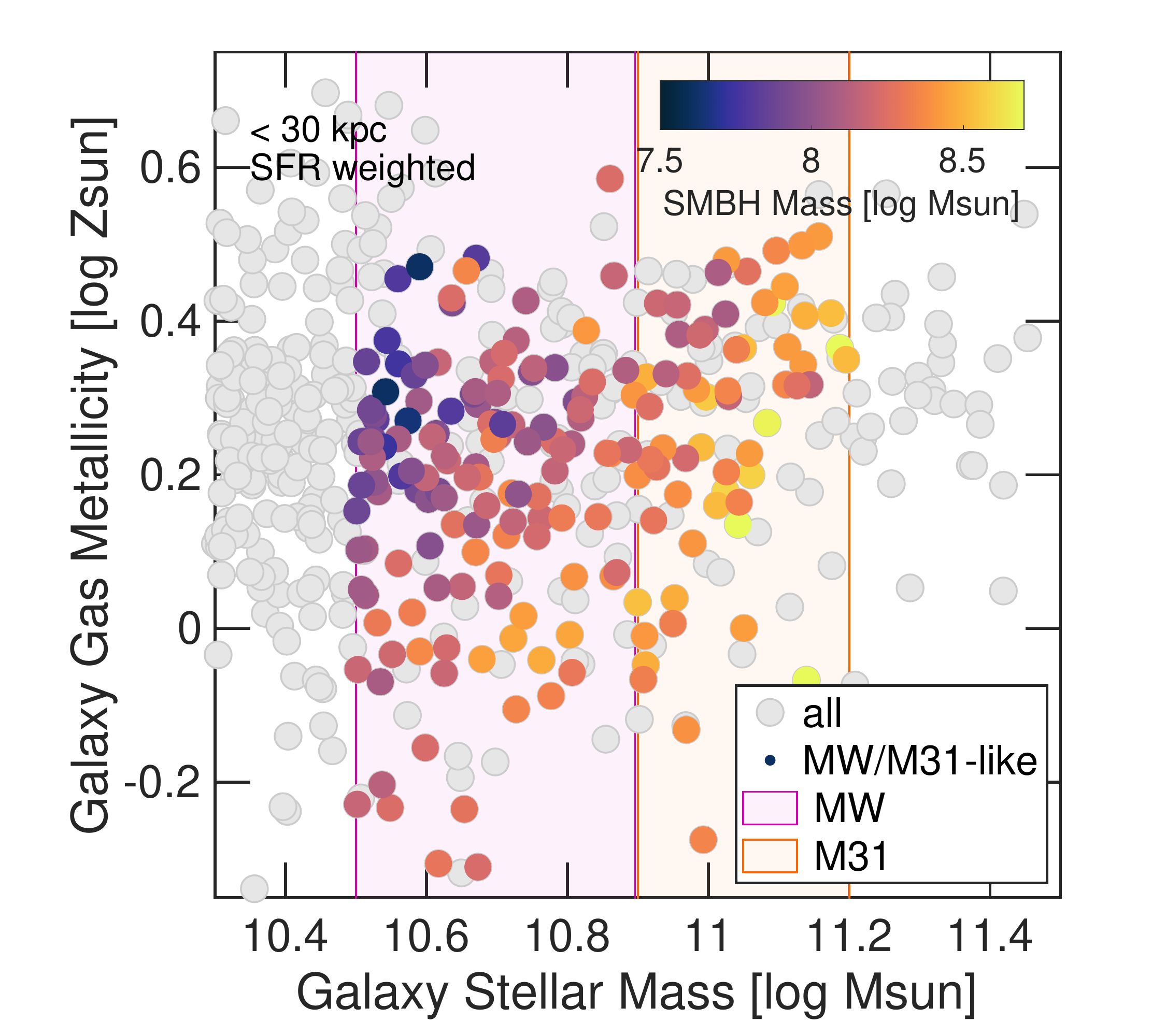}
\caption{{\bf Overall stellar population properties of TNG50 MW/M31-like galaxies.} We show luminosity-weighted mean stellar age (right) and stellar metallicity (middle) of each galaxy at $z=0$ as a function of its galaxy stellar mass, juxtaposed at face value against extra-galactic constraints from SDSS \citep[given in terms of medians across hundreds  of galaxies:][omitting the large scatters]{Gallazzi.2005, Gallazzi.2021}. Although each galaxy includes stars that encompass the whole range of ages and a much wider range of metallicities than the depicted ones, the galaxy-wide averages reflect underlying trends: galaxies with current smaller mass fractions of cold disk gas (left, colors) are older and these are somewhat less enriched globally (middle, colors). The range of average stellar metallicities is much narrower than the corresponding present-day gas metallicities (right, colorcoded by the mass of the central SMBH).}
\label{fig:stellarpops}
\end{figure*}

\subsection{Global stellar populations}
\label{sec:stellarpops}
The diverse SF histories of the previous section manifest themselves in rather diverse global, i.e. average, $z=0$ stellar population properties, namely stellar ages and stellar metallicities.

In Fig.~\ref{fig:stellarpops} we show, from left to right, r-band weighted mean stellar age, r-band weighted mean stellar metallicity, in addition to SFR-weighted mean gas metallicity of each TNG50 MW/M31 analog (colored circles) as a function of its galaxy stellar mass. All are averaged including all the gravitationally-bound stars or gas cells within a 30 kpc radial aperture; each galaxy is further color coded by the mass fraction of cold disk mass \citep[HI+H2, from][]{Diemer.2018} to stellar mass (left), the average stellar age (middle), and the mass of the central SMBH (right).

The globally-average properties of Fig.~\ref{fig:stellarpops} do not do justice to the complexity and width of the stellar populations in each individual galaxy, whose ages and especially metallicities can be many orders of magnitude diverse (see e.g. \textcolor{blue}{Sotillo-Ramos et al. submitted}), also depending on location within each system. However, this allows us to contrast the TNG50 outcome to results from extra-galactic surveys (see gray bands, representing median values across multiple galaxies in the local and low-redshift Universe). Even though we do not replicate in detail the reported measurements for external galaxies, it is encouraging to see that TNG50 returns overall realistic galaxy stellar properties: this is in line with what we found and quantified more extensively for the case of TNG100 galaxies \citep{Nelson.2018}, but we refer the reader to dedicated studies, including forward modeling and spectral fitting (\textcolor{blue}{Boecker et al. in prep.}).

Interestingly, the average stellar ages and metallicities of Fig.~\ref{fig:stellarpops}, even though global quantities, still reflect other global or past-related properties. Among the TNG50 MW/M31-like sample, the oldest galaxies are those with the lowest $z=0$ SFRs (not shown) or, correspondingly, very low relative amounts of cold gas (left panel) -- the cold gas mass can be in certain systems orders of magnitudes smaller than the stellar mass. The average stellar metallicities of TNG50 MW/M31 analogs congregate in a rather narrow super-solar range of $\sim0.3$ dex, and yet at fixed galaxy stellar mass older galaxies tend to be, on average, more metal poor than typical, at least at the M31-mass range (middle). Noticeably, the metal enrichment of TNG50 galaxies is connected to the activity and properties of their central SMBH, which, as already pointed out, modulate their SF history and $z=0$ status. At fixed stellar mass, both average stellar and gas metallicities are somewhat lower in galaxies with more massive SMBHs: we show this in the right-most panel of Fig.~\ref{fig:stellarpops}, where it can also be appreciated that present-day stellar metallicities correspond to wider ranges of present-day (SFR-weighted) metallicities of the gaseous component.

\begin{figure*}
	\includegraphics[trim=0 4.5cm 0 3.5cm, clip,width=0.95\textwidth]{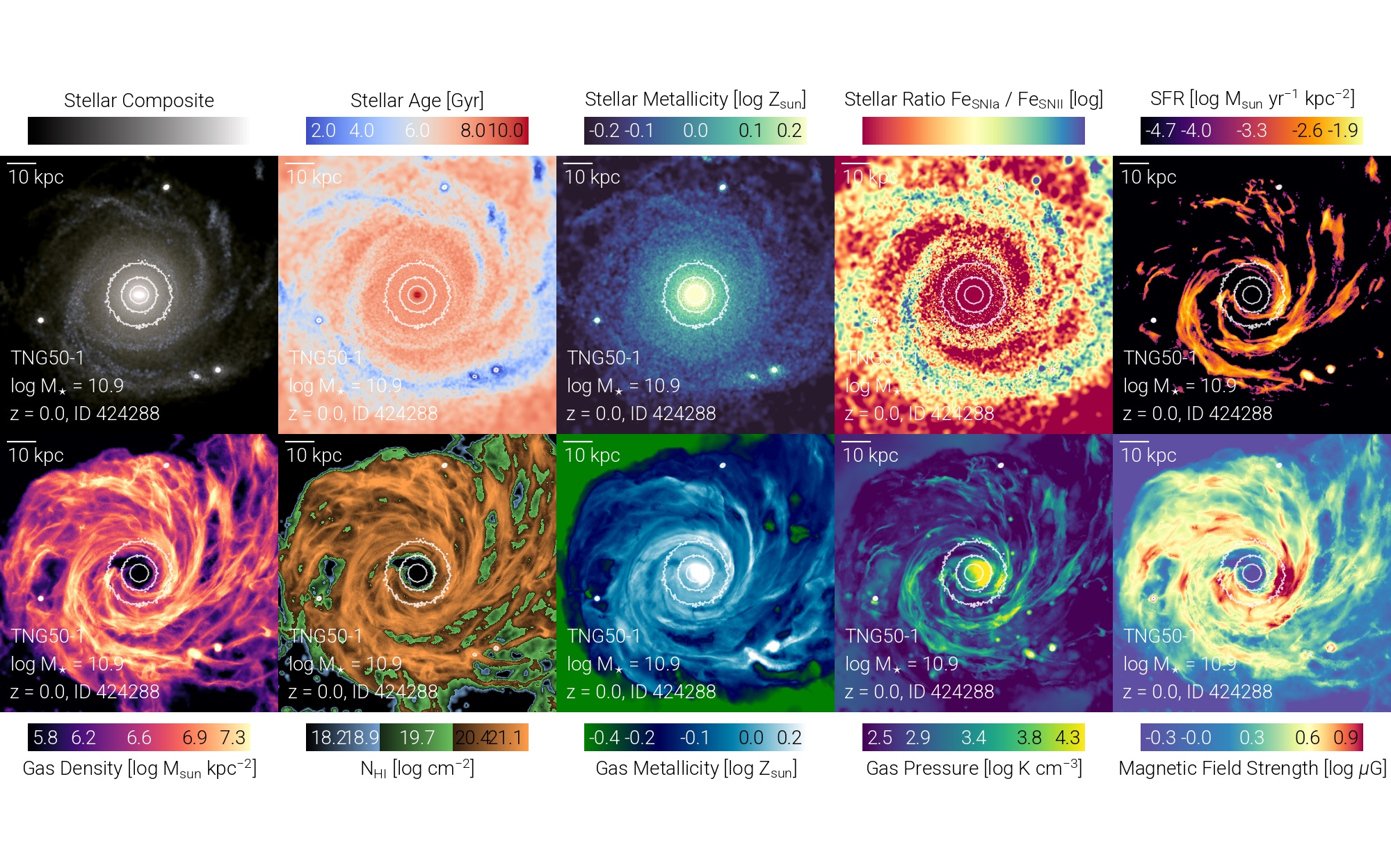}
	\includegraphics[trim=0 4.5cm 0 3.5cm, clip,width=0.95\textwidth]{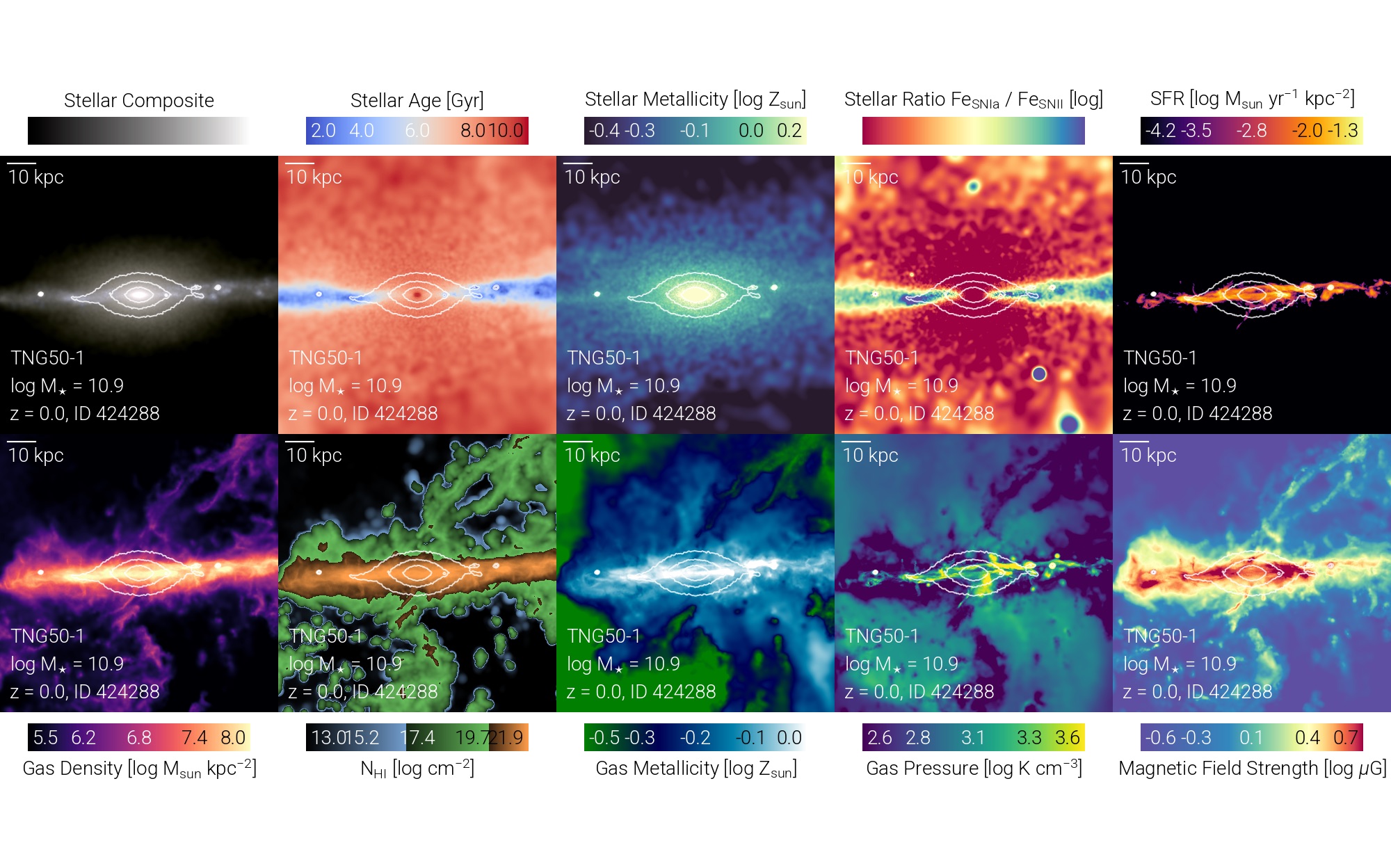}
    \caption{{\bf The many manifestations of matter and the rich phenomenology in the disk region of an example TNG50 MW/M31-like galaxy.} All stamps depict the same galaxy (SubhaloID 424288) at $z=0$, including matter in a cube of 100 kpc a side, in face-on (top) and edge-on (bottom) projections, including white contours denoting the distribution of the galaxy stellar mass. In each set of panels, the top row shows quantities pertaining to the stellar component: from left to right, stellar light for the JWST NIRCam f200W, f115W, and F070W filters (rest-frame), mass-weighted stellar ages, mass-weighted stellar metallicity, the ratio between the mass of Iron produced by SNII and SNIa, respectively, and the star formation surface density. The bottom set of panels show properties of the gas: from left to right, surface density of all gas irrespective of phase and temperature, column density of neutral atomic Hydrogen, mass-weighted gas metallicity, mass-weighted gas thermal pressure, and the strength of the magnetic fields. These represent only a portion of the information content available within the TNG50 simulation (see Section~\ref{sec:data} for a data overview). All other TNG50 MW/M31-like disks can be inspected at \url{https://www.tng-project.org/explore/gallery/pillepich23b}.}
    \label{fig:disk}
\end{figure*}

\subsection{The disks}
\label{sec:disks}

The most notable and prominent characteristics of TNG50 MW/M31 analogs is that they are disky, by selection, at least in their stellar component. In fact, the majority of them also exhibit a gaseous disk, which may be however vanishing in the case of those MW/M31-like systems with vanishing current SF activity (see Section~\ref{sec:sf}).
In the following, we provide an overview of the properties of the TNG50 MW/M31-like galaxies in their stellar disk regions. 

We start here by giving in Fig.\ref{fig:disk} a visual overview of the rich phenomenology that can be studied with the TNG50 simulation. There we show one of the 198 TNG50 MW/M31-like galaxies seen face on (top set of panels) and edge-on (bottom set of panels), as an example (SubhaloID 424288). The spiral arms that are clearly visible in stellar light (top left) are the sites of ongoing star formation (top right), with younger stellar populations, a higher fraction of metals produced by SNII compared to the delayed SNIa, and with dense and over-pressurized gaseous regions, characterized by magnetic fields of many $\mu$G. In the edge-on projections, a stream of lower-metallicity gas seems to feed the innermost region of the galaxy, which exhibits a non vanishing bulge component, and a clear inside-out and upside-down formation \citep[][]{Bird.2013}, with younger stars located closer to the mid plane and dominating progressively at larger radii within the disk.

\begin{figure}
		\includegraphics[width=\columnwidth]{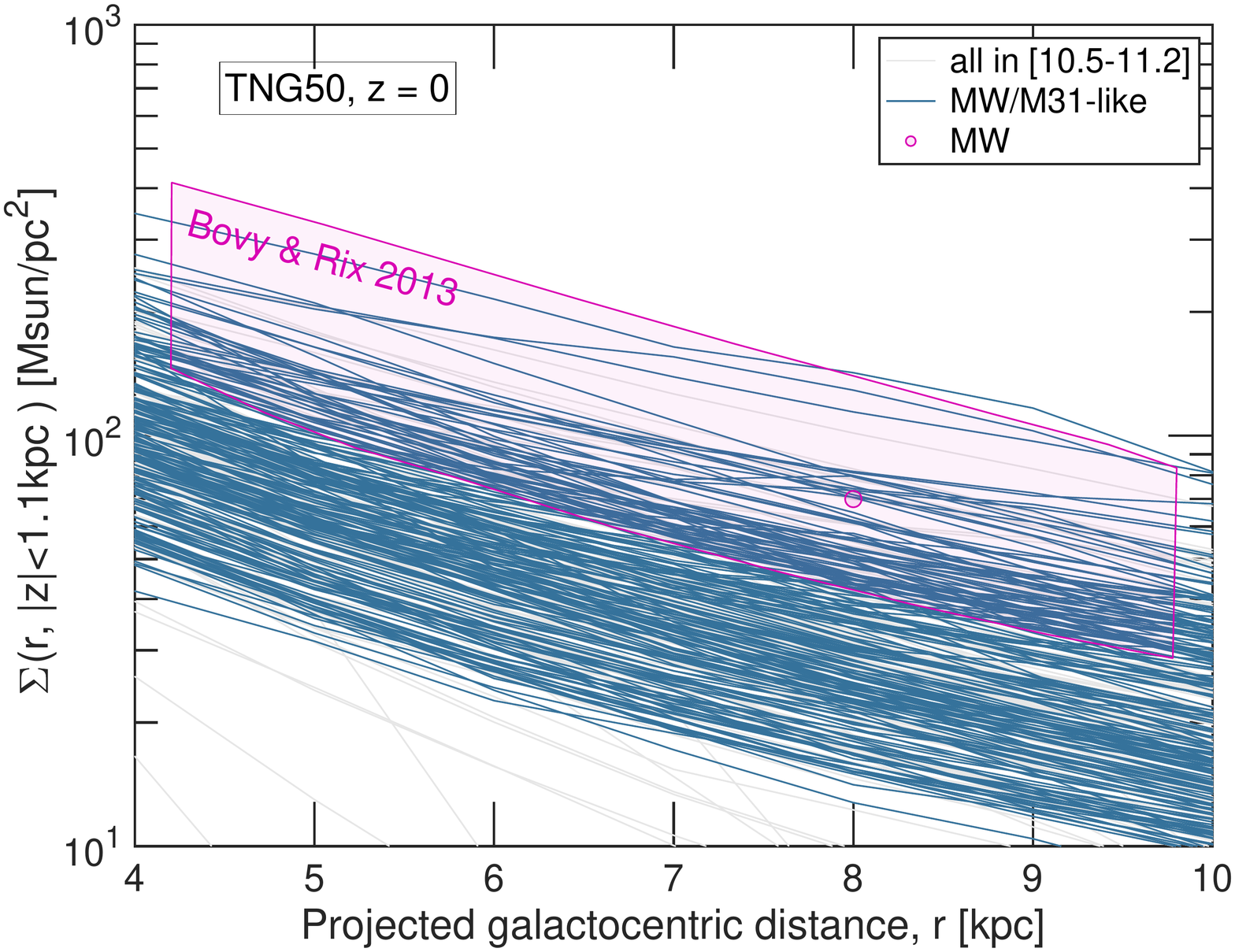}
		\includegraphics[width=\columnwidth]{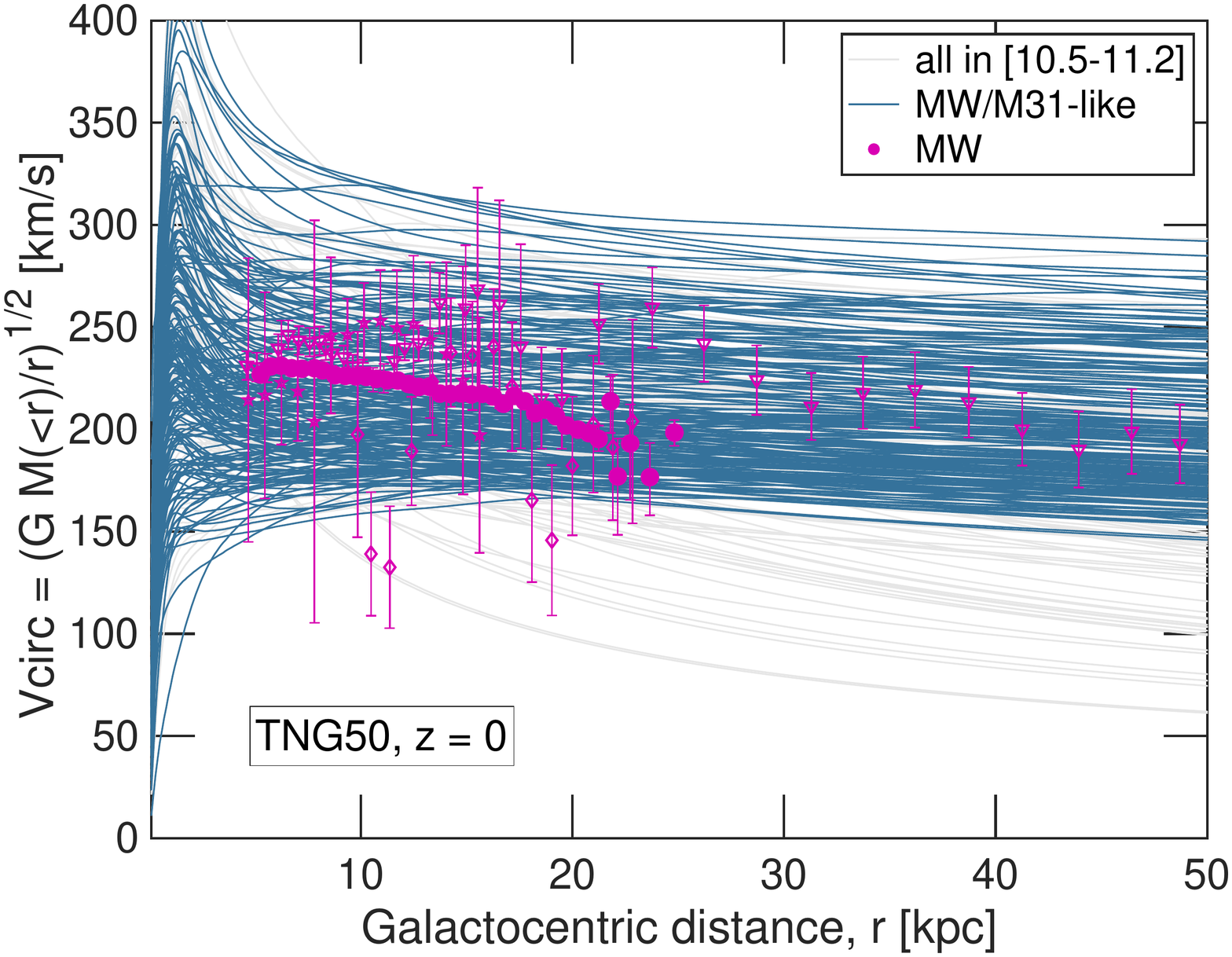}
		\includegraphics[width=\columnwidth]{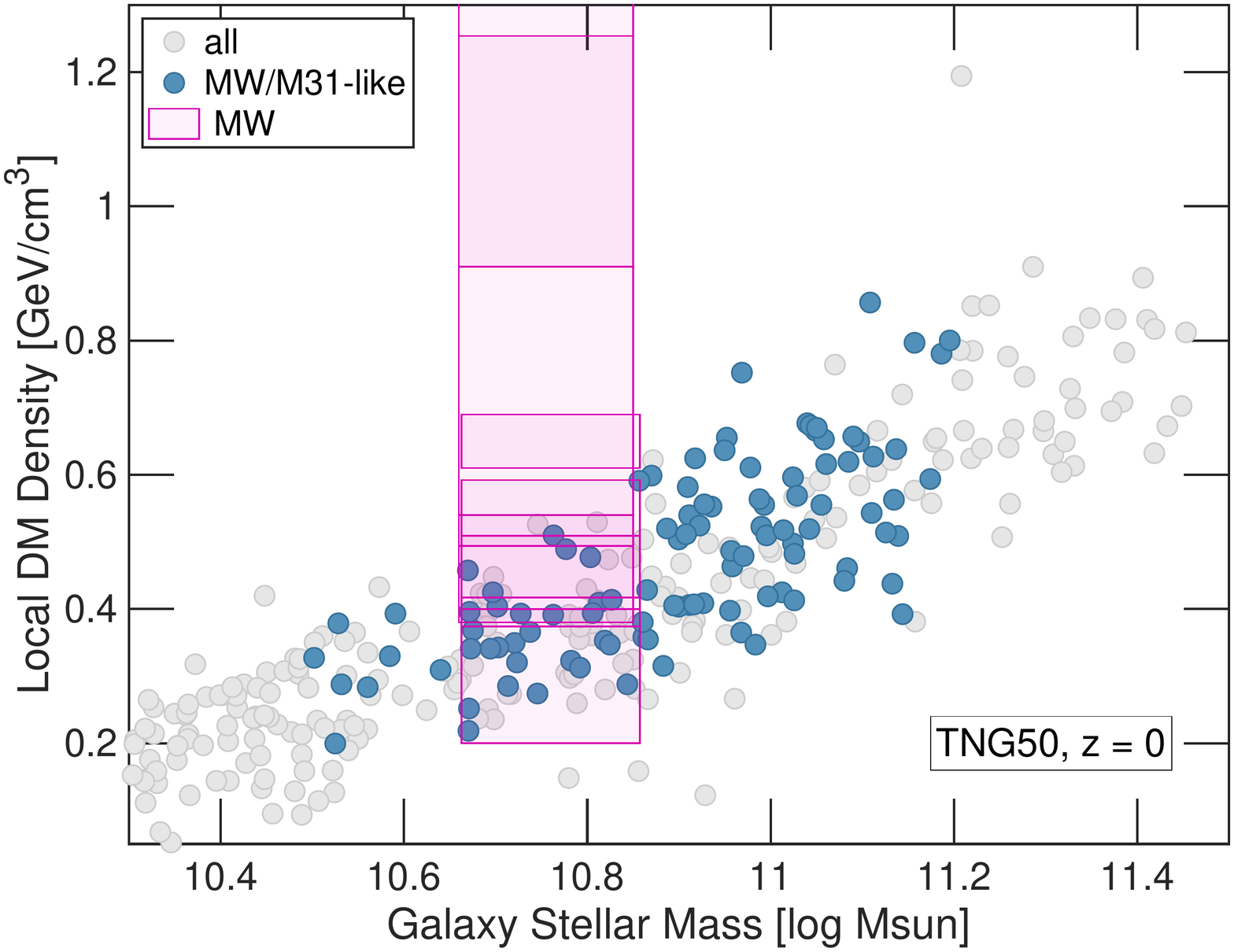}
    \caption{{\bf Matter content in the disk regions of the 198 MW/M31-like galaxies of TNG50, at $z=0$, including local DM density}. From top to bottom we show total matter surface density in the regions occupied by the stellar disks, circular velocity profiles, and (cold) DM density in the stellar disks at the nominal Sun location ($7-9$ kpc). Blue curves and circles denote the results for the MW/M31 analogs, gray symbols those for the general TNG50 galaxy population, and magenta shaded area and data points indicate constraints for our Galaxy (see text for details). The magenta data points for the circular velocity curve of the Galaxy are from a compilation by \citet{Eilers.2019}.}
    \label{fig:totaldisk}
\end{figure}

\subsubsection{Matter content in the disk regions, including local DM density}

Fig.~\ref{fig:totaldisk} shows the total matter surface density in the disk region, the circular velocity profile, and the local volume density of DM at the Sun-like location in all 198 TNG50 MW/M31-like galaxies at $z=0$.

We define the disk region as a cylindrical volume aligned with the stellar disk and extending up to a certain height ($\pm1.1$ kpc in the top panel and $\pm0.1$ or $\pm1$ kpc in the bottom panel) above and below the disk's midplane and calculate density profiles by binning particles and resolution elements in evenly-spaced linear cylindrical annuli.

In the top and middle panels, each galaxy is represented by one curve (blue) and is contrasted to the general galaxy population in TNG50 within the same stellar mass range as our sample (gray curves). Current observational constraints or observationally-based inferences for the Milky Way are given as magenta shaded areas or data points: from the models of \citet{Bovy.2013} for the surface density profile and from the measurements (filled symbols) and compilation (empty symbols) by \citet{Eilers.2019} for the circular velocity curve, via a combination of tracers (e.g. luminous red giant stars, red clump stars in the Galactic disk, blue horizontal branch stars in the halo).

The surface density of all matter components within $\pm 1.1$ kpc in the vertical direction from the mid plane (Fig.~\ref{fig:totaldisk}, top) can vary by almost one order of magnitude from galaxy to galaxy at fixed galactocentric distance and by 0.5 dex between the inner and outer disk regions in any single simulated galaxy. A number of TNG50 MW/M31-like galaxies are consistent with the available constraints for the Galaxy.

The majority of the TNG50 MW/M31 analogs exhibit a peaked central circular velocity curve (Fig.~\ref{fig:totaldisk}, middle), representing more or less pronounced stellar bulges (see Section~\ref{sec:inner}), and a flat curve at large distances. Also in this case, it is possible to identify many TNG50 MW/M31-like galaxies with $V_{\rm circ} = 225-235$ km s$^{-1}$ at the nominal Sun location ($7-9$ kpc from the center) similar to the constraints on our Galaxy. 

It should be however kept in mind that the normalization of the profiles in both top and middle panels of Fig.~\ref{fig:totaldisk} is higher the higher the galaxy's stellar mass and that, as we will see in the next Section, the disk sizes of the simulated objects can be very diverse.

Finally, the bottom panel of Fig.~\ref{fig:totaldisk} shows that also the density of the DM within a galaxy's disk and at a given galactocentric distance is higher for more massive galaxies, even in the relatively narrow range of MW/M31 analogs (blue circles, in comparison to gray circles representing all galaxies in TNG50). Here we plot the DM density averaged in an annulus at the nominal location of the Sun within a volume extending 0.1 kpc above and below the disk's midplane -- very similar results hold within $|z| < 1$ kpc, albeit with somewhat enhanced scatter. At the MW-mass scale, TNG50 returns local DM densities in the range $0.2-0.5$ GeV cm$^{-3}$ (about $0.005 - 0.013 \, \MSUN$ pc$^{-3}$), bracketing most of the available inferences for our Galaxy e.g. by \citet{Bovy.2012, Sivertsson.2018, Salomon.2020, Buch.2020, Guo.2020} but not \citet{Hagen.2018, Garbari.2012}. This occurs amid a pronounced contraction of the DM haloes, as described in Section~\ref{sec:haloes}.

\subsubsection{Structural properties of the stellar disks}
\label{sec:stellardisks}

\begin{figure*}
\centering
\includegraphics[width=11cm]{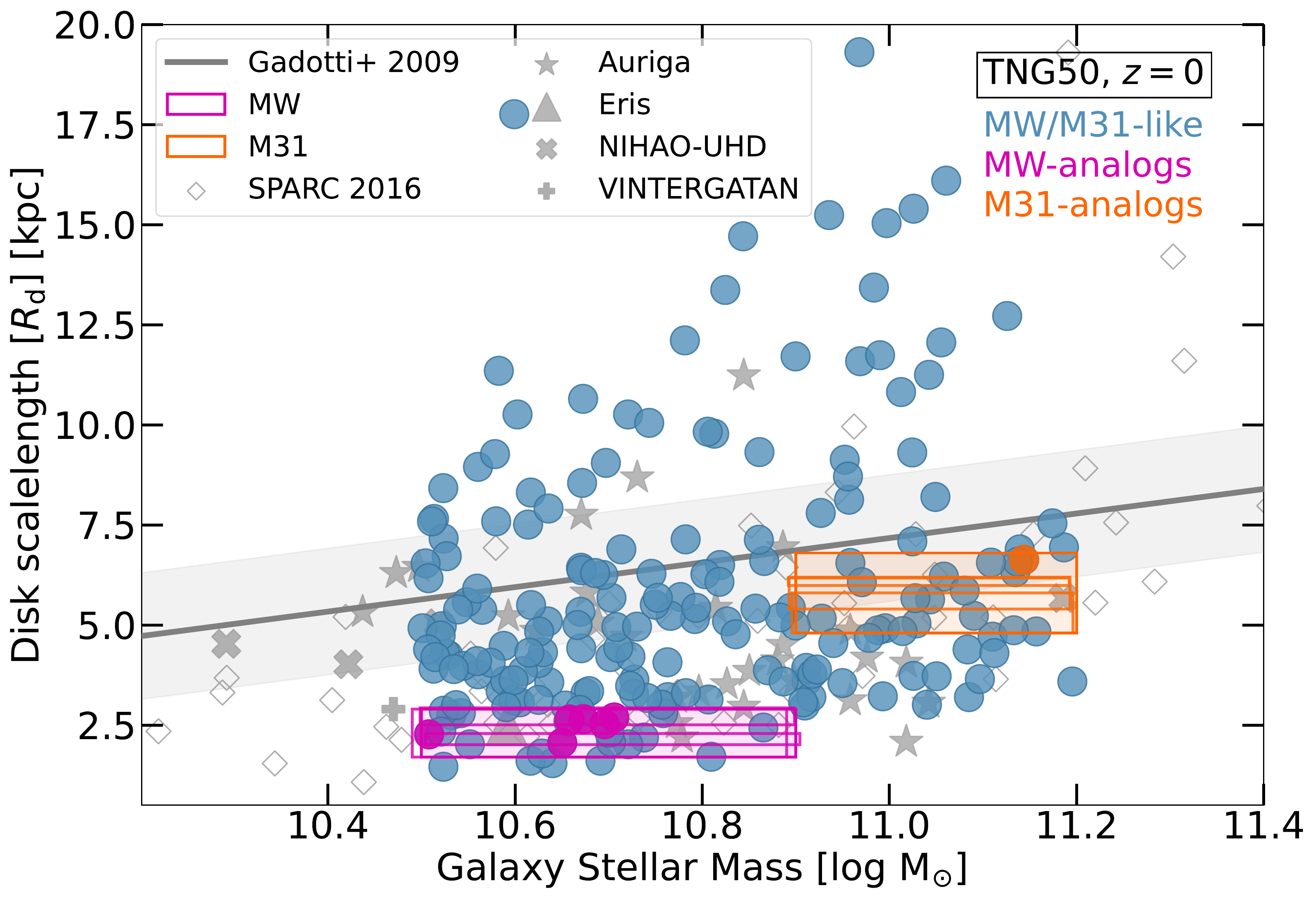}
\includegraphics[width=0.48\textwidth]{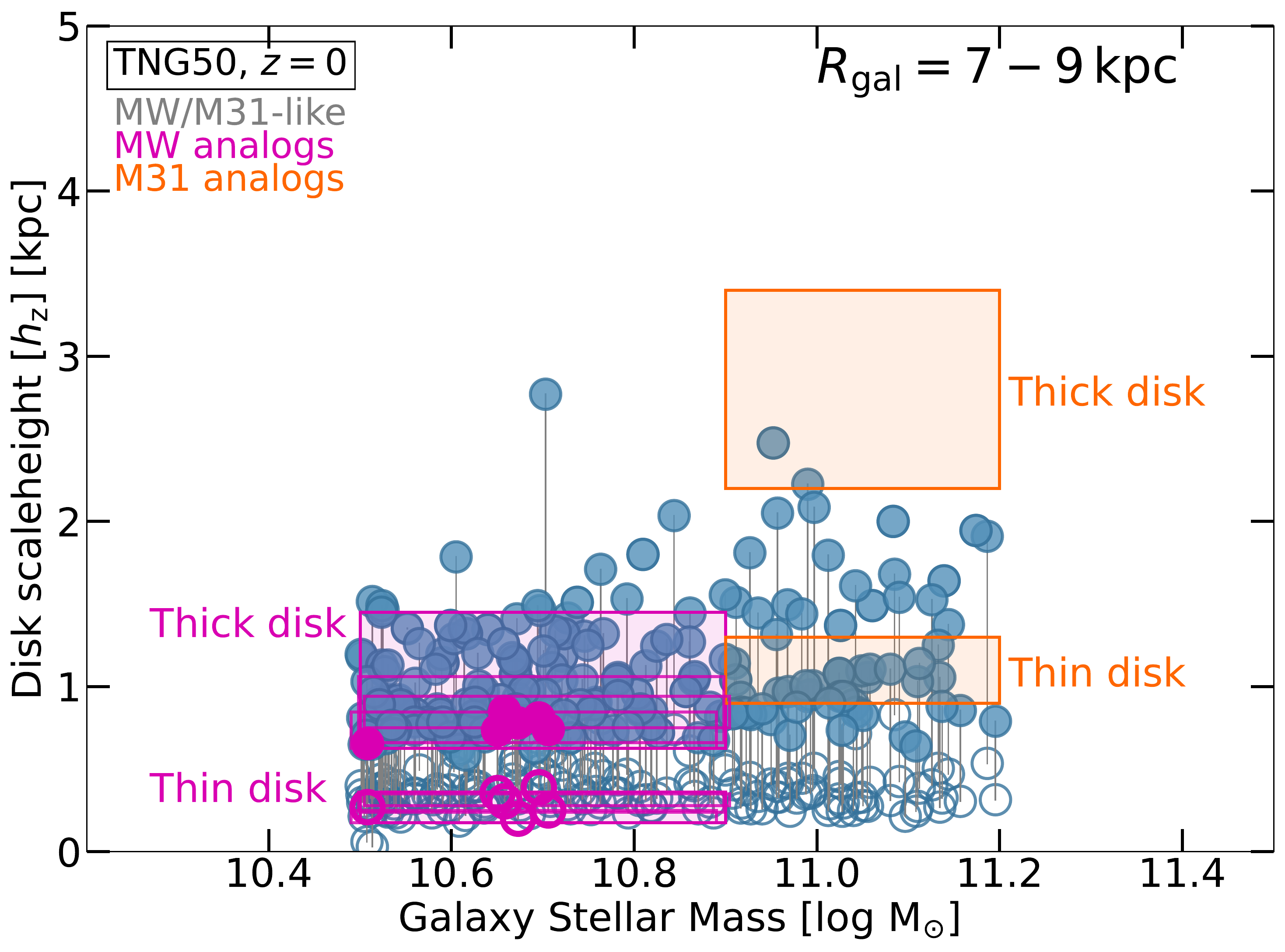}
\includegraphics[width=0.48\textwidth]{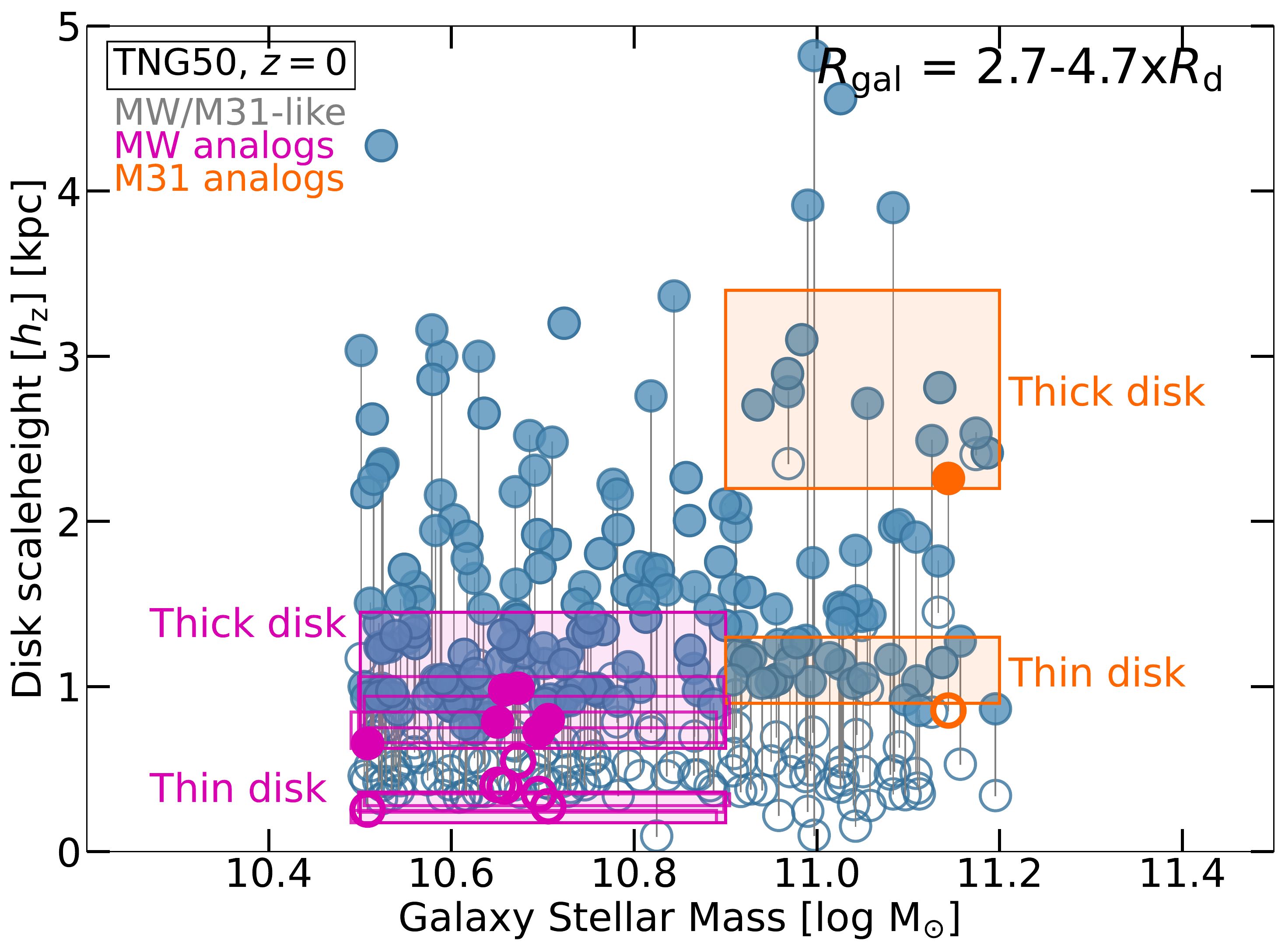}
\caption{ \textbf{The structural properties of the stellar disks of TNG50 MW/M31 galaxies.} We showcase disk scalelengths (top panel) and disk scaleheights (left and right bottom panels) as a function of the galaxy stellar mass, for the 198 MW/M31-like galaxies at $z=0$.
The scaleheights are evaluated at cylindrical shells located from the center at $7-9$ kpc (bottom left) and at $(2.7-4.7)\times$ the scalelength (bottom right, respectively): see text details.
The magenta and orange rectangles in the top panel represent observational estimates of the scalelength for the Galaxy \citep[1.7-2.9 kpc,][]{Hammer.2007,Juric.2008,Bovy.2013} and for Andromeda \citep[4.8--6.8 kpc,][]{Worthey.2005,Barmby.2006,Hammer.2007}, respectively.
In the bottom panels we show the scaleheights of the geometrically thin (empty symbols) and geometrically thick (full symbols) stellar disks; two sets of rectangles indicate the values of thin and thick disk inferred for the Milky Way \citep[magenta:][]{Gilmore.1983, Siegel.2002, Juric.2008, Rix.2013, Bland-Hawthorn.2016} and for Andromeda \citep[orange:][]{Collins.2011}. To guide the eye the thin and thick disk of each galaxy are also connected with a solid line. A number of TNG50 galaxies are identified as MW and M31 analogs, according to their mass, disk length and disk height values: their SubhaloIDs are given in the text.}
\label{fig:stellardisks}
\end{figure*}

Despite the diskyness-based selection and the relatively narrow mass range, TNG50 returns MW/M31-like galaxies that are very diverse also in terms of stellar disk structure, including diverse stellar disk lengths and heights -- see example stellar light images in Fig.~\ref{fig:examples}.

Firstly, we notice that the sub-galactic stellar morphologies are diverse, in that at $z=0$, the selected TNG50 MW/M31 analogs can appear flocculent, grand design or multi arm. By visual inspection, we estimate that among the MW-like subset, about 60, 40, and 40 of the simulated objects at $z=0$ belong to the above categories, respectively (\textcolor{blue}{Bisht et al. in prep}). Furthermore, the number of spiral arms can also vary, from no spiral arms (the already-discussed S0) to systems with up to six discernible spiral arms.\\

{\it Extent of stellar disks.} More quantitatively, the top panel of Fig.~\ref{fig:stellardisks} shows the disk scalelength $R_{\rm{d}}$ of the 198 TNG50 MW/M31-like galaxies (blue circles) as a function of the galaxy stellar mass (within 30 kpc). Here by $R_{\rm{d}}$ we mean the exponential disk length obtained by fitting the stellar mass surface density of stellar particles in disky orbits between 1 and 4 times the stellar half-mass radius (i.e. excluding the bulge region -- see \textcolor{blue}{Sotillo-Ramos et al submitted} for details on the fitting procedure, their Section 2.3.2). For comparison, we show with gray symbols the findings from selected zoom-in simulations, see also Table~\ref{tab:sims}: Auriga \citep[][star symbols]{Grand.2017}, Eris \citep[][triangle]{Guedes.2011}, NIHAO-UHD \citep[][crosses]{Buck.2020} and VINTERGATAN \citep[][plus sign]{Agertz.2021}. Current constraints on the stellar disk length of the Milky Way and Andromeda are given as shaded magenta \citep{Hammer.2007,Juric.2008,Bovy.2013} and orange \citep{Worthey.2005,Barmby.2006,Hammer.2007} areas, respectively, accounting for errors. Moreover, for qualitative comparison only, as based on stellar light rather than stellar mass, we include also the observed relation of local disky galaxies analyzed by \citealt[][gray line and shaded area, for median and galaxy-to-galaxy variation, respectively,]{Gadotti.2009} and the sizes of local spiral galaxies as observed in the SPARC survey \citep[][gray diamonds]{Lelli.2016}. 

As it can be clearly appreciated, TNG50 predicts a wide range of stellar disk sizes, also at fixed stellar mass. Within the TNG50 MW/M31 sample, the stellar disk scalelengths vary between $\sim 1.5$ and $\sim$20 kpc, denoting a remarkable variety of disk extents, also in even narrower ranges of stellar mass. Those galaxies with very large exponential disk length are truly extended in that they look more extended than the average in the images and their large exponential disk lengths correspond to large disk half-mass radii. Whether they can be considered low-surface brightness galaxies, similar to e.g. Malin 1, shall be assessed in dedicated studies \citep[as done e.g. by][in TNG100]{Perez.2022}: here we notice that, within the TNG model, we have shown that one known physical driver of galaxy sizes (although not the only one) is the spin of the underlying DM halo \citep{Rodriguez-Gomez.2022}.

Overall, TNG50 MW/M31-like galaxies and previous zoom-in simulations of $\sim10^{12}\,\MSUN$ haloes seem to return consistent stellar disk sizes\footnote{It should be kept in mind, however, that the one in Fig.~\ref{fig:stellardisks} is not an apple-to-apple comparison. Both disk length and galaxy stellar mass may be based on different operational definitions across the cited works. Furthermore, different galaxy selections may bias, or under sample, the resulting disk lengths.}. And, at face value, TNG50 MW/M31-like galaxies lie on the same ranges of parameter space as local observed spiral galaxies.

Importantly, we note that a number of TNG50 MW/M31 analogs fall within the observed values for the scalelength and stellar mass of the Galaxy and Andromeda, whereas the rest have more or less extended stellar disks for their mass.
For the case of the Galaxy and assuming that the TNG50 stellar disk lengths are well consistent with those of the real galaxy populations both in the median and in the galaxy-to-galaxy variation \citep[see e.g.][for additional comparisons of TNG50 galaxy sizes to data]{Pillepich.2019}, the top of Fig.~\ref{fig:stellardisks} indicates that the Milky Way has a rather compact stellar disk given its mass, as it settles at the lower end of the TNG50 distribution, while Andromeda appears a rather average galaxy for its mass. 

{\it Vertical stellar structure.} In the bottom panels of Fig.~\ref{fig:stellardisks} we show the scaleheights of thin and thick stellar disks. By thin and thick stellar disk heights we mean the scaleheights from fitting the vertical stellar mass density distribution of disk stars with a double squared hyperbolic secant functional form, so thin and thick disks here are to be intended as geometrical -- see \textcolor{blue}{Sotillo-Ramos et al submitted} for details on the fitting procedure, their Section 2.3.3. 
On the left bottom panel of Fig.~\ref{fig:stellardisks}, the scaleheights are evaluated at the radius of 8 kpc ($\pm1$kpc), as representative of the Sun location in our Galaxy -- roughly $2.7-4.7$ times the MW's disk scalelength, (for the estimations limiting values of 2.9 and 1.7 kpc, respectively). However, due to the broad range of stellar disk sizes (top panel), a fixed 8 kpc radius may fall well inside the the stellar disk of the more extended MW/M31-like galaxies. Therefore, in the bottom right panel we also show stellar disk scaleheights evaluated at $2.7-4.7$ times the disk scalelength of each TNG50 MW/M31-like galaxy.

To guide the eye, the heights of the thin (empty dot symbols) and thick (full  dot symbols) disks of each TNG50 MW/M31-like galaxy are connected with a solid vertical thin line. In fact, among 198 galaxies, for 48 of them a single vertical formula would be a better descriptor of the vertical mass distribution at $(2.7-4.7)\times R_{\rm{d}}$ (or 27 for the measurements at 7-9 kpc).
For comparison, the magenta and orange rectangles show the corresponding observational estimates for the Milky Way \citep{Gilmore.1983, Siegel.2002, Juric.2008, Rix.2013, Bland-Hawthorn.2016} and Andromeda \citep{Collins.2011}\footnote{Although the estimates for the MW are based on measurements at a quite specific radius, for M31 the estimates were performed considering a wider radial range: this must be taken into account when we examine these panels.}.

From the bottom panels of Fig.~\ref{fig:stellardisks} and confirming previous results \citep{Pillepich.2019, Sotillo.2022}, it can be appreciated that the scaleheights of TNG50 galaxies, evaluated at galactocentric distances of a few times the disk length, can be as small as $\simeq$ 200 pc (lowest 10th percentiles), excluding dramatic limitations due to the numerical resolution of the simulation. Yet, stellar disks of TNG50 MW/M31-like galaxies (as selected in Section~\ref{sec:selection}) can be also as thick as a few kpc. We think that this diversity is physical and not driven by the numerical resolution of the simulation. For example, in \citet{Sotillo.2022} we have shown that MW/M31-like galaxies with recent major mergers have, on average, somewhat thicker and decidedly hotter stellar disks (their Figs. 16 and 17).  We also note that TNG50 disk heights are consistent with those of zoom-in simulations of comparable or better resolution \citep[e.g.][not shown]{Guedes.2011,Grand.2017,Buck.2020}.

As is the case for the disk extent, TNG50 MW/M31-like galaxies have typically thicker {\it thin} disks than the Galaxy but not necessarily than of Andromeda. As we extensively quantify in \textcolor{blue}{Sotillo-Ramos et al. submitted}, there are also a few systems that exhibit the same level of {\it flaring} (i.e. increase of stellar disk height with galactocentric distance) as the one inferred for our Milky Way.

Importantly, a number of TNG50 MW/M31-like galaxies indeed exhibit thin and thick disks with similar heights as the observational estimates of both the Galaxy and Andromeda.
\begin{figure*}
		\includegraphics[width=\textwidth]{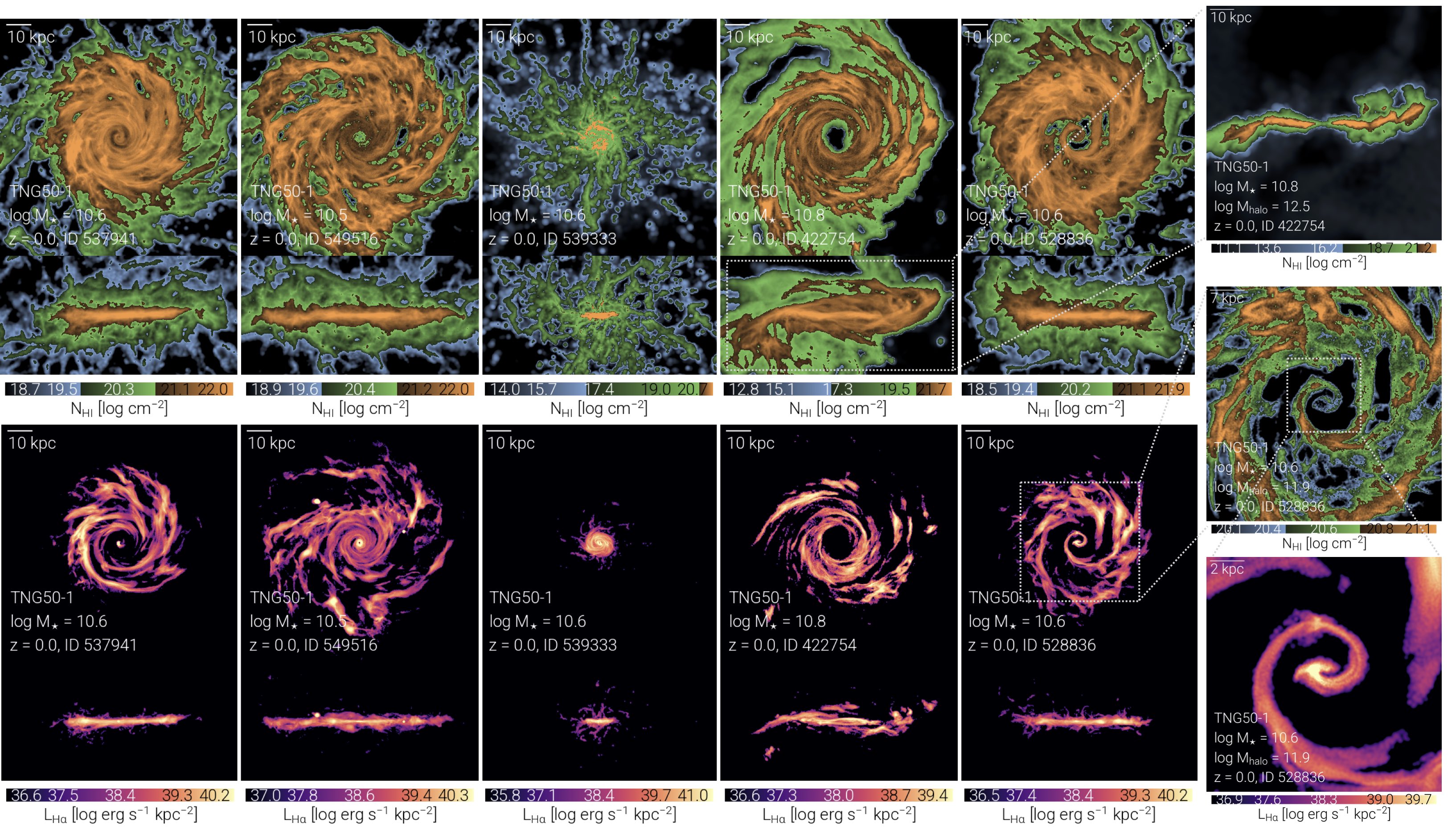}
    \caption{{\bf HI and H$\alpha$ maps of example TNG50 MW/M31-like galaxies.} We show HI column density and H$\alpha$ surface brightness of selected representative galaxies across 100 kpc a side. The gaseous disks of the MW/M31 analogs are diverse, as their stellar counterparts. On the top right, the panel shows the edge-on view of galaxy with SubhaloID 422754 but accounting only for 20 kpc in depth, to highlight the central mid-plane thinning of its HI disk. The rightmost middle and bottom panels are zoom-in views of the galaxy with SubhaloID 528833, to highlight the gas configuration in the innermost regions, where it feeds the SMBH.}
    \label{fig:disk_HI_Halpha}
\end{figure*}
We in particular highlight and take note of six galaxies (SubhaloIDs 516101, 535774, 538905, 550149, 552581, 536365), which we refer to as \textit{MW-analogs}, and one galaxy (SubhaloID 432106), \textit{M31-analog}, whose stellar disk properties, including disk lengths and total stellar mass, are within the observational estimates for the Galaxy (magenta dots) or M31 (orange dots), respectively. The MW-analogs are chosen among the TNG50 MW/M31-like galaxies that have: a) thin and thick disk heights consistent with those of the Galaxy (approximately in the range $175-360$ pc and $625-1450$ pc, respectively), measured at either 7-9 kpc or $2.7-4.7\times R_{\rm{d}}$; b) and both disk scale length and stellar mass in the ranges $1.7-2.9$ kpc and $10^{10.5-10.9}\,\MSUN$, encompassing available literature contraints.
The one TNG50 galaxy identified to have mass and disk properties consistent with observational inferences for Andromeda (M31-analog) is found within the following measurement ranges: $900-1300$ pc and $2200-3400$ for the thin and thick disks at $(2.7-4.7)\times R_{\rm{d}}$; $4.8-6.8$ kpc and $10^{10.9-11.2}\,\MSUN$ for the disk length and stellar mass, respectively, again encompassing available observational estimates in the literature. Additional TNG50 analogs of the Galaxy and Andromeda could be identified if considering errors in the measurements of the stellar disk properties also on the side of the simulated galaxies.

\subsubsection{Gaseous disks and kinematics}
\label{sec:disks_gas}
The diversity of stellar disk structures that we have described above is mirrored by an even richer phenomenology of the underlying gaseous disks. 

We expand upon Fig.~\ref{fig:disk} by showing a few more examples of gaseous disks of TNG50 MW/M31-like galaxies in Fig.~\ref{fig:disk_HI_Halpha}, for HI column density (top) and H$\alpha$ surface brightness (bottom, from the star formation rate via \citealt{Kennicutt.1998}). Firstly, the spatial distribution of the atomic hydrogen can vary from homogeneous and smooth (first galaxy from the left) to flocculent (second galaxy from the left) and fragmented. Secondly, by comparing the top with the bottom rows -- or in Fig.~\ref{fig:disk} by comparing gas density with star formation rate surface density (bottom left vs. top right) --, it is manifest that not all gas is star forming, with the cooler and denser gas being typically more centrally concentrated than HI, as expected. In fact, there are cases where there is very little star-forming gas, and thus in our maps very little H$\alpha$ light, as is the case for galaxy with SubhaloID 539333, which is below the SFMS (Section~\ref{sec:sf}) and exhibits a very compact and disturbed gaseous component. Thirdly, TNG50 MW/M31-like galaxies on the SFMS may nevertheless exhibit signatures of ongoing inside-out quenching, with more or less extended depressions of gas in their inner regions: see the cases with SubhaloID 422754 and galaxy of Fig.~\ref{fig:disk}, which also clearly shows an innermost region dominated by $\gtrsim10$ Gyr-old stars.

We have already discussed this inside-out quenching predicted by TNG50 in Section~\ref{sec:sf}, noticing that in our model it is driven by the feedback of the central SMBH. In fact, it remains unclear whether this central-gas phenomenology is supported by observations, i.e. across large samples of MW/M31-like galaxies 
On the other hand, observations by \citet{Lockman.2016} provide evidence for a vertical thinning of the 21cm HI emission centered approximately on the Galactic Center of the Milky Way, i.e. of a $\pm2-3$ kpc wide ``hole'' in the HI column density, which appears to map the boundaries of the Galactic nuclear winds associated to the Fermi and eROSITA bubbles. Suggestively, in TNG50 we not only have eROSITA-like bubbles in the CGM of most MW/M31 analogs \citep[][and Section~\ref{sec:cgm} and Fig.~\ref{fig:bubbles}]{Pillepich.2021} but also cases of similar HI-disk morphologies towards their center and mid plane: see top right panel in Fig.~\ref{fig:disk_HI_Halpha}. This is not rare but is not an ubiquitous occurrence across our sample either. In fact, the same TNG50 model also returns MW/M31-like systems on the SFMS whereby the central SMBH has not (yet) cleared out (completely) the gas from the central kpcs, and the feeding of the SMBH proceeds in a spiral-like fashion all the way to the inner tens of parsecs (below which we lose spatial resolution): see middle and bottom right-most panels in Fig.~\ref{fig:disk_HI_Halpha}, across 50 and 20 kpc, respectively.\\

\begin{figure}
	\includegraphics[width=1\columnwidth]{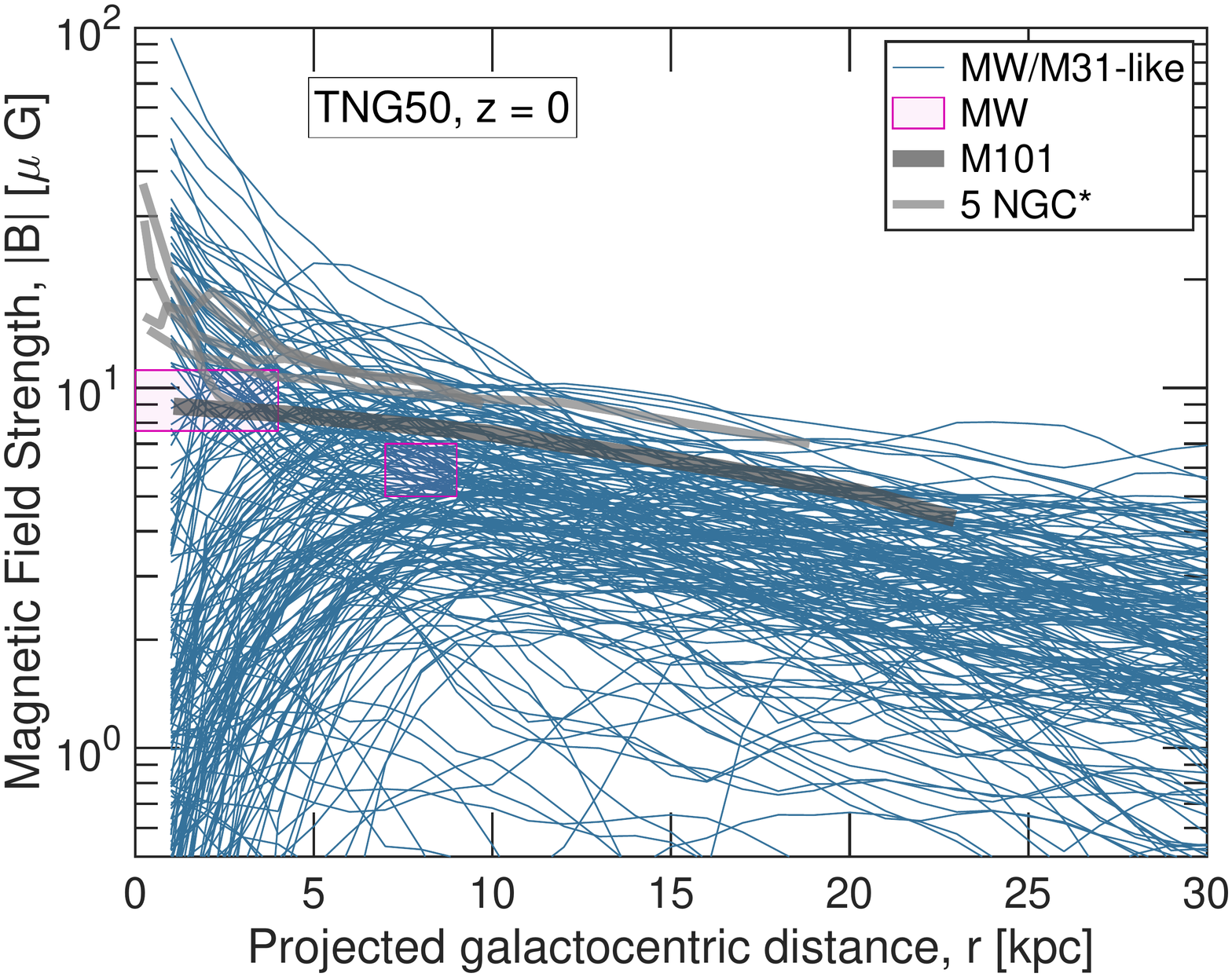}
	\includegraphics[width=1\columnwidth]{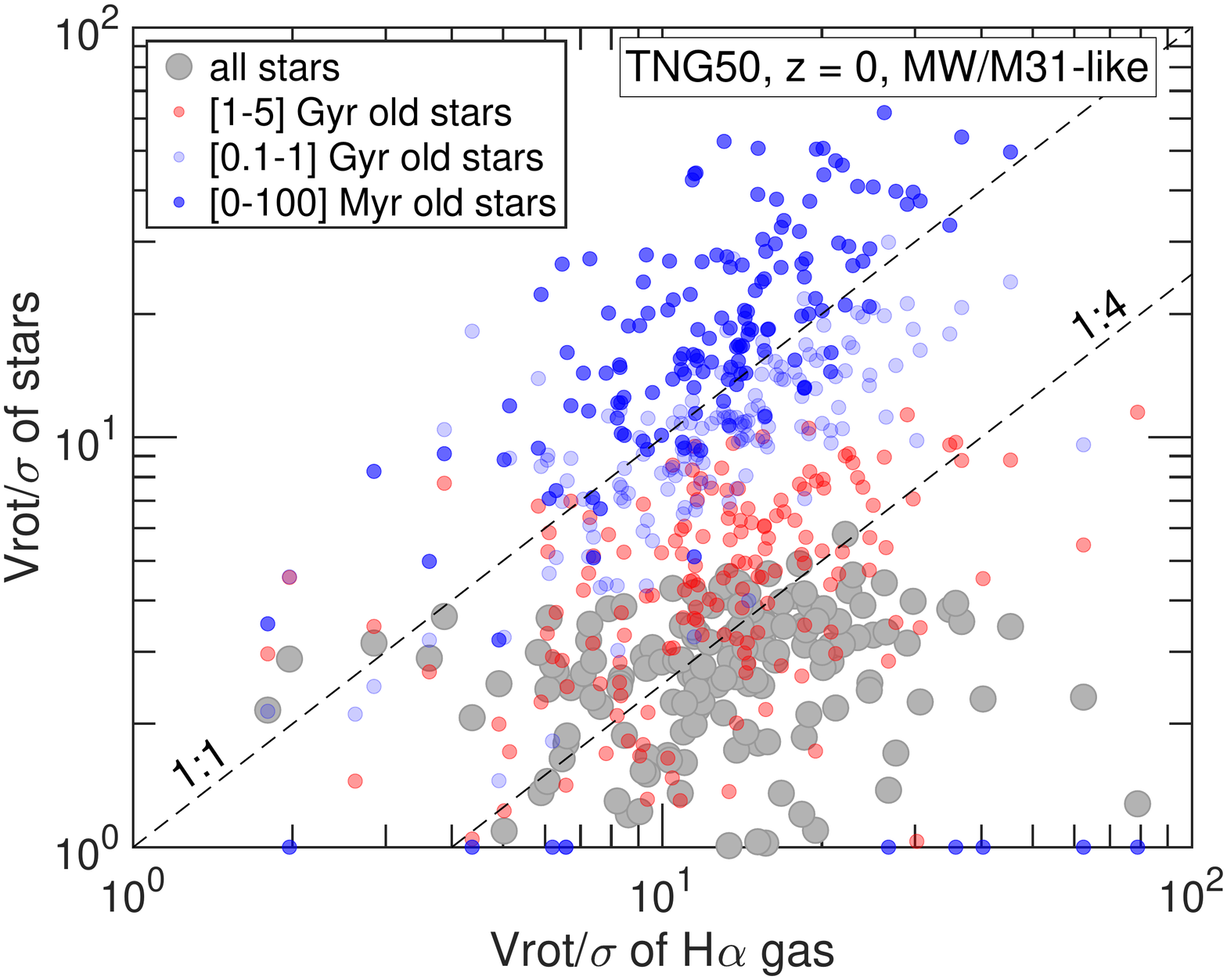}
    \caption{{\bf Magnetic field strength in the disks (top) and kinematics of TNG50 MW/M31-like galaxies (bottom).} In the top panel, the radial profiles of the mean intensity of the magnetic fields within $\pm2$ kpc from the mid plane span from $\lesssim \mu$G levels (particularly in the outer disks or in the innermost regions of a subset of galaxies) to tens of $\mu$G. Many TNG50 MW/M31 analogs are characterized by B-field strengths that appear compatible with those inferred for the galaxy (magenta shaded areas) and nearby galaxies (gray curves, see text for details). In the bottom panel, we quantify the degree of ordered versus turbulent motion (V$_{\rm rot}/\sigma$) of the stellar vs. gaseous component. The gray large circles denote values for each galaxy obtained considering all its stars, irrespective of age; the smaller colored circles show the V$_{\rm rot}/\sigma$ of stars of progressively younger age (from red to blue): younger stars are in colder orbits, as cold as those of the H$\alpha$-emitting i.e. star-forming gas.}
    \label{fig:disk_more_gas}
\end{figure}

Much could be quantified about the gaseous disks of MW/M31 analogs, also thanks to the bounty of the simulated data. We close this overview by highlighting two aspects of the disk regions of our sample.\\

{\it Magnetic field strength in the disk region.} Starting from vanishingly-small values at the initial conditions \citep[$10^{-14}$ comoving Gauss at $z>100$,][]{Pillepich.2018}, magnetic fields in TNG50 galactic disks grow exponentially at early
times owing to a small-scale dynamo effect \citep{Pakmor.2017} in addition to adiabatic compression and the combined effects of shear flows and turbulence due to structure formation and stellar and SMBH feedback processes \citep{Marinacci.2018}. As a result, the strength of magnetic fields in TNG50 MW/M31-like galaxies can be as high as many tens of $\mu$G, particularly in their innermost regions.

In Fig.\ref{fig:disk_more_gas}, top, we quantify the $z=0$ magnetic field radial profiles of TNG50 MW/M31-like galaxies in terms of mean strength in cylindrical annuli of $\pm2$ kpc thickness around the stellar disk plane. Each galaxy is shown with one curve (blue lines) and all are compared to available inferences of the magnetic field strength in our Milky Way's disk \citep[magenta shaded areas, from][]{Strong.2000, Crutcher.2010}, in M101 \citep[dark gray profile][]{Berkhuijsen.2016}, and in five nearby normal NGC galaxies \citep[lighter gray curves][]{Basu.2013}. Firstly, although these inferences are uncertain and the comparison is made at face value, it is reassuring to see that many TNG50 MW/M31 analogs share similar profiles or B-field values with observed ones, as is the case for the Auriga simulated galaxies \citep[not shown here but see][]{Pakmor.2017}. Secondly, and differently than in the case of the Auriga galaxies, not all TNG50 galaxies show declining magnetic fields strength with distance: many TNG50 systems actually show suppressed magnetic fields in the inner few kpc from the centers, in tandem with the gas mass density suppression and inside-out quenching discussed above. In fact, we showed in previous TNG-based works that stronger magnetic fields organized in large-scale, ordered, toroidal configurations naturally arise in gas-rich disk galaxies, i.e. with topologies reflective of the underlying rotationally-supported gaseous disks \citep{Marinacci.2018} -- we quantify the latter next.\\

{\it The kinematics of the gaseous and stellar components.} Even if the quantification of e.g. the HI disk heights needs to be postponed to a dedicated analysis, Fig.\ref{fig:disk_more_gas}, bottom panel, demonstrates that the gaseous disks of TNG50 MW/M31 analogs are by far dominated by ordered motions. The maxima of the rotational velocities exceed the vertical velocity dispersions of the  H$\alpha$-emitting i.e. star-forming gas by factors of many, up to two orders of magnitude. This is shown with the gray circles (one per galaxy) and with the values on the x-axis, whereby the V$_{\rm rot}/\sigma$ are estimated as in \citet{Pillepich.2019}.

By contrast, the stellar components of TNG50 MW/M31 analogs are much hotter than the gas, with V$_{\rm rot}/\sigma$ ratios about four times lower (gray circles): these are obtained by measuring the kinematics at $z=0$ of all the stars in each galaxy, irrespective of age. We had shown that, according to TNG50, stars are less rotationally-supported than the star-forming gas across all galaxy stellar masses and epochs \citep{Pillepich.2019}, and we reiterate this here for the case of MW/M31-like galaxies at the present epoch. In fact, smaller colored circles in Fig.\ref{fig:disk_more_gas} show that, in each galaxy and on average, younger stars are in colder orbits than older stars at the same time of inspection (blue vs. red): stars born in the last 100 million years are as rotationally supported, if not more, than the star-forming gas out of which they form.

It is now understood both observationally and theoretically that, at fixed stellar mass, galaxies were on average dynamically hotter at earlier epochs ($z\gtrsim1.5$) than today \citep[see][and references therein]{Pillepich.2019}. However, Fig.\ref{fig:disk_more_gas} alone cannot provide clues as to whether the hotter components of the stellar disks are such because they were born hotter or because they underwent heating after birth due to e.g. secular processes or merger or flyby events. We are fully convinced that both phenomena are, and have been, at play in TNG50 MW/M31-like galaxies across their evolution: this is quantified in detail by \textcolor{blue}{Frankel et al. in prep.}

\subsubsection{Chemical properties of the disks}
Whereas stellar radial migration has been often, albeit arguably, evoked as a possible heating process affecting the stellar kinematics discussed in the previous Section, its role in determining the metallicity and abundance make up of the galactic disks is not disputed. In \textcolor{blue}{Bisht et al. in prep} we quantify that, in the average TNG50 MW/M31 analog, disk stars that are currently located at the Sun-like location have migrated over the last 2 billion years at a rate of 0.5 kpc Gyr$^{-1}$. In the following, we briefly provide inputs on the chemical properties of TNG50 MW/M31-like disks, adding to the zeroth-order characterization of Fig.~\ref{fig:stellarpops}.

The bulk of the TNG50 MW/M31-like galaxies exhibit negative metallicity gradients, with progressively metal-poorer stars (\textcolor{blue}{Lian et al. 2023}) and gas phases \citep{Hemler.2021} at larger galactocentric distances, consistently with extra-galactic observations but with large galaxy-to-galaxy variations. In fact, recently, \textcolor{blue}{Lian et al. (2023)} have provided the first measurement of the global stellar metallicity profile of the Milky Way's disk, finding it non monotonic when all stars of different ages are considered: it has a mildly positive gradient inside a galactocentric radius of 7 kpc and a steep negative gradient outside. Intriguingly, a few of TNG50 MW-like galaxies exhibit a similar broken stellar metallicity profile (\textcolor{blue}{Lian et al. 2023}).

At any given galactocentric radius, however, the metallicity distribution of TNG50 MW/M31-like stars can be extremely wide \citep{Chen.2023}. Even though extremely metal-poor stars (e.g. [Fe/H]$<-3$) preferentially occupy the
central regions of MW/M31-like systems, three dozens among the TNG50 MW/M31-like galaxies actually have more than $10^{6.5-7}\,\MSUN$ of extremely metal-poor stars in cold circular orbits \citep{Chen.2023}.
This could provide hints to understand the origin of the very metal-poor stars with near-circular orbits that have been recently observed in the Galaxy \citep[][and references therein]{Carollo.2023}, also by investigating their various chemical element abundances. The TNG50 simulated data offers in fact the possibility of directly following back in time the origin and birth of each stellar particle in the simulation, in addition to the modeling of stellar evolution and chemical enrichment: a vast richness of phenomenology awaits to be uncovered.

\subsection{The inner regions}
\label{sec:inner}

In our journey from large to small spatial scales, we have finally reached the inner regions of TNG50 MW/M31-like galaxies (i.e. $\lesssim$ a few kpc from the centers).

With their SMBHs, stellar bulges and stellar bars, these centers are the result too of billions of years of evolution: namely, of mergers with other galaxies and their contribution of stars, gas and/or SMBHs, of gas inflows from the disk and from the CGM, of gas outflows triggered by SMBH or stellar feedback, of stellar radial migration, of dynamical instabilities, etc. They hence encapsulate enormous information about the overall assembly and history of their galaxy.
For example, according to TNG50, there can be cases among MW/M31 analogs whereby even the innermost (spherical) 500 pc of the galaxies are made of up to 20 per cent ex-situ stellar mass, i.e. star accreted from mergers \citep{Boecker.2023}, with another 20 per cent, for the average MW/M31-like galaxy, composed of stars migrated inward from the outer regions of the disk \citep{Boecker.2023}. 

\subsubsection{On supermassive black holes (SMBHs)}
\label{sec:smbhs}

All TNG50 MW/M31-like galaxies host a SMBH at their center. 

\begin{figure*}
	\includegraphics[width=\columnwidth]{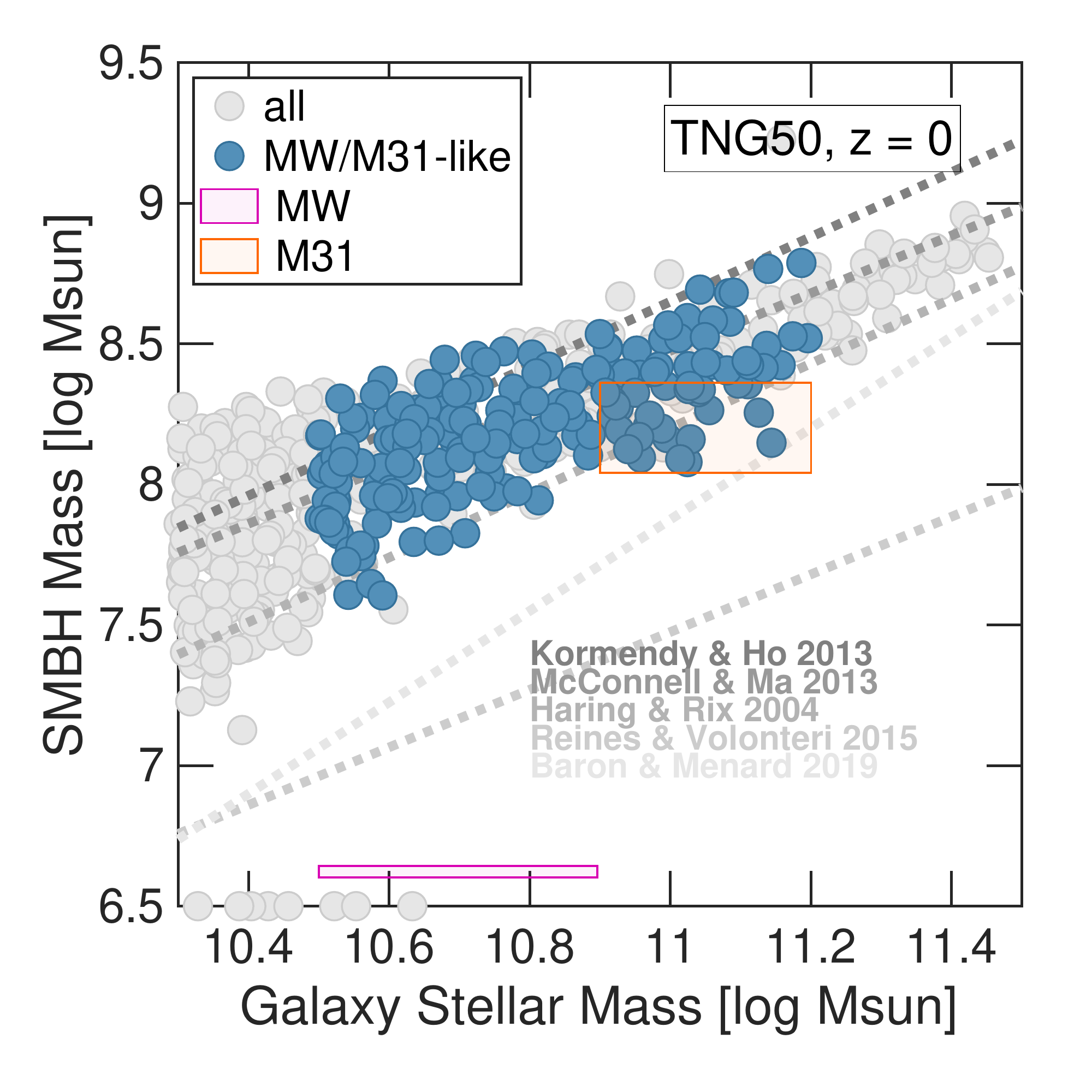}
	\includegraphics[width=\columnwidth]{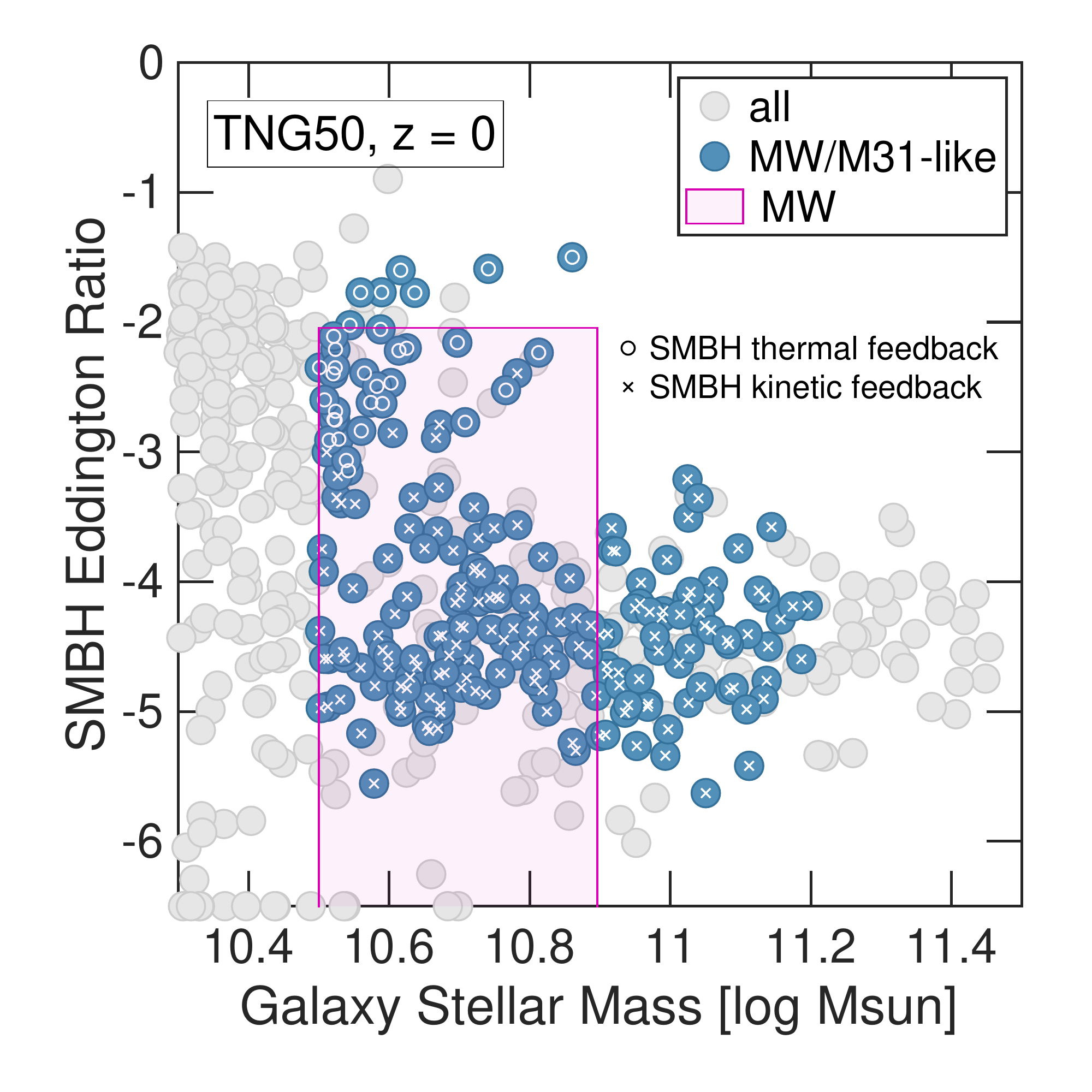}
	\includegraphics[width=1\columnwidth]{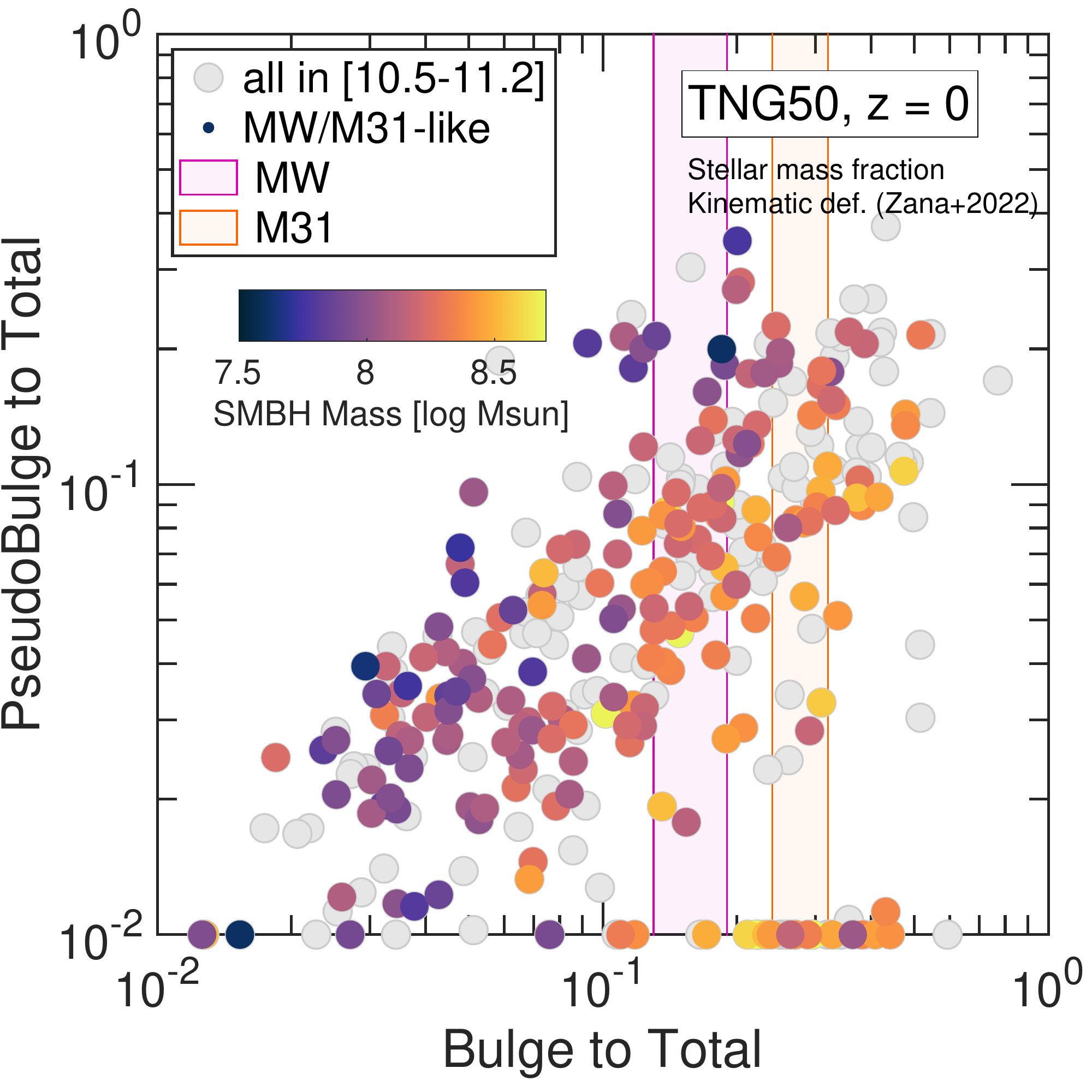}
    \includegraphics[width=1\columnwidth]{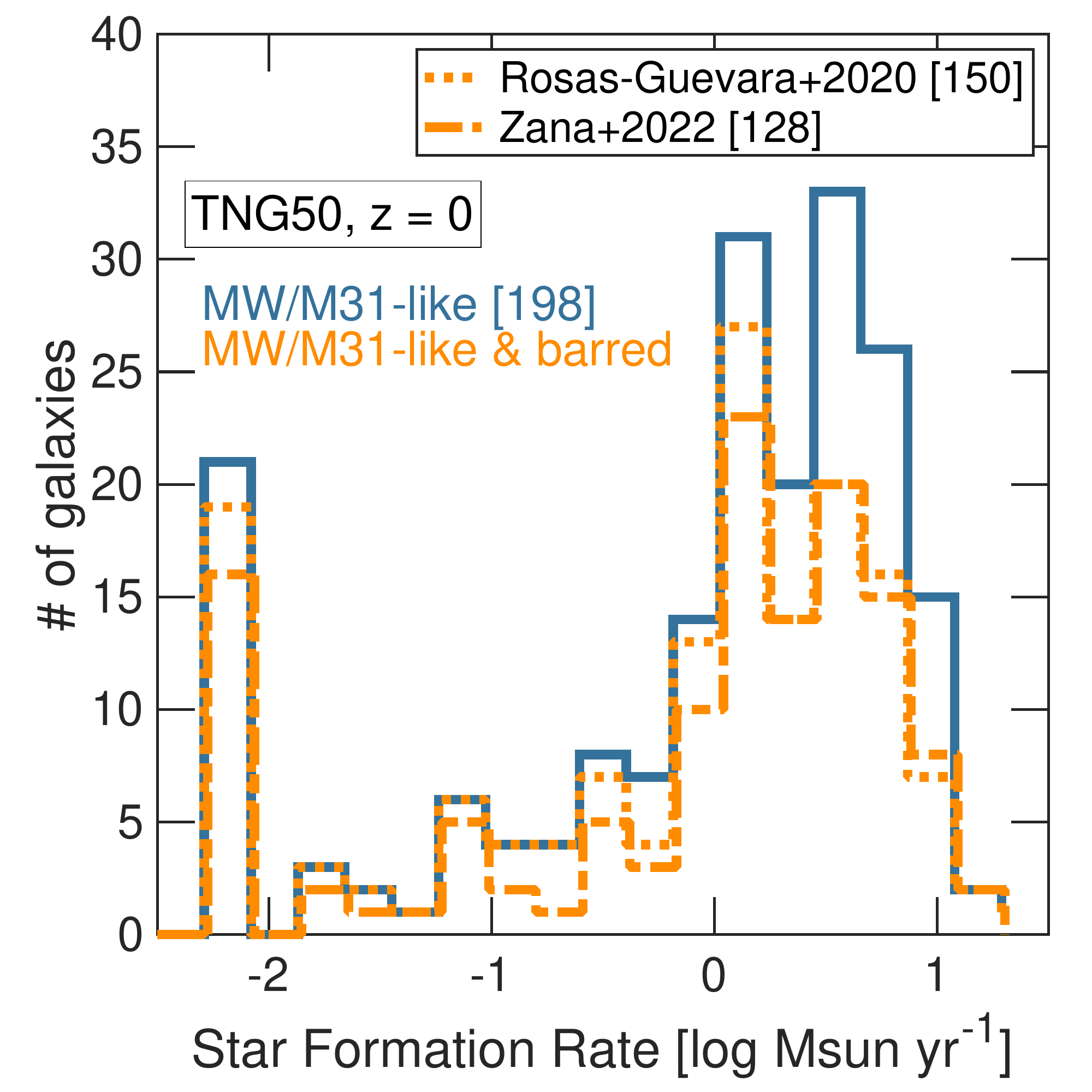}
    \caption{{\bf The fundamental characteristics of the central regions of TNG50 MW/M31 galaxies.} In the top panels, we showcase the SMBH populations of MW/M31 analogs (blue circles), including their mass (left) and their gas accretion rates, i.e. Eddington ratios (right). We indicate with circles (crosses) the SMBHs that are in the so-called thermal (kinetic) feedback mode at $z=0$. In the bottom left, we quantify the bulge-to-total stellar mass ratio of the sample, contrasted to the stellar mass ratio of pseudo-bulges and connected to the mass of the central SMBHs (colors). The bottom right panel shows that at least 2/3 of TNG50 MW/M31-like galaxies exhibit a bar.}
    \label{fig:center}
\end{figure*}

As it can be seen from the top left panel of Fig.\ref{fig:center}, the TNG50 simulation returns a relationship between SMBH mass and galaxy stellar mass that falls in the ballpark of many, albeit not all, observational constraints for the general galaxy population (see dotted gray empirical scaling relations). The SMBHs of TNG50 MW/M31-like galaxies (blue circles) span the range $10^{7.5-8.7}\,\MSUN$. Therefore, whereas the case of Andromeda with its $\sim10^8\,\MSUN$ SMBH \citep{Bender.2005} is well represented by the TNG50 outcome, no MW/M31 analog hosts a SMBH as small as that of Milky Way, i.e. $\sim4\times10^6\,\MSUN$ \citep[][and references therein]{Bland-Hawthorn.2016}. This is undesirable and reflects a too tight Magorrian relation in TNG50, in turn possibly due to numerical-implementation choices for SMBH seeding and feeding that lead to non fully-realistic outcomes. However, it shall not be considered a dramatically-limiting failure of the model. Firstly, it is important to point out that, within the larger-volume but lower-resolution run TNG300, the TNG model does return many example galaxies of MW-like mass with SMBH masses compatible with that of SgrA*. Furthermore, on the one hand, even though less unbiased observational samples seem to return also much smaller SMBHs at fixed galaxy stellar mass than earlier compilations, to what extent SgrA* of our Galaxy is an outlier remains a possibility: at face value, it has almost two orders of magnitude smaller mass than the SMBH of Andromeda, which is only $\sim0.2-0.3$ dex more massive in stars than the Milky Way. On the other hand, the impact of SMBHs on galaxies is mostly driven, in the TNG model, by their feedback and it is hence the energetic of such feedback the essential element of the SMBH-galaxy interaction. We refer the interested reader to the in-depth discussion on this issue in \citet{Pillepich.2021}, Sections 5.3 and 6.5 therein.

In terms of accretion and hence feedback, in fact, the majority of TNG50 MW/M31 analogs host SMBHs that accrete material from the surroundings at very low rates, typically $10^{-5}-10^{-3}\,\MSUN$ yr$^{-1}$. As it can be appreciated from Fig.\ref{fig:center}, right top panel, these SMBHs exhibit Eddington rates of $\sim10^{-4}$ on average \citep[see also Fig.11 of][]{Pillepich.2021}, and thus are consistent with the estimates for the Milky Way, with upper limits of current/past accretion rates of SgrA* at $8\times10^{-5}\,\MSUN$ yr$^{-1}$ \citep{Quataert.1999}. 

Within the TNG model, such low Eddington ratios imply that the majority of SMBHs of MW/M31 analogs exercise feedback in the kinetic mode: 165 of the 198 SMBHs impart high-velocity, intermittent, isotropic, accretion-disk winds (or small-scale jets: crosses in Fig.\ref{fig:center}, right top panel). The remaining ones, mostly at the lower galaxy and SMBH mass end, are in the thermal feedback mode. The different SMBH feedback in MW/M31-like have been shown to leave starkly different imprints into the physical properties of the gas in and around the galaxies \citep{Pillepich.2021, Ramesh.2023}, with the kinetic feedback ultimately being the physical cause for star-formation quenching (see also previous Sections). Interestingly, however, even though more than 80 per cent of the TNG50 MW/M31-like galaxies are impacted at $z=0$ by the SMBH kinetic feedback, only about half of them are quenched, as the feedback effects are not instantaneous. We note that the median SMBH of TNG50 MW/M31-like galaxies transitioned for the first time from thermal to kinetic feedback mode about $5-6$ billion years ago, but with an enormous galaxy-to-galaxy variation. Two thirds of the SMBHs that are currently in kinetic mode have transitioned to thermal mode only once, i.e. have never got back to sufficiently high accretion rates.

\subsubsection{On bulges}
\label{sec:bulges}
A detailed analysis of the properties and origin of photometrically-defined stellar bulges in TNG50 MW/M31-like galaxies has been presented by \citet{Gargiulo.2022}, to which the reader is referred. There it was pointed out, among others, that less than one fifth of the analogs exhibit bulges with high S\'ersic indeces ($n>2$) and that those show, on average, a higher fraction of ex-situ stars in their kinematically-selected bulges. On the other hand, the largest majority of TNG50 MW/M31-like galaxies host low S\'ersic-index bulges, and show properties more akin to {\it pseudo-bulges}.

We expand here on kinematically-defined bulges of the TNG50 MW/M31 analogs and plot in Fig.~\ref{fig:center}, bottom left, the bulge-to-total stellar mass fraction of both bulges and pseudo-bulges, according to the kinematic decomposition of \citet{Zana.2022}. MW/M31 analogs are represented by colored circles, whereas gray circles denote all TNG50 galaxies within the relevant stellar mass range.

More massive pseudo-bulges seem to cohabitate with more massive classical bulges. The median bulge-to-total (pseudo bulge-to-total) ratio for TNG50 MW/M31-like galaxies is 0.13 (0.05), well consistent with the constraints for the Milky Way and Andromeda \citep[magenta and orange shaded regions, from ][]{Boardman.2020b}. We find (not shown) that such ratios in TNG50 do not seem to depend on galaxy stellar mass (at least within the MW/M31-like range). On the other hand, as the color code demonstrates and as expected, galaxies with more massive SMBHs tend to exhibit also more massive bulges, a relationship that is reflected also on that with SFRs -- galaxies with low levels of star formation have higher-than-average bulge-to-total mass ratios (not shown). Suggestively, the bottom left panel of Fig.~\ref{fig:center} also shows that galaxies with more massive SMBHs, within the MW/M31-mass bin and within the MW/M31-like selection, have relatively less massive (or even vanishing) pseudo bulges.

On average and for what it pertains bulges, the kinematic decomposition by \citet{Zana.2022} is well consistent with that by \citet{Du.2020}, according to which the median bulge-to-total mass fraction for TNG50 MW/M31-like is about 15, instead of 13, per cent \citep{Sotillo.2022}. However, whereas there can be galaxies without pseudo-bulges, the former decomposition does not return MW/M31-like galaxies without any bulge component. Instead and because of a different technique to identify kinematically-defined morphological components, according to the decomposition by \citet{Du.2020}, 25 of 198 TNG50 MW/M31-like galaxies can be labelled as bulgeless: see Fig. 13 of \citet{Sotillo.2022}. There, we have also shown that a recent rich merger history does {\it not} imply more massive bulges for galaxies that are nevertheless MW/M31 analogs at $z=0$, contrary to expectations. 

Finally, we find no manifest connections between the relative mass of pseudo/bulges and the presence of stellar bars, discussed next.


\begin{figure*}
        \includegraphics[width=1\textwidth]{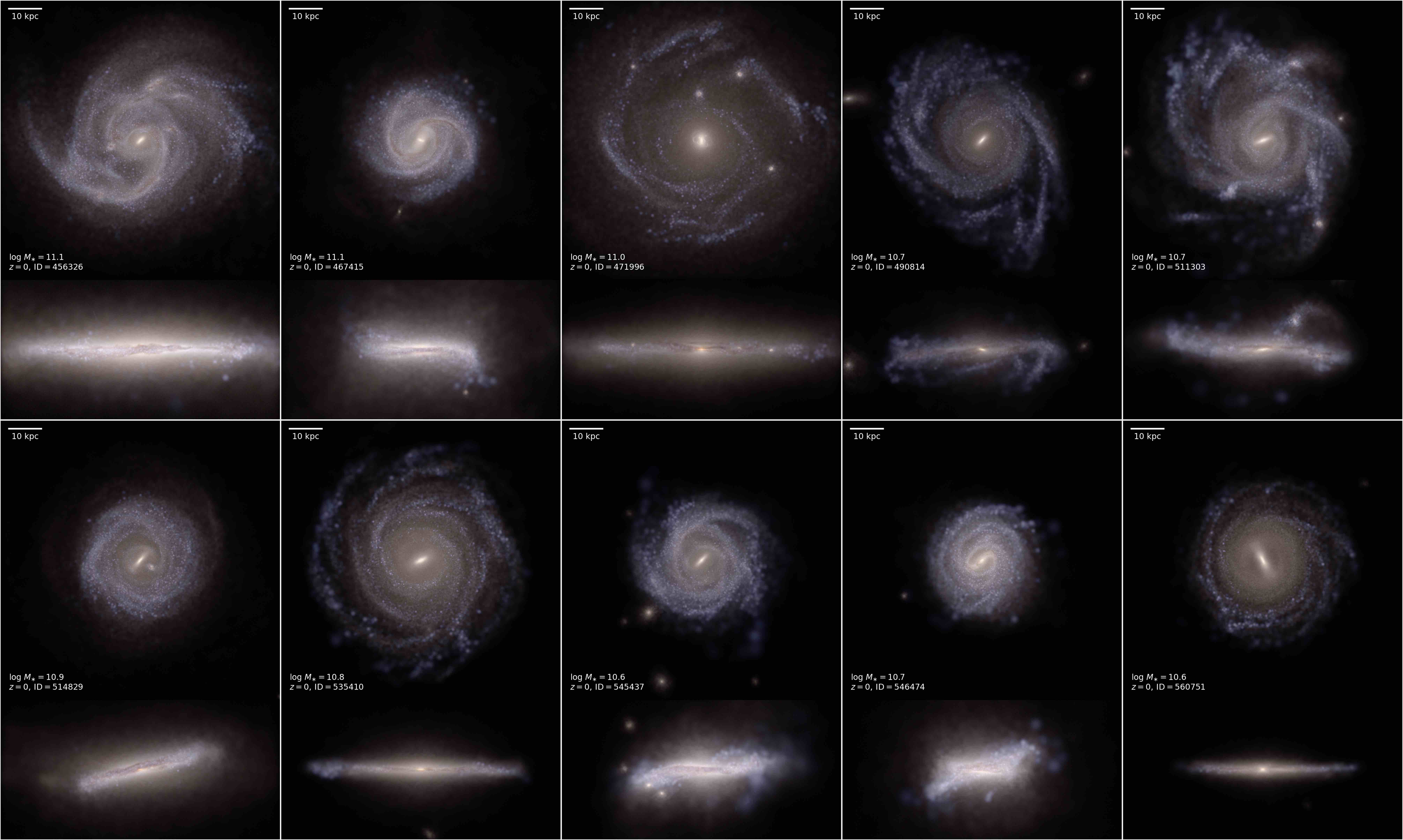}
    \caption{{\bf Examples of TNG50 MW/M31-like galaxies with a stellar bar}. At least 60 per cent of the 198 galaxies in the sample are barred, depending on how the identification is conducted. Here we show a few examples in face-on and edge-on projections, in composite stellar light images as in Fig.~\ref{fig:examples}.}
    \label{fig:barred}
\end{figure*}
\subsubsection{On stellar bars}
\label{sec:bars}
At least 60 per cent of the TNG50 MW/M31-like galaxies exhibit a stellar bar at $z=0$: see Fig.~\ref{fig:center}, bottom right panel, and a few examples in Fig.~\ref{fig:barred}.

We count 128 TNG50 MW/M31-like galaxies with bars if we adopt the criteria of \citet{Zana.2022} and 150 (among strong and weak ones) if we adopt the criteria of \citet{Rosas-Guevara.2020}, of which 118 are in common -- we provide the flags of the bar identification in Section~\ref{sec:data_MWM31scat}. 

Such a high fraction of barred galaxies is consistent with previous results based on the TNG simulations and with observations, whereby it has been quantified that galaxies with $\MS \sim 10^{11}\,\MSUN$ exhibit the highest relative occurrence of bars at $z=0$ (in comparison to the global galaxy population, with or without stellar disk). In fact, among disk galaxies, the fraction of barred
galaxies increases with stellar mass both in observations and in the TNG100 simulation \citep{Rosas-Guevara.2020, Zhao.2020}: among the TNG50 MW/M31 analogs, bars are relatively less frequent at the MW-mass scale ($\lesssim 10^{10.9}\,\MSUN$) than at the M31-mass scale. 

We also find a suggestive connection between bar occurrence and galaxy's SF activity, with the fraction of barred TNG50 MW/M31-like galaxies being higher towards lower levels of SF: Fig.~\ref{fig:center}, bottom right panel. Moreover and not surprisingly, the fraction of barred TNG50 MW/M31 analogs is higher among galaxies with lower gas mass fractions in the disk regions (not shown). This in line with the TNG100-based results by \citet{Rosas-Guevara.2020, Rosas-Guevara.2022} and \citet{Zhou.2020}, with the fraction of barred galaxies being higher among quenched and gas-poor systems. These relationships have been also seen with observational data, including intriguing correlations between bar occurrence and {\it past} SF activity \citep{Fraser-McKelvie.2020} and hence possibly interesting connections among bar formation, quenching and SMBH activity \citep[and mergers:][]{Lokas.2020}.

The bar sizes of TNG50 MW/M31-like galaxies vary between 1 and 10 kpc (also depending on definition); many bars appear ``decoupled'' from the spiral arms, whereas a few exhibit a more typical SBb morphology (i.e. a bar +two-arm spiral morphology, like NGC1300). Overall, the stellar bars produced by the TNG50 simulation and their physical properties are reasonably realistic, also thanks to its enhanced numerical resolution compared to previous simulations: via a comparison of MaNGA barred galaxies to a selection of barred galaxies from TNG50, we have shown that TNG50 bars may be shorter but are not slower than those observed \citep{Frankel.2022}, differently than what claimed in other studies and/or in the case of previous numerical simulations\footnote{The \citet{Frankel.2022} results stem from a quantitative and apples-to-apples comparison between simulated and observed galaxies: firstly, the simulated galaxies were matched in galaxy stellar mass and size distributions to the observed ones; furthermore, we have measured bar properties as closely as possible to what is done with the observed data, including measuring bar sizes from r-band SDSS-mocked images.}.

The large sample and the realism of barred MW/M31-like galaxies from TNG50 offer the opportunity to further and conclusively disentangle physical causes in the bar-quenching-SMBH loop, and to quantify bar formation scenarios and bar evolution pathways.


\section{Summary and conclusions}
\label{sec:summary}

About a decade after the first realistic cosmological hydrodynamical simulations of single Milky Way-like systems (Fig.~\ref{fig:comparison} and Table~\ref{tab:sims}), TNG50 marks a new level of maturity and scope for $\Lambda$CDM cosmological simulations of galaxies. 

TNG50 \citep{Nelson.2019b, Pillepich.2019} is a gravity+magnetohydrodynamics simulation of a large cosmological volume of 50 comoving Mpc per side. It follows the co-evolution of thousands of galaxies across the mass spectrum, cosmic epoch, and environment (Fig.~\ref{fig:environment}). With about 200 systems that can be considered analogs of our Galaxy and Andromeda (Fig.~\ref{fig:examples}), TNG50 provides a rich and large theoretical dataset with which to learn about our closest galaxies, and what they tell us about galaxy formation and evolution in general. Importantly, this is achieved with a physical model that was not designed nor intended to replicate in detail the properties of our Galaxy, Andromeda or any other specific observed galaxy, but rather of the average galaxy population.

In this paper, we have introduced, motivated, and discussed our fiducial selection for MW/M31 analogs, which we hence propose either for direct usage or for reference for future analyses. These are identified as those TNG50 galaxies with stellar disky morphology, with a stellar mass in the range of $M_* = 10^{10.5 - 11.2}~\MSUN$, and within a MW-like Mpc-scale environment at $z=0$, namely without nearby massive companions and hosted by a dark-matter halo less massive than $10^{13}\,\MSUN$ (Fig.~\ref{fig:selection} and Section~\ref{sec:selection}). 

TNG50 returns 198 of such MW/M31 analogs at $z=0$. These are resolved with stellar (dark matter) particles of $8.5\times10^4\MSUN$ ($4.5\times10^5\MSUN$) and with a gravitational softening at low redshift of about 300 pc. However, thanks to the quasi-lagrangian nature of the underlying code (\textsc{AREPO}), the spatial resolution in the gas is fully adaptive, with gas cells as small as $\sim10-20$ pc (Fig.~\ref{fig:res}) and computation time steps as short as 50 kyr in the innermost dense regions of the simulated galaxies. The $\sim150$ pc {\it average} spatial resolution in the star-forming regions (Fig.~\ref{fig:res}) means that astrophysical processes such as star formation and stellar feedback are subgrid only below the scale of giant molecular clouds and that gaseous and stellar disks as thin as $100-200$ pc are actually realized in TNG50 (Figs.~\ref{fig:stellardisks}, \ref{fig:disk_HI_Halpha} and \ref{fig:barred}, and \citealt{Pillepich.2019}). At the same time, the hot gaseous haloes surrounding MW/M31-like galaxies (Fig.~\ref{fig:LGlike}, Section~\ref{sec:cgm} and \citealt{Ramesh.2023}) are filled with numerous small, $\lesssim$\,kpc-sized cold clouds (\citet{Nelson.2020} and \textcolor{blue}{Ramesh et al. 2023b}). Therefore, thanks to its unique combination of volume and resolution, TNG50 expands by many factors (2 orders of magnitude) the sample size of cosmologically-simulated MW/M31 analogs with similar ($\times$10 better) numerical resolution (Fig.~\ref{fig:comparison}).

Our MW/M31-like sample is deliberately designed to be as complete as possible -- i.e. we did not want to miss any disky galaxies in TNG50 within the target mass range (Section~\ref{sec:criterion_diskyness}). However, it may contain a few systems, such as S0 galaxies, that are not in fact analogs of the Galaxy nor Andromeda. In this paper, we have provided the tools and suggestions on how to make the selection more stringent, depending on the chosen scientific application. 

In fact, any galaxy (real or simulated) looked at in ever-specific and in an ever-growing number of details will be, to some extent, unique. This is apparent when contrasting the measured or inferred properties of our Galaxy and Andromeda (magenta vs. orange shaded areas in many Figures). Even though our sample selection was designed to accommodate both types of systems (chiefly via the stellar mass selection, Section~\ref{sec:criterion_mstars}) and even though the two galaxies can be considered similar in the broad context of cluster galaxies and ultrafaint dwarfs, it is clear that these two galaxies are not alike in many details. In fact, because our selection is agnostic to the past assembly history (Fig.~\ref{fig:progenitors} and Section~\ref{sec:mergers}), we can identify many Andromeda-like galaxies from TNG50. These are, by choice or necessity, rare among zoom-in simulation efforts to date \citep[see also a discussion on this by][]{Sotillo.2022}.

Remarkably, within the TNG50 MW/M31-like sample, it is possible to identify several simulated galaxies whose many $z=0$ integral and structural properties are consistent, singly or jointly, with those measured for the Galaxy and/or Andromeda. The majority are hosted by haloes whose total mass is compatible with current inferences for the Milky Way (Fig.~\ref{fig:stellarhalomass}); the sample includes 3 Local Group (LG)-like systems (Section~\ref{sec:LG} and Fig.\ref{fig:LGlike}); many of them have satellite populations consistent, in abundance and galaxy properties, with those of the Galaxy and Andromeda (Section~\ref{sec:sats}, Fig.~\ref{fig:sats}, \citealt{Engler.2021, Engler.2023}); and a substantial number of TNG50 MW/M31 analogs host one or more satellites as massive as e.g. the Magellanic Clouds (Section~\ref{sec:massivesats}).

The TNG50 MW/M31 analogs are surrounded by gaseous haloes (Section~\ref{sec:cgm}) whose total CGM mass is consistent with observational estimates for the Milky Way \citep{Ramesh.2023} and that even exhibit X-ray emitting eROSITA/Fermi-like bubbles similar in morphology to those detected towards our Galactic Center (Fig.~\ref{fig:bubbles} and \citealt{Pillepich.2021}). Their stellar haloes are also diverse (Section~\ref{sec:stellarhaloes}), with many of them showing 3D power-law slopes of the stellar mass density profiles consistent with those measured to be very steep for the Galaxy or more shallow for Andromeda \citep{Sotillo.2022}. Moreover, 60 per cent of the TNG50 MW/M31 analogs are star-forming (like the Milky Way) or just below the star-forming main sequence (like Andromeda; Fig.~\ref{fig:sf}). And their average stellar ages and metallicity are in the ball park of extra-galactic observations (Fig.~\ref{fig:stellarpops}), albeit varying widely from system to system.

Even within such a relatively narrow selection, TNG50 reveals a large galaxy-to-galaxy diversity, not only in terms of global and halo properties, but also in disk structures (Figs.~\ref{fig:disk}, \ref{fig:disk_HI_Halpha} and Section~\ref{sec:disks}) and the innermost regions (Section~\ref{sec:inner}). At least 60 per cent of the TNG50 MW/M31analogs exhibit a bar, with varying degrees of strength and lengths (Figs.~\ref{fig:center}, bottom right, and \ref{fig:barred}). The total matter surface density at the Sun-like location varies between 20-120 $\MSUN$\,kpc$^{-2}$ (16th-84th percentiles; Fig.~\ref{fig:totaldisk}, top) and yet local dark-matter densities of $0.2-0.5$\,GeV cm$^{-3}$ are a natural outcome of $10^{10.5-10.9}\MSUN$ TNG50 analogs (larger at larger galaxy masses, and overall consistent with observational estimates; see Fig.~\ref{fig:totaldisk}, bottom). Magnetic fields in the inner disk regions can be up to a few tens of $\mu$G strong (Fig.\ref{fig:disk_more_gas}), and are overall in accordance with estimates for the Galaxy, but their average strength profiles are not always monotonic.

It is important to emphasize that such a galaxy-to-galaxy diversity is predicted by TNG50 to be in place also at fixed galaxy stellar mass, i.e. it is not only the result of a 0.7 dex-wide stellar mass selection. For example, the stellar disk length (heights) of TNG50 MW/M31-analogs varies by up to a factor 6 (3) in 0.1 dex bins of stellar mass at the MW-like scale (Fig.~\ref{fig:stellardisks}). Suggestively, while we do not find a trend between galaxy stellar mass and bulge-to-total ratio within the sample, galaxies that host more massive SMBHs are also characterized by more massive bulges, comparable to those of Andromeda or the Galaxy, respectively (Fig.~\ref{fig:center}, bottom left). 

By combining the outcome of TNG50 with observational estimates taken at face value, it appears that Andromeda is a rather average disk galaxy for its stellar mass, whereas the Galaxy lies on the lower envelope of stellar disk lengths, stellar disk heights, and SMBH masses. TNG50 returns a few galaxies whose detailed stellar disk structural properties are compatible with the available measurements for the Galaxy and Andromeda (Section~\ref{sec:stellardisks} and Fig.~\ref{fig:stellardisks}), e.g. in stellar disk length (at least as measured so far -- see also \citealt{Boardman.2020}). However, no TNG50 galaxy exhibits a SMBH mass as small as that of the Milky Way (Fig.~\ref{fig:center}, top left).

As this summary implies, in this paper we have deliberately provided a general and holistic overview of the MW/M31-like galaxies realized by the TNG50 simulation, with the goal of showcasing the diverse number of possible scientific and astrophysical applications. More in-depth investigations on specific topics have been highlighted throughout, and many more are ongoing (see Table ~\ref{tab:papers}).

With this paper, we release a series of broadly useful data outputs and products ($\sim$ 10 TB in total) that build upon the public release of the TNG simulations \citep{Nelson.2019}. These are described in detail in Section~\ref{sec:data} and include: the list of IDs and the key properties of our selected TNG50 MW/M31-like galaxies (Section~\ref{sec:data_MWM31scat} and Table~\ref{tab:files_ids}); the list of IDs and main properties of their galaxy satellites at $z=0$ (Section~\ref{sec:data_MWM31ssatellites}; and Table~\ref{tab:files_satsids}); and, importantly, datacubes of 800 comoving kpc per side of {\it all} particle-level simulation data, i.e. of cold dark matter, gas and magnetic fields, stars, tracer particles, and super massive black holes (Section~\ref{sec:data_cutouts} and Tables~\ref{tab:files_cutouts_1} and \ref{tab:files_cutouts_2}). These are provided for each of the 198 TNG50 MW/M31-like galaxies at $z=0$ and at all available previous points in time along their main progenitor branch. This public data release aims to maximally facilitate access and analysis by public users, enabling the community-wide exploitation of the costly TNG50 simulation for years to come.

\section*{Acknowledgements}

AP thanks Neige Frankel, Hans-Walter Rix, Christina Eilers, and Tommaso Zana for useful inputs; Gergo Popping, Gandhali Joshi, and Nhut Truong for producing their HI/H2, stellar assembly times, and X-ray emission catalogs, respectively; and Yetli Rosas-Guevara and Tommaso Zana for making their post-processing catalogs publicly available. The authors also thank all the simulator colleagues who have provided the exact number of MW/M31-like galaxies in their simulations for Table~\ref{tab:sims}.
This work was funded by the Deutsche Forschungsgemeinschaft (DFG, German Research Foundation) -- Project-ID 138713538 -- SFB 881 (``The Milky Way System'', subprojects A01 and A06).
RR and DN acknowledge funding from the Deutsche Forschungsgemeinschaft (DFG) through an Emmy Noether Research Group (grant number NE 2441/1-1).
RR and DSR are Fellows of the International Max Planck Research School for Astronomy and Cosmic Physics at the University of Heidelberg (IMPRS-HD).
The TNG50 simulation was run with compute time granted by the Gauss Centre for Supercomputing (GCS) under Large-Scale Projects GCS-DWAR on the GCS share of the supercomputer Hazel Hen at the High Performance Computing Center Stuttgart (HLRS). This analysis has been carried out on the VERA supercomputer of the Max Planck Institute for Astronomy (MPIA), operated by the Max Planck Computational Data Facility (MPCDF), where the released data are also stored.

\section*{Data Availability}

All the simulation data of the IllustrisTNG runs, including TNG50, are publicly available and accessible at \url{https://www.tng-project.org/data} \citep[][]{Nelson.2019}. With this paper, we release added-value catalogs and products for the MW/M31-like galaxies of TNG50, as described in Appendix~\ref{sec:data}: these are stored and accessible at the permanently available webpage: \url{https://www.tng-project.org/data/milkyway+andromeda/}.


\bibliographystyle{mnras}
\bibliography{MWM31like}


\appendix

\section{Data products}
\label{sec:data}

With this paper, we release a series of value-added and supplementary data products that build upon the original data release of all the TNG simulations\footnote{ \url{https://www.tng-project.org/data/}} \citep{Nelson.2019}. With this new material, we aim at providing the community with even richer and easier-to-use simulation outputs. By design, this release focuses exclusively on the TNG50 MW/M31-like galaxies selected and studied here.

\subsection{Preliminary considerations about the existing TNG50 data products and infrastructure}
\label{sec:fullvolumedata}
The TNG50 data is stored and organized exactly as all other TNG simulations and is already fully publicly accessible: see \citet{Nelson.2019b, Pillepich.2019} and \citet{Nelson.2019}, their Section 3, for details. Here we summarize salient features to orient the novel readers and users.\\

{\it Snapshots and catalogs.} Data associated to the whole TNG50 volume is available at 100 snapshots, i.e. it has been stored at 100 points in time, between $z=15$ and $z=0$ \citep[see e.g. Table 3 of][or online]{Nelson.2019}. The separation between two subsequent snapshots at low redshift is about 150 Myr. 
At any of these points in time, data is provided in a few main flavours:
\begin{itemize}
\item particle data, i.e. snapshots;
\item halo and galaxy catalogs;
\item merger trees;
\item supplementary data catalogs.
\end{itemize}

By particle data we mean properties of all gas cells and of all the DM, stellar, tracer, and SMBH particles throughout the ($\sim$50 com Mpc)$^3$ simulated volume. These include, for example, their 3D coordinates and velocities in the reference system of the comoving box, the total gravitational potential they sample, but also e.g. the internal energy and magnetic field content in the case of the gas and the mass, metallicity and metal abundances at birth for the stellar particles, etc. The complete list, including description and units, is given e.g. in Tables A.4-A.8 of \citet{Nelson.2015}, with full updates and descriptions in \citet{Nelson.2019} and online. For all TNG runs, there are two different types of snapshots: `full' and `mini', which both encompass the entire volume. However, the `mini' snapshots (80 in total) provide only a subset of particle/cell fields available in the full ones (20 in total). 

The halo and galaxy catalogs provide global and integral properties of FoF haloes and of their central and satellite subhaloes identified by \textsc{subfind}: if the latter have non-vanishing stellar masses, they are called galaxies. These catalogs contain quantities like the galaxy stellar mass, the total halo mass, the instantaneous SFR, and others, on an object-by-object basis. Again, the full list of provided quantities is given online, at the same 100 points in time as the snapshots above.

The merger trees allow to connect individually-identified (sub)haloes and galaxies across cosmic time, i.e. across the snapshots at which data is stored. 

Additional files include technical inputs for I/O, i.e. for accessing subsets of the snapshot data, which can amount otherwise up to 2.7 TBytes for each TNG50 snapshot: for example, it is possible to access only the gravitationally-bound stellar particles of a given galaxy with a given SubhaloID. 

Finally, over the course of the past years, analyses of the simulations data have allowed the collection and documentation of a plethora of additional quantities and derived properties, either at the particle or at the halo/galaxy level. These include, for example, a diversity of measurements of galaxy sizes and morphological properties based on the particle data or on photometry mock images, results of kinematic-decomposition algorithms, more detailed summary statistics of the galaxies' merger histories, but also the characterization of the large-scale cosmic web environment. These are all listed and documented online\footnote{\url{https://www.tng-project.org/data/docs/specifications/}}.\\

{\it Other resolution runs.} The flagship TNG50 simulation that is the focus of this paper is also referred to as TNG50-1, to denote that it represents the highest-resolution run of the given sampled comoving volume (see Section~\ref{sec:res}). The same volume (and hence the MW/M31-like galaxies presented here) is also realized at three lower (i.e. poorer) resolution levels, in the runs called TNG50-2, TNG50-3, TNG50-4, with mass (spatial) resolution getting 8 (2) times larger at progressively lower-resolution levels \citep[see Table 1 of][]{Nelson.2019}. 

These additional runs allow us to quantify the effects of numerical resolution on the scientific and quantitative findings, if needed: see examples of such studies in the Appendix of \citet{Pillepich.2018, Pillepich.2018b, Pillepich.2019, Engler.2020} and others. For most scientific applications, we recommend using the TNG50 MW/M31-like data as provided, i.e. to consider the outcome and the properties of the TNG50 MW/M31-like galaxies as the best possible realizations of the TNG model at the TNG50 resolution -- this is because model choices and numerical resolution are to some extent degenerate with each other and not easy to interpret. In fact, numerical convergence is a complex issue and working with simulations across multiple resolution levels can be challenging. However, should numerical-resolution tests be needed, then each TNG50 MW/M31-like galaxy can be contrasted to its lower-resolution counterparts. Importantly, the SubhaloIDs of the TNG50 MW/M31-like galaxies in TNG50 are {\it not} the same as in the lower-resolution simulations: the mapping is available from the ``Subhalo Matching Between Runs'' catalog.\\

{\it DM-only runs.} Moreover, each run of the TNG50 volume has also been performed including only cold DM and gravity: these are called DM-only runs and allow us to quantify the effects of baryonic physics and galaxy formation on the phase-space properies of DM (see e.g. results described in Section~\ref{sec:haloes}). Also in this case, the SubhaloIDs of the TNG50 MW/M31-like galaxies between full-physics and DM-only realizations are not the same, but are provided. 

\subsubsection{TNG50 MW/M31-like galaxies in TNG50 subboxes}
\label{sec:subboxes}
For analysis studies that require finer temporal spacing across snapshots than about $100-200$ Myr, additional and separate cutouts of the full simulated volume exist for all TNG runs, including TNG50. These are called `subboxes' and are spatial cutouts of fixed comoving location and size. 

There are three subboxes in the TNG50 volume, whereby the whole particle data therein are stored 3600 times between $z=49$ and today, i.e. every few Myr at recent times. These are visualized also in Fig.~\ref{fig:environment}, as colored shaded cubes in projection. Subbox0 is meant to include a handful of MW-mass haloes at $z=0$, with a cubic volume of about 6 comoving Mpc a side; Subbox1 was chosen to sample a low-density region, with many dwarf galaxies at $z=0$ and a volume of about 6 comoving Mpc a side; Subbox2 is centered around the most massive TNG50 cluster at $z=0$ and encompasses a cube of about 7.5 comoving Mpc a side. More details are given in Tables 4 and 5 of \citet{Nelson.2019} and online.

\begin{table}
\input{table_subboxes}
\end{table}

Now, as these are comoving spatial cubes out of the whole TNG50 periodic-boundary volume, individual galaxies may move in and out of these subboxes along their cosmic evolution. At $z=0$, 4 of the 198 TNG50 MW/M31-like galaxies are located in subboxes and have been in there since billions of years: these are listed in Table \ref{tab:subboxes} and are all centrals. An example of time-evolution study of two of these galaxies exhibiting eROSITA/Fermi-like bubbles in their CGM can be found in \citet{Pillepich.2021}.

\subsection{IDs and properties of TNG50 MW/M31-like galaxies}
\label{sec:data_MWM31scat}
\begin{table*}
\input{table_files_ids}

\end{table*}

With this paper, we release the list of Subhalo i.e. \textsc{Subfind}IDs of the TNG50 MW/M31-like galaxies identified in Section~\ref{sec:selection} and presented throughout this paper. With this, we invite colleagues to further explore the simulated galaxies, to apply additional selection criteria as required according to their targeted scientific application, or to compare to other choices. As discussed in Section~\ref{sec:selection}, the fiducial selection for MW/M31 analogs from TNG50 aimed at returning an as-complete-as-possible sample, even though it may not be as pure as certain analyses may require. 

The table is given in hdf5 format and includes a number of flags and galaxy properties extracted for this or other analyses, at $z=0$: the list and basic description are provided in Table~\ref{tab:files_ids}. Additional integral properties of the selected galaxies can be obtained by using the particle-level data -- see next.

\subsection{Snapshot data in cutouts for each TNG50 MW/M31 analog}
\label{sec:data_cutouts}
\begin{table*}
\input{table_files_cutouts_1}

\end{table*}
\begin{table*}
\input{table_files_cutouts_2}
\end{table*}

For each of the 198 MW/M31-like galaxies at $z=0$, with this paper we provide a replica of all the TNG50 snapshot data by selecting {\it all particles and cells} within a cube of $\pm$400 comoving kpc from the center of each galaxy. The latter is denoted as SubhaloPos and is defined as the location of the most gravitationally bound particle or cell associated to the galaxy, according to the \textsc{Subfind} algorithm. 

We do so for each selected galaxy at both the $z=0$ snapshot as well as across all previous snapshots up to $z=7$ (snapshot 011), keeping the size of the cube fixed at $\pm$400 comoving kpc and choosing the center of the box according to the position of the main branch progenitor of the galaxy -- the main progenitor branch is obtained by using the {\textsc SublinkGal} merger tree \citep{Rodriguez-Gomez.2015}. Redshift $z=7$ is the first snapshot where {\it all} the progenitors of the $z=0$ MW/M31 analogs have a non-zero stellar mass.

Crucially, these cutouts offer a few advantages with respect to the way data can be accessed with the standard TNG I/O approaches: 
\begin{itemize}
\item Firstly, we provide {\it all} the particle data of, and around, each MW/M31-like galaxy, namely irrespective of the functioning and results of the \textsc{Subfind} and FoF halo finders.

\item We provide the data of these particular galaxies as a function of time, without the need to get access to the merger tree files. 

\item Finally, and most importantly, we augment these cutouts with fields, i.e. properties, that are either not publicly available or are distributed across disjoint supplementary catalogs.
\end{itemize}

In fact, the snapshots of the TNG50 MW/M31 analogs contain all the fields present in the original snapshot files (full or mini), as mentioned in Section~\ref{sec:fullvolumedata} and described online.

In addition to the standard fields, we provide essential data of all particles and cells (coordinates, velocities, magnetic field strength, etc) also in the coordinate system of the main galaxy, namely we implement a coordinate translation + rotation such that the center of the cutout is at the center of the galaxy and the z-axis is perpendicular to the plane of the disk i.e. aligned with the spin vector of the galaxy. This is measured as the angular momentum of star-forming gas and stars within twice the stellar half mass radius of each galaxy. 

Additionally, we collect in these cutouts also fields like the post-processed neutral atomic and molecular hydrogen mass per gas cell, and many quantities that describe the origin and assembly of stellar particles, such as their in-situ or ex-situ flags. 

All datasets of the snapshot cutouts of the TNG50 MW/M31-like galaxies are listed and described in Tables~\ref{tab:files_cutouts_1} and \ref{tab:files_cutouts_2}.


\subsection{Properties of the TNG50 MW/M31-like satellites}
\label{sec:data_MWM31ssatellites}
With this paper and in conjunction with \citealt{Engler.2023}, we also release a TNG50 MW/M31-like satellite catalog, namely a catalog with the SubhaloIDs and a selection of properties of all ``satellite'' galaxies around TNG50 MW/M31-like galaxies at $z=0$.

This satellite sample comprises all galaxies with $\MS\ge 5\times10^6\,\MSUN$ and located within a 3D radius of 300 physical kpc from any of the MW/M31 analog. Objects with SubhaloFlag$\neq$1 are excluded \citep[][and Section~\ref{sec:remarks}]{Nelson.2019}. This is the fiducial galaxy selection we have used in Section~\ref{sec:sats}, Fig.~\ref{fig:sats} and \citealt{Engler.2021, Engler.2023} and comprises a total of 1237 galaxies. It should be noted that this selection does not depend on the outcome of the FoF and \textsc{Subfind} halo finders and it is purely based on stellar mass and spatial location. All the particles and cells of these galaxies are automatically included in the custom data cutouts of Section~\ref{sec:data_cutouts}.

In Table~\ref{tab:files_satsids}, we describe the provided quantities: these are listed according to the MW/M31-like galaxy to which they are closest, which is defined as their host.

\begin{table*}
\input{table_files_satsids}

\end{table*}

\subsection{Scientific remarks and other practical inputs}
\label{sec:remarks}
Any user of the data that is released with this paper must be aware of the possible limitations and areas of caution that have been identified in past analyses. In fact, the TNG50 simulation, and the TNG model in general, have been shown to return realistic outcomes in many astrophysical regimes and to various levels of qualitative or quantitative agreement (see Sections~\ref{sec:intro} and \ref{sec:tng50} and references therein). The results put forward and referenced to in this paper add to this encouraging evidence for the particular case of MW/M31-like galaxies: in particular, they demonstrate that TNG50 is capable of returning systems whose many integral and structural properties are consistent, one or a few at the time, with those measured for two particular galaxies, the Milky Way and Andromeda. However, numerous areas still require additional study: the comparison with observational results is inherently complex, and the underlying galaxy formation model is necessarily simplified and hence involves limitations and uncertainties.

In the TNG data release paper \citep{Nelson.2019}, we have identified regimes of caution for the usage of the data and the interpretation of the ensuing scientific results: see their Section 5. Below we stress upon already-known tensions and limitations and include additional considerations relevant to the scientific exploration of MW/M31-like analogs with TNG50: 

\begin{itemize}
    \item {\it Cold and dense gas.} In the TNG simulations, and hence in TNG50, gas colder than $10^4$ K is not modeled and hence not realized. Gas denser than the effective threshold for star formation (0.13 atoms cm$^{-3}$) is labeled as star forming and its temperature is `effective' i.e. intermediate between cold and hot phases (see Section~\ref{sec:tng50} and references therein). Studies with TNG50 of the cold and dense ISM are therefore limited. Forms of molecular hydrogen and the atomic-to-molecular transitions can be however modeled in post processing \citep[see e.g.][]{Diemer.2018, Popping.2020} and we have included those estimates in this release (see Table~\ref{tab:files_cutouts_1}).
    \item {\it Stellar particles.} Stellar particles in TNG50 represent mono-age stellar populations of about $8\times10^4\,\MSUN$: although the model accounts for their aging, mass return and metal production and return, stellar particles do not represent individual stars. Mocks of stellar catalogs targeting specific surveys, such as with Gaia, can be however constructed in post processing \citep[see e.g.][for other simulations]{Grand.2018, Sanderson.2020}.
    \item {\it Subhaloes of non-cosmological origin.} As extensively discussed by \citet[][Section 5.2 therein]{Nelson.2019}, not all objects identified by \textsc{Subfind} and stored in the corresponding catalogs should be considered 'galaxies'. This is particular relevant when considering satellite galaxies, as some of them may not have formed in their own DM halo and have not a ``cosmological'' origin. Such objects can be easily discarded with the SubhaloFlag provided in the official TNG data release and in the cutouts of Section~\ref{sec:data_cutouts}. In stellar and gas surface mass density maps and stellar light maps, these objects appear like massive star clusters (e.g. in Fig.~\ref{fig:disk}). However, it is not clear whether their formation is physical or due to numerical artifacts and so users should treat them with caution. A quantitative analysis of their occurrence can be found in \citet{Boecker.2023}.
    \item {\it UV background at high z.} In all the TNG simulations, and hence also in TNG50, the UV background is switched on at $z = 6$ instead of following the gradual build-up since higher redshifts. This may impact the progenitors of TNG50 MW/M31-like galaxies and their satellites during the epoch of Reionization. 
    \item {\it Alpha abundances.} As shown and discussed by \citet{Naiman.2018}, the simulated stars in MW-like galaxies of the TNG100 simulation are too alpha-enhanced (specifically for [Mg/Fe]) in comparison to observations of our Galaxy. This is the case also in TNG50 and may be due to either too-low SNIa rates or too high $\alpha$-element yields from SNII adopted in the fiducial TNG model. As SNIa only contribute to the metal abundances, as yield stellar tables are notoriously uncertain, and as metals in the TNG model are passive scalars, i.e. they do not dynamically affect the physical evolution of the systems, the mismatch does not impact the functioning and outcome of the TNG50 simulation in general. In fact, analyses relying or focusing on abundance ratios can still be worthy and useful, especially when studied in relative terms and keeping in mind that they depend on the underlying choices of the yield tables \citep[see][and Table 2 and Fig. 1 therein]{Pillepich.2018}. A global rescaling in post processing of the alpha abundances is a possible way forward, especially given that the needed modifications to bring TNG50 stars in agreement with the measurements of the Milky Way are all acceptable within current observational constraints for SNIa rates and stellar yield tables \citep{Naiman.2018}.
    \item {\it SMBH masses.} As shown and discussed in Section~\ref{sec:smbhs}, there are no galaxies within the TNG50 MW/M31 analogs whose SMBH have a mass similar to that of SgrA*: scientific applications that rely on the specific mass value of the central SMBH may be limited.
\end{itemize}

To streamline and guide the usage of the data released here, we also note the following points\footnote{Many additional practical inputs can be obtained also in the Discussion Forum and FAQs at \url{https://www.tng-project.org/data/}.}: 

\begin{itemize}
    \item {\it Units.} The quantities in the data cutouts (Section~\ref{sec:data_cutouts}, Tables\ref{tab:files_cutouts_1} and \ref{tab:files_cutouts_2}) are given in code/simulation units for consistency with the particle-level fields present and documented in the original snapshot files: useful notes on conversions can be found online.
    \item {\it Metal abundances.} Individual abundances of nine species (H, He, C, N, O, Ne, Mg, Si, Fe) are provided for both gas cells and stellar particles. These are all metal mass fractions, i.e. the dimensionless ratios of mass in the given species to the total gas-cell or stellar-particle mass (see online documentation). An additional conversion is required to obtain number fractions in solar units, e.g. [Fe/H] or [Mg/Fe], where for example [Fe/H] = $\mathrm{log}_{10} (N_\mathrm{Fe}/N_\mathrm{H}) - \mathrm{log}_{10} (N_\mathrm{Fe,\odot}/N_\mathrm{H,\odot})$, where $N_\mathrm{Fe}$ and $N_\mathrm{H}$ are the fractional abundances of iron and hydrogen, respectively, and $N_\mathrm{Fe,\odot}$ and $N_\mathrm{H,\odot}$ correspond to the solar abundances.
    
    \item {\it Metal tagging.} The production and mixing of the sum of all metals produced by AGB stars, SNIa and SNII are {\it separately} followed in the TNG simulations and recorded for both gas cells and stellar particles \citep{Pillepich.2018}. Additionally, we keep track of the iron mass separately ``produced'' by SNIa and SNII (see e.g. maps in Fig.~\ref{fig:disk}). These are convenient features of the released data but should be used and interpreted with caution, as these tags do not track where a given heavy element was created, but rather identify the last star it was ejected from \citep[see notes in][]{Nelson.2019}.
    \item {\it Europium.} The TNG model also includes the injection of r-process material from neutron stars–neutron star mergers, which are modeled similarly as SNIa \citep{Pillepich.2018, Naiman.2018}. The total metal mass produced by NSNS is also tracked and stored for both gas cells and stellar particles. This allows to extract estimates of e.g. Europium: see \citet{Naiman.2018} and online documentation.
    \item {\it Stellar quantities at birth and stellar ages}. To circumvent the limitations of the time-wise sparsity of the data outputs, a few quantities are stored and provided for each stellar particle, such as the spatial position and velocity at the time of birth, particularly relevant to study e.g. radial migration. The latter is also provided so that the ages i.e. the formation times of the stellar particles are available irrespective of the time cadence of the output data. The temporal resolution of the stellar ages in TNG50 is as good as $\sim0.03$ Myr, and permits to extract star formation histories with exquisite time resolution (as in Figs.~\ref{fig:sf} and \ref{fig:progenitors}).
    
    \item {\it X-ray emission quantities.} The novel and additional X-ray emission quantities provided in the cutouts (Table~\ref{tab:files_cutouts_1}) are given as cooling rates. Here is an example of how to compute, e.g., the X-ray luminosity of each gas cell in the [0.5-2] keV band: $L_x[\rm{erg~s}^{-1}] = {\rm cooling~rate}[\rm{erg~ cm}^3$ s$^{-1}] \times n_e[cm^{-3}] \times n_h[$cm$^{-3}] \times V[$cm$^3]$, where $n_e$ and $n_h$ are hydrogen and electron number densities, and V is the volume of the gas cell.
    
    \item {\it Shock finder in subboxes.} Another attractive code feature of the TNG simulations, including TNG50, is the on-the-fly shock finder, which returns the energy dissipated via shocks and their Mach number. However, although the two shock finder fields exist also in the subboxes (Section~\ref{sec:subboxes}), they are not updated for each subbox output, but only occasionally: hence these two fields in the subboxes should {\it not} be used at present.
    \item {\it Additional useful catalogs.} A number of post-processing catalogs exist for the TNG50 simulation in general: these include additional quantities related to the TNG50 MW/M31-like sample of this paper. For example: various kinematic/morphological decompositions, SDSS-like spectra, MaNGA-like intregral-field-unit mocks, photometry for tens of broadbands from FUV to submillimetre, etc. These are available online.
\end{itemize}

\label{lastpage}
\end{document}

%% file: table_sims.tex
 \begin{center}
    \footnotesize
    \caption{{\bf Cosmological (magneto-)hydrodynamical simulations of MW/M31-like galaxies or Local Group-like (LG) systems in a $\Lambda$CDM scenario} currently pursued by the community and referenced in this paper. The list deliberately omits idealized and isolated-disk simulations, dark-matter only simulations -- even if they have paved the road for the subsequent hydrodynamical ones, especially in the context of MW-mass haloes --, and cosmological simulations with too poor resolution for galaxies and their stellar disks to be studied. Namely, here we include only cosmological models, i.e. simulations that start from cosmologically-motivated initial conditions on $>>$ 10s Mpc scales, which are run to $z=0$. We hence also list large-volume uniform-resolution projects (specifically those with baryonic mass resolution better than a few $10^7\MSUN$), as they naturally include MW/M31-mass galaxies. As it can be seen from the second column, it is only since about a decade that it has been possible to simulate MW-like galaxies, i.e. with relatively thin 
    stellar disks. The underlying numerical codes are varied, and mostly differ for the adopted  schemes to discretize the collisional component and to numerically solve for the (coupled) equations of (gravity and) hydrodynamics. They span lagrangian and smooth particle hydrodynamics (GASOLINE, GADGET-3, ChaNGa), eulerian and adaptive mesh refinement (RAMSES) and meshless finite-mass (GIZMO) and moving-mesh (AREPO) codes. Also the assumptions on the precise value of the cosmological parameters have evolved through the years. However, what really distinguishes the outcome of these simulations, and hence their galaxies, are the different implementations of the models for star formation, stellar evolution and heavy element production, radiative (metal) cooling, stellar feedback, and SMBH feedback. The details and sophistication of these models differ across simulations, as do their results. Not all simulations include feedback from SMBHs: see 5th column. Only the Auriga, Hestia, and the TNG simulations (TNG100, TNG300, TNG50) include MHD. None include cosmic rays \citep[but see][]{Buck.2020b}. By ``$\#$ MW/M31-like galaxies'', here we mean: for zoom-in simulations, the number of targeted haloes/galaxies (see text for more details on the variously adopted selection criteria); for uniform-resolution box simulations, we quote the number of central galaxies with stellar mass in the $10^{10.5-11.2}\MSUN$, irrespective of morphology.}
\label{tab:sims}

    \begin{tabular}{l c c c c c c c}
        \hline
        Simulation(s) & Year & Code & Technique & SMBH & $\#$  MW/M31-like &$M_{\rm baryon}^{\bigstar}$ & Simulation/Method \\
         & & & & feedback & galaxies & $[\MSUN]$ & Reference(s)\\
        \hline
        
        & & & & & & & \\
        Eris 				&2011 	&GASOLINE 	&zoom 		& no  		& 1 							& $2 \times 10^{4}$ 					&\cite{Guedes.2011} \\
        Agertz's			&2011	&RAMSES	&zoom 		& no 		& 1							& --								&\cite{Agertz.2011}\\
        Few's 				&2012	&RAMSES 	&zoom  		& no  		& 19 							& --								&\cite{Few.2012} \\
        MaGICC 			&2013 	&GASOLINE 	& zoom 		& no  		& 1 							& $2.2 \times 10^{5}$				&\cite{Stinson.2013} \\
        CLUES 			&2014 	&GADGET-3 	&zoom 		& no  		& 1 $^{\vartriangle,\Diamond}$ 		& $5.5 \times 10^{5}$ 				&\cite{Nuza.2014}\\
        Aquarius with Gas	&2014 	&AREPO 		&zoom 		& yes  	& 8  							& $4.1 \times 10^{5}$				&\cite{Marinacci.2014a}\\
        APOSTLE 		&2016  	&GADGET-3 	& zoom 		& yes  	& 12 $^{\vartriangle}$ 			& $1.0 \times10^{4}$ 				&\cite{Sawala.2016, Fattahi.2016}\\
        Latte 			&2016 	&GIZMO 		& zoom 		& no  		& 1 							& $7.1\times10^3$							&\cite{Wetzel.2016} \\
        Auriga L4		&2017 	&AREPO 		&zoom 		& yes  	& 30  						& $5 \times10^{4}$ 					&\cite{Grand.2017} \\        
        Auriga L3		&2017 	&AREPO 		&zoom 		& yes  	& 3  						& $6 \times10^{3}$ 					&\cite{Grand.2017} \\        
        FIRE-2 Suite 		&2017  	&GIZMO 		& zoom 		& no  		& 7 							& $(4.2-7.1)\times10^3$ 					&$\blacklozenge$\\
        ELVIS on FIRE 		&2019  	&GIZMO 		&zoom 		& no  		& 3 $^{\vartriangle}$ 				& $(3.5-4.0)\times10^3$						&\cite{Garrison-Kimmel.2019,Garrison-Kimmel.2019b}\\        
        ARTEMIS 			&2020  	&GADGET-3	&zoom 		& yes  	& 42  						& $3.2 \times 10^{4}$ 				&\cite{Font.2020}\\        
        NIHAO-UHD 		&2020  	&GASOLINE2 	&zoom 		& no  		& 6  							& $(2.0-9.4)\times 10^{4}$ 				&\cite{Buck.2020}\\        
        VINTERGATAN 	&2020 	&RAMSES 	&zoom 		& no  		& 1  							& 7070 								&\cite{Agertz.2021} \\        
        Hestia Int. Res		&2020  	&AREPO 		&zoom 		& yes  	& 13 $^{\vartriangle,\Diamond}$ 	& $1.8 \times 10^{5}$ 				&\cite{Libeskind.2020}\\       
        Hestia High Res	&2020  	&AREPO 		&zoom 		& yes  	& 3 $^{\vartriangle,\Diamond}$ 		& $2.2 \times 10^{4}$ 				&\cite{Libeskind.2020}\\       
        DC Justice League 	&2020  	&ChaNGa 	&zoom 		& yes  		& 2  							& $3.3 \times 10^{3}$				&\cite{Applebaum.2021}\\   
        Auriga L2		&2021 	&AREPO 		&zoom 		& yes  	& 1  						& $800$ 					&\cite{Grand.2021} \\        
        & & & & & & \\
        & & & & & & \\
        Illustris			&2014	&AREPO 		&unif.res. 		& yes  	& $1518$					& $1.3 \times 10^{6}$				&$\spadesuit$\\
        Magneticum 4uhr	&2014	&GADGET-3 	&unif.res. 		& yes  	& 389						& $1.0 \times 10^{7}$				&\cite{Hirschmann.2014}	\\
	Eagle			&2015	&GADGET-3 	&unif.res. 		& yes  	& 812					& $1.8 \times 10^{6}$				&\cite{Schaye.2015, Crain.2015}	\\
	MassiveBlack-II		&2015	&GADGET-3 	&unif.res. 		& yes  	& 1427					& $3.1 \times 10^{6}$				&\cite{Khandai.2015}	\\
	Horizon-AGN		&2015	&RAMSES 	&unif.res. 		& yes   & $\sim{2000}$					& $1.4\times10^6$					&\cite{Dubois.2014}	\\
	Romulus 25		&2016	&ChaNGa 	&unif.res. 		& yes 	& $\sim{20}$ 					& $2.1 \times 10^{5}$				&\cite{Tremmel.2017}	\\
	MUFASA			&2016	&GIZMO 		&unif.res. 		& yes 	& 237						& $1.8 \times 10^{7}$				&\cite{Dave.2016}	\\
	TNG100			&2017	&AREPO 		&unif.res. 		& yes  	& 1606					& $1.4 \times 10^{6}$				&$\clubsuit$	\\
	TNG300			&2017	&AREPO 		&unif.res. 		& yes  	& 23470				& $1.1 \times 10^{7}$				&$\clubsuit$	\\
	FABLE			&2018	&AREPO 		&unif.res. 		& yes  	& 217						& $9.4 \times 10^{6}$				&\cite{Henden.2018}	\\
	Simba 100		&2019	&GIZMO 		&unif.res. 		& yes  	& 2075						& $1.8 \times 10^{7}$				&\cite{Dave.2019}	\\
	\bf{TNG50	}		&\bf{2019}	&\bf{AREPO} 	&\bf{unif.res.} 	& \bf{yes} 	& \bf{198$^{\square}$}  					& \bf{$8.5 \times 10^{4}$}				&\bf{\cite{Pillepich.2019, Nelson.2019b}}\\     
	{\textsc NewHorizon} 		&2020	&RAMSES 	&unif.res. 		& yes 	& $\sim{10}$					& $1.3\times10^4$								&\cite{Dubois.2021}	\\	
	FIREbox 			&2022	&GIZMO 		&unif.res. 		& no		& 53					& $6.3 \times 10^{4}$				&\cite{Feldmann.2022}	\\
	& & & & & & \\       
        \hline
    \end{tabular}   
 \end{center}
$^{\bigstar}$ If constant, the gas particle mass; if variable, the target/average gas cell mass for Lagrangian-type simulations; for grid/Eulerian simulations, this value is not generally comparable. For the latter, when available, we provide the stellar mass resolution or an equivalent estimate.  \\

$^{\vartriangle}$ Local-Group systems, i.e. each with one MW-mass and one M31-mass galaxy \\

$^{\Diamond}$ Constrained initial conditions \\
$\blacklozenge$ \cite{Garrison-Kimmel.2017, Hopkins.2018,Samuel.2020,Garrison-Kimmel.2019}\\

$\spadesuit$ \cite{Vogelsberger.2014a,Vogelsberger.2014b, Genel.2014,Sijacki.2015,Nelson.2015}\\

$\clubsuit$ \cite{Pillepich.2018b, Nelson.2018, Springel.2018, Marinacci.2018,Naiman.2018, Nelson.2019}\\

$^{\square}$ 198 is the number of MW/M31-like galaxies of TNG50 presented throughout this paper and that satisfy the conditions advocated in Section~\ref{sec:selection} and Fig.~\ref{fig:selection}. The total number of galaxies with stellar mass in the $10^{10.5-11.2}\MSUN$ range is 324, including also non-disky and galaxies in group or cluster hosts: of these 325 in TNG50, 221 are centrals.


%% file: table_papers.tex
 \begin{center}
   
    \caption{{\bf Existing and upcoming scientific analyses that focus on TNG50 MW/M31-like galaxies}, in addition to this reference paper and focusing on more specific and selected topics. We give them in chronological order and include only works based on the galaxy selection proposed here. This list will be kept up to date online, for future reference. All this is in addition to analyses of MW/M31-like galaxies in TNG100 and TNG300 and to scientific publications that use TNG50 MW/M31-like systems that either adopt different selections or do not fully focus on them: a full list of scientific results based on the IllustrisTNG data can be found online.}
\label{tab:papers}

    \begin{tabular}{l c }
        \hline
        Reference & Focus/topic \\
        \hline
        & \\
        \textcolor{blue}{Pillepich et al. (2023)} & overall introduction \\
            (this paper) & \\
        & \\
        \cite{Engler.2021} & abundance of satellites \\
        \cite{Pillepich.2021} & X-ray eROSITA-like CGM bubbles\\
        \cite{Gargiulo.2022} & stellar bulges and bars \\
        \cite{Sotillo.2022} & disk survival through mergers \\
        \cite{Carollo.2023} & formation of very metal-poor disky stars \\
        \cite{Engler.2023} & satellites' star formation and gas content \\
        \cite{Ramesh.2023} & properties of the CGM \\
        \cite{Chen.2023} & location of extremely metal-poor stars \\
        \textcolor{blue}{Ramesh et al. (2023b)} & high-velocity and cold clouds in the CGM\\
        \textcolor{blue}{Sotillo-Ramos et al. submitted} & flaring of the stellar disk \\
        \textcolor{blue}{Bisht et al. in prep} & stellar radial migration \\
        \textcolor{blue}{Sotillo-Ramos et al. in prep} & ex-situ disks \\
        
        & \\
        \hline
        \end{tabular}   
 \end{center}


%% file: table_subboxes.tex
\begin{center}
    \caption{Subsample of the 198 TNG50 MW/M31-like galaxies that are located within the boundaries of the high time-resolution subboxes of the TNG50 volume at $z=0$. In the subboxes, the data are stored every few million years (see Tables 4 and 5 of \citealt{Nelson.2019}), allowing time evolution analyses that require high time output cadence.} 
    \label{tab:subboxes}
    \begin{tabular}{l c c c}
        \hline
        SubhaloID & Subbox & In Subbox since $z$ & Notes \\
        \hline
        537941& 0& 15& -\\
        543114& 2& 0.26& barred, S0\\
        565089& 2& 0.42& barred\\
        613192& 2& 1.36& barred, S0\\
        \hline
    \end{tabular}   
\end{center}

%% file: table_files_ids.tex
\begin{center}
    \caption{Description of all fields in the ``TNG50 MW/M31-like catalog'' released with this paper. All fields are float32 unless otherwise specified. The Flag fields are all either 1 or 0. All fields are 198 long: namely, for each field, there is at least one entry per TNG50 MW/M31 analog. All properties refer to $z=0$ and are listed here in conceptual order. Many more properties of the TNG50 MW/M31-like galaxies can be either found in the official \textsc{Subfind} catalogs (by matching the IDs/indexes) or can be computed and measured from the particle-level data also released here (Section~\ref{sec:data_cutouts}).} 
    \label{tab:files_ids}
    \begin{tabular}{p{3cm} p{1cm} p{12.5cm}}
        \hline
        Field & Units & Description \\
        \hline
        \textsc{Subfind}ID &  - & $z=0$ ID of the galaxy, in the space of the {Subfind} i.e. Subhalo catalog of the whole TNG50 simulation. It is 0 indexed (int32). \\
        FlagCentral &  - & 1 if the galaxy is the central i.e. the first/main/most massive Subhalo object of its FoF halo (int8).\\
        FlagDiskyVisual &  - & 1 if the galaxy appears disky and with spiral arms based on the authors' visual inspection of 3-band stellar light images, both face-on and edge-on. See Section~\ref{sec:criterion_diskyness} (int8).\\
        FlagDiskyMinor2Major &  - & 1 if the stellar-mass minor-to-major axis ratio is smaller than a certain value: $c/a \le 0.45$. Based on galaxy shapes from \citet{Pillepich.2019}. See Section~\ref{sec:criterion_diskyness} (int8)\\
        FlagDisky & - & 1 if FlagDiskyVisual=1 OR FlagDiskyMinor2Major=1. Namely 1 if the galaxy has a stellar disky morphology either based on visual inspection or based on the stellar axis ratios. All entries in this catalog should be equal to 1. See Section~\ref{sec:criterion_diskyness} (int8)\\
        FlagIsolated &  - & 1 if there is no galaxy with $M_*(<30{\rm kpc})\ge 10^{10.5}\,\MSUN$ within 500 kpc distance \& if $\MTWOC ({\rm host}) <10^{13}\,\MSUN$. See Section~\ref{sec:criterion_environment} (int8)\\
        FlagMstars &  - & 1 if $M_*(<30{\rm kpc}) = 10^{10.5-11.2}\,\MSUN$. See Section~\ref{sec:criterion_mstars} (int8)\\
        FlagM200c &  - & 1 if $\MTWOC ({\rm host}) = 6\times10^{11}-2\times10^{12}\,\MSUN$. See Section~\ref{sec:selection_comp} (int8)\\
        FlagNoVirgo &  - & 1 if there is no FoF/Group halo with $\MTWOC \gtrsim 10^{14}\,\MSUN$ within 10 Mpc distance. For TNG50, this includes a constraint on the two most massive systems in the box at $z=0$. See Fig.~\ref{fig:environment} and Section~\ref{sec:environment} (int8)\\
        FlagS0 &  - & 1 if the galaxy appears as an S0 per visual inspection, namely it has a disky stellar morphology but with no manifest spiral arms. See Section~\ref{sec:criterion_diskyness} (int8)\\
        FlagBarredZana22 &  - & 1 if the galaxy has a bar according to \citet{Zana.2022}. See Section~\ref{sec:bars} (int8)\\
        FlagBarredRosas22 &  - & 1 if the galaxy has a bar according to \citet{Rosas-Guevara.2020, Rosas-Guevara.2022}. See Section~\ref{sec:bars} (int8)\\
        & & \\
        FlagMWM31 &  - & 1 if FlagDisky=1 \& FlagMstars=1 \& FlagIsolated=1. This is the fiducial selection proposed and discussed in this paper. All entries in this catalog should be equal to 1. See Fig.~\ref{fig:selection}. Other more restrictive selections can be imposed using the provided additional flags (int8)\\
        & & \\
        StellarMass\_30kpc & $\MSUN$& Galaxy stellar mass evaluated by summing up the mass of all stellar particles that are gravitationally bound according to \textsc{Subfind} and within a 3D radius of 30 kpc from the galaxy center. It is the fiducial measure used throughout this paper. \\
        StellarMass\_2r1/2 & $\MSUN$& Galaxy stellar mass evaluated by summing up the mass of all stellar particles that are gravitationally bound according to \textsc{Subfind} and within a 3D radius of twice the stellar half mass radius. This is equivalent to the 5th entry of SubhaloMassInRadType in the official \textsc{Subfind} catalogs \citep{Nelson.2015, Nelson.2019}.\\
        StellarMass\_all& $\MSUN$& Galaxy stellar mass evaluated by summing up the mass of all stellar particles that are gravitationally bound according to \textsc{Subfind}, with no distance limit. This also includes the mass in e.g. the stellar halo.\\
        HaloMass\_M200c & $\MSUN$& Total halo mass in terms of the spherical-overdensity measure $\MTWOC$. For non central galaxies, this is the halo mass of the host. \\
        HaloMass\_Mdyn & $\MSUN$& Total halo mass obtained by summing up the mass of all particles and cells that are gravitationally bound according to \textsc{Subfind}. \\
        HaloVirialRadius\_R200c & kpc & Virial radius of the underlying halo in terms of $\RTWOC$, in analogy with HaloMass\_M200c. \\
        SFR\_inst & $\MSUN$yr$^{-1}$ & Star formation rate of the galaxy based on the instantaneous SFR of the gas, i.e. sum of the individual star formation rates of all gravitationally-bound gas cells in this galaxy. It is equivalent to SubhaloSFR in the official \textsc{Subfind} catalogs \citep{Nelson.2015, Nelson.2019}. \\
        SFR\_50Myr & $\MSUN$yr$^{-1}$ & Time-averaged star formation rate of the galaxy including stars within a 3D aperture of 30 kpc, based on the stellar particles actually produced over the last 50 million years and using their initial mass at birth; as in \citet{Pillepich.2019}.\\
        DiskScaleLength & kpc & Exponential disk length of the galaxy obtained by fitting the stellar mass surface density of stellar particles in disky orbits between 1 and 4 times the stellar half mass radius. See errorbars and fitting procedure in \citet{Sotillo.2022} and \textcolor{blue}{Sotillo et al. submitted}. \\
        DiskScaleHeightThin\_8kpc & pc & Thin stellar disk height of this galaxy obtained by fitting the vertical stellar mass density distribution of disk stars with a double squared hyperbolic secant functional form in an annulus of $7-9$ kpc from the center. See errorbars, fitting procedure, and additional different measurements in \citet{Sotillo.2022} and \textcolor{blue}{Sotillo et al. submitted}. \\
        DiskScaleHeightThick\_8kpc & pc & Same as above, for the thick component. Thin and thick disk heights here are meant as geometrical. \\
        SMBH\_Mass & $\MSUN$& Mass of the central (i.e. most massive) SMBH in this galaxy. The measurement is based on the PartType5/BH\_Mass of the official snapshot data \citep{Nelson.2015, Nelson.2019}. For MW/M31-like galaxies, this mass is to all effects equivalent to SubhaloBHMass in the official \textsc{Subfind} catalog.  \\
        SMBH\_AccretionRate & $\MSUN$yr$^{-1}$& Instantaneous gas mass accretion rate into the SMBH of this galaxy, based on SubhaloBHMdot from the official \textsc{Subfind} catalog \citep{Nelson.2015, Nelson.2019}. \\
        SMBH\_EddingtonRatio & - & Instantaneous Eddington ratio of the SMBH of this galaxy.\\
        SMBH\_FeedbackMode & $\MSUN$& 0 if the SMBH of this galaxy is in thermal feedback mode, 1 if it is exercising kinetic feedback mode. See Sections~\ref{sec:tng50}, \ref{sec:sf}, and \ref{sec:smbhs} and references therein (int8).\\
        SMBH\_CumEnergy\_QM & erg & Cumulative amount injected into the surrounding gas by the central SMBH in the high accretion-state (quasar) mode, total over its entire lifetime. \\
        SMBH\_CumEnergy\_RM & erg & As SMBH\_CumEnergy\_QM but for the energy injected in the low accretion-state (wind or kinetic) mode.\\
        \hline
    \end{tabular}   
\end{center}

%% file: table_files_cutouts_1.tex
\begin{center}
    \caption{Description of the fields in the snapshot cutouts released with this paper (continues in Table~\ref{tab:files_cutouts_2}). There is one hdf5 file for each of the 198 TNG50 MW/M31-like galaxies at each redshift: we provide one cutout per galaxy, each from snapshot 011 to snapshot 099 (i.e. $z=0$) along the main progenitor branch of each galaxy: see text for details. Each cutout contains all particle-level fields present in the original snapshot files across the entire simulated volume (full or mini snapshots depending on the redshift; Section~\ref{sec:fullvolumedata}) and contained within a cube centred on the galaxy position (SubhaloPos) and extending $\pm400$ ckpc along each direction. However, with this release we provide additional quantities for the galaxies under focus that require substantial analysis or post-processing of the whole-simulation data. These additional fields are listed and described below. They are organized by particle type, i.e. PartType = 0, 1, 3, 4, and 5 for gas cells, DM particles, tracers, stars and wind particles, and SMBH particles, respectively. N is the number of each PartType in each file.} 
    \label{tab:files_cutouts_1}
    \begin{tabular}{p{2.5cm} c c c p{9.4cm} }
        \hline
        Field & Dimensions & Units & PartType(s) & Description \\
        \hline
        \textsc{Subfind}ID & N & - & all & The ID of the subhalo that a given particle/cell is a part of at the given redshift, in the space of the {Subfind} i.e. Subhalo catalog of the whole TNG50 simulation. It is set to -1 if the particle/cell does not belong to any subhalo (int32). \\
        MainSnapshotIndex & N & - & all & Index into the main snapshot file(s) of this particular particle/cell. This is useful if one wants to load data from one of the post processing quantities that are available online but are not included in these cutouts (int64). \\
        RotatedCoordinates & N,3 & ckpc$/h$ & all & Spatial position of each particle/cell in the reference system of the main galaxy: a coordinate transformation is first performed to shift to the frame of reference of the (centre of the) main galaxy (SubhaloPos); thus a rotation is performed by diagonalizing the moment of inertia tensor (or mass tensor) such that projecting along the z-axis yields a face-on view of the central galaxy. The mass tensor is computed from stars and star-forming gas within a given aperture: minimum of once (twice) the stellar half mass radius for stars (gas).\\
        RotatedCenterOfMass & N,3 & ckpc$/h$& PartType0 & Spatial position of the center of mass of each gas cell, after having applied the translation+rotation transformation as for RotatedCoordinates. \\
        RotatedVelocities & N,3 & km $\sqrt(a)/s$ & all & Spatial velocity of each particle/cell in the reference system where the main galaxy is at rest: a velocity transformation is first performed with respect to the bulk/peculiar velocity of the main galaxy through the simulated volume (SubhaloVel). Then a rotation is applied as for RotatedCoordinates. \\
        RotatedMagneticField & N,3 & see online
        & PartType0 & Magnetic field strength of each gas cell after having applied the transformation (rotation only) as for RotatedCoordinates. For the units see description online.\\
        MH & N & $10^{10}\MSUN$/$h$ & PartType0 & Total neutral hydrogen mass of each gas cell. It equals the sum of the mass of atomic HI plus molecular H2: MH = MHI + MH2 \citep{Popping.2020}\\
        MH2* & N & $10^{10}\MSUN$/$h$ & PartType0 & Molecular hydrogen mass of each gas cell, according to three different models and hence three different partitioning schemes (denoted in the names with * = 'GK', 'BR', 'KMT'), based on \citet{Popping.2020}; see also \citet{Diemer.2018}. \\
        Xray\_Emission\_03\_2keV & N & erg cm$^3$/s & PartType0 & Intrinsic and instantaneous X-ray emission of the gas in the soft broad band [0.3-2] keV. This includes contributions from both the continuum and the lines and is given as X-ray cooling rate. The cooling rate is physically similar to the gas field ``GFM\_CoolingRate'' and it is computed assuming an emission model ``APEC'' from the XSPEC package \citep{Smith.2001} using the element abundances traced by the simulation (practically it is done by using the VAPEC model in the XSPEC package).Available only at the $z=0$ snapshot. Based on \textcolor{blue}{Truong et al. 2023 to be submitted}.\\
        Xray\_Emission\_05\_2keV & N & erg cm$^3$/s & PartType0 & Same as above but for the [0.5-2] keV soft X-ray broad band. \\
        Xray\_Emission\_03\_2keV\_C & N & erg cm$^3$/s & PartType0 & Same as Xray\_Emission\_03\_2keV but for continuum emission only. \\
        Xray\_Emission\_05\_2keV\_C & N & erg cm$^3$/s & PartType0 & Same as Xray\_Emission\_05\_2keV but for continuum emission only.\\
        Xray\_Emission\_Line\_X &N & erg cm$^3$/s & PartType0 & Intrinsic and instantaneous X-ray emission of the gas (as Xray\_Emission\_03\_2keV) in narrow bands at nine specific lines, with X = CV (298.97 eV), CVI (367.47 eV), NVI (419.86 eV), NVII (500.36 eV), OVIIf (560.98 eV), OVIIr (573.95 eV), OVIII (653.49 eV), FeXVII (725.05 eV), and NeX (1021.5 eV). Based on \textcolor{blue}{Truong et al. 2023 to be submitted}.\\
        \hline
    \end{tabular}   
\end{center}

%% file: table_files_cutouts_2.tex
\begin{center}
    \caption{Continues from Table~\ref{tab:files_cutouts_1}, and focuses on the origin and assembly of stellar particles (PartType4). Here the dimension of all fields is equal to N = number of stellar particles in each cutout (we hence omit the column for brevity).} 
    \label{tab:files_cutouts_2}
    \begin{tabular}{p{3.7cm} c c p{9.8cm} }
        \hline
        Field & Units & PartType(s) & Description \\
        \hline
        SubfindIDAtFormation &  - & PartType4 & \textsc{Subfind}ID of the subhalo i.e. galaxy in which the stellar particle first appeared; $-1$ if it was formed outside of any subhalo (int16). \\
        SnapNumAtFormation &  - & PartType4 & Snapshot number in which the stellar particle first appeared (int16). \\
        InSitu &  - & PartType4 & $1$ if the stellar particle was formed in situ, $0$ if it was formed ex situ, and $-1$ if it does not currently belong to any subhalo. A stellar particle is considered to have been formed in situ if the subhalo in which it was formed lies along the ``main branch'' of the subhalo in which the stellar particle is currently found. This is decided using the SubLink ``baryonic'' merger trees (int8). Based on \citealt{Rodriguez-Gomez.2016}. \\
        AfterInfall &  - & PartType4 & 1 if the subhalo in which the stellar particle first appeared had already ``infalled'' into the halo \textsc{FoF group} where it is currently found; 0 otherwise; $-1$ if not applicable i.e., if the particle was formed in situ or if it was formed outside of any subhalo (int16). Based on \citealt{Rodriguez-Gomez.2016}.\\
        AccretionOrigin &  - & PartType4 &  This dataset can take the following integer values: 0, 1, and 2 for ex-situ stellar particles that were accreted from completed mergers (i.e., when the subhalo in which the stellar particle formed has already merged with the current subhalo), ongoing mergers (i.e., when the subhalo in which the stellar particle formed has not yet merged with the current subhalo, but will do so at a later snapshot in the simulation), and flybys (i.e., when the subhalo in which the stellar particle formed has not merged with the current subhalo, and will not do so at any future snapshot in the simulation), respectively; and $-1$ if not applicable i.e., if the particle was formed in situ or if it was formed outside of any subhalo (int8). Based on \citealt{Rodriguez-Gomez.2016}.\\
        MergerMassRatio & - & PartType4 & The stellar mass ratio of the merger in which a given ex-situ stellar particle was accreted (if applicable). The mass ratio is measured at the time when the secondary progenitor reaches its maximum stellar mass. NOTE: this quantity was calculated also in the case of flybys, without a merger actually happening. Based on \citealt{Rodriguez-Gomez.2016}.\\
        DistanceAtFormation &  - & PartType4 & The galactocentric distance when the stellar particle was formed, given in units of the stellar half-mass radius of the parent galaxy at the formation time.\\
        SnapNumAtStripping &  - & PartType4 & The snapshot in which the stellar particle last switched galaxies, and has hence remained in its present galaxy -- i.e. time of last stripping.  This field is $-1$ for all in-situ stellar particles and for particles that aren't part of any subhalo (int16).\\
        ProgGalaxyMassAtStripping &  $10^{10}\MSUN$/$h$ & PartType4 & Total stellar mass (as identified by \textsc{Subfind}) of the secondary progenitor galaxy of this stellar particle, from snapshot 'SnapNumAtStripping - 1'.\\
        ProgGalaxyMassInRadAtStripping &  $10^{10}\MSUN$/$h$ & PartType4 & As ProgGalaxyMassAtStripping but for the stellar mass within twice the stellar half mass radius of the secondary progenitor galaxy.\\  
        ProgSubhaloMassAtStripping &  $10^{10}\MSUN$/$h$ & PartType4 & As ProgGalaxyMassAtStripping but for the total subhalo mass of the secondary progenitor galaxy, i.e. the sum of all gravitationally-bound mass. \\ 
        DistanceToHostAtStripping &  ckpc$/h$ & PartType4 & Distance to the centre of the primary galaxy at snapshot SnapNumAtStripping. Here, the centre of the galaxy refers to the position of the most-bound particle of PartType = 4.\\
        \hline
    \end{tabular}   
\end{center}

%% file: table_files_satsids.tex
\begin{center}
    \caption{Description of all fields in the ``TNG50 MW/M31-like satellite catalog'' released with this paper. All fields are float32 unless otherwise specified. All fields are 1237 long: namely, for each field, there is one entry per satellite of the TNG50 MW/M31 analogs. All properties refer to $z=0$ and are listed here in conceptual order. The same considerations apply as for the TNG50 MW/M31-like main galaxies.} 
    \label{tab:files_satsids}
    \begin{tabular}{l l p{12cm}}
        \hline
        Field & Units & Description \\
        \hline
        \textsc{Subfind}IDSat &  - & $z=0$ ID of the satellite galaxy, in the space of the {Subfind} i.e. Subhalo catalog of the whole TNG50 simulation. It is 0 indexed (int32). \\
        \textsc{Subfind}IDHost &  - & $z=0$ ID of the host galaxy of this satellite, in the space of the {Subfind} i.e. Subhalo catalog of the whole TNG50 simulation. These can only take the values of the SubhaloIDs of the TNG50 MW/M31 analogs of Table~\ref{tab:files_ids}. The host is the closest main galaxy. It is 0 indexed (int32). \\
        & & \\
        DistanceToHost3D  & kpc & Three-dimensional distance of the satellite galaxy to the closest MW/M31-like hosts.\\ 
        DistanceToHost2D & kpc & Two-dimensional distance of the satellite galaxy  in a random projection to the closest MW/M31-like hosts (along z axis of the simulation box).\\   
        3DVelocityRelToHost & km/s & Three-dimensional relative velocity of the satellite galaxy with respect to the closest MW/M31-like hosts.\\
        LoSVelocityRelToHost & km/s & Line-of-sight relative velocity of the satellite galaxy  with respect to the closest MW/M31-like hosts, in a random projection (along z axis of the simulation box).\\
        StellarMass\_2r1/2 & $\log \MSUN$& Galaxy stellar mass evaluated by summing up the mass of all stellar particles that are gravitationally bound according to \textsc{Subfind} and within a 3D radius of twice the stellar half mass radius. This is equivalent to the 5th entry of SubhaloMassInRadType in the official \textsc{Subfind} catalogs \citep{Nelson.2015, Nelson.2019}.\\
        Magnitude\_VBand & mag & Absolute V-band Buser's magnitude of the satellite galaxy  based on the summed-up luminosities of all the gravitationally-bound stellar particles (Vega magnitudes), from the GFM\_StellarPhotometrics field of the official snapshot data \citep{Nelson.2015, Nelson.2019}.\\
        Magnitude\_rBand & mag & As for Magnitude\_VBand, but in the SDSS r band (AB magnitudes).\\
        Mdyn & $\MSUN$& Total halo mass of the satellite galaxy obtained by summing up the mass of all particles and cells that are gravitationally bound according to \textsc{Subfind}. \\
        GasMass\_2r1/2  & $\log \MSUN$&  Total gas mass of the satellite galaxy evaluated by summing up the mass of all stellar particles that are gravitationally bound according to \textsc{Subfind} and within a 3D radius of twice the stellar half mass radius. This is equivalent to the 1st entry of SubhaloMassInRadType in the official \textsc{Subfind} catalogs \citep{Nelson.2015, Nelson.2019}.\\
        HIGasMass\_2r1/2 & $\log \MSUN$& Atomic hydrogen mass of the satellite galaxy  evaluated by summing up the HI mass of all gas cells that are gravitationally bound according to \textsc{Subfind} and within a 3D radius of twice the stellar half mass radius. This is based on the 'GK' HI+H2 partitioning by \citet{Popping.2020}. \\
        H2GasMass\_2r1/2 & $\log \MSUN$& As HIGasMass\_2r1/2 but for the molecular hydrogen. \\
        HalfLightRadius2D\_VBand  & kpc & Two-dimensional circularized stellar half-light radius of the satellite galaxy  in the V-band, from a face-on projection. Based on \citet{Pillepich.2019}.\\
        StellarVelDisp\_3D & km/s & Three-dimensional standard deviation of the velocities of all stellar particles of the satellite galaxy within twice the stellar half-mass radius weighted by their respective stellar mass. \\
        SurfaceBrightness2D\_rBand & mag arcsec$^{-2}$& Two-dimensional surface brightness of the satellite galaxy in the SDSS r band (see above). \\
        Vmax & km/s & Maximum of the circular velocity profile of the satellite galaxy, accounting for {\it all} matter components. \\
        SFActivity & - & Flag denoting the star-formation activity of the satellite gaalxy: 1, 0, -1 for star-forming, quenched and green-valley galaxies, respectively. This is based on \citet{Pillepich.2019} and on the distance from the ridge of the star-forming main sequence (int8).\\  
        t10 & Gyr ago & Stellar assembly time $\tau_{10}$, i.e. time when the galaxy assembled 10 per cent of its $z=0$ stellar mass. Based on \citet{Joshi.2021}.\\   
        t50 & Gyr ago & As t10 but for the time when the galaxy assembled 50 per cent of its $z=0$ stellar mass. \\   
        t90 & Gyr ago & As t10 but for the time when the galaxy assembled 90 per cent of its $z=0$ stellar mass.\\
        \hline
    \end{tabular}   
\end{center}